\documentclass[prx,twocolumn,showpacs,amsmath,amssymb,superscriptaddress,floatfix, longbibliography]{revtex4-1}
\pdfoutput=1
\usepackage{amsmath}
\usepackage{amsfonts}
\usepackage{ graphicx } 
\usepackage{ epstopdf }
\usepackage{ bm } 
\usepackage{ color }
\usepackage{float}

\newcommand{\bra}[1]{\langle #1 |}
\newcommand{\ket}[1]{| #1 \rangle}

\begin{document}


\author{Jirawat Tangpanitanon}
\email{cqtjt@nus.edu.sg}
\affiliation{Centre for Quantum Technologies, National University of Singapore, 3 Science Drive 2, Singapore 117543}
\author{Supanut Thanasilp}
\affiliation{Centre for Quantum Technologies, National University of Singapore, 3 Science Drive 2, Singapore 117543}
\author{Marc-Antoine Lemonde}
\affiliation{Centre for Quantum Technologies, National University of Singapore, 3 Science Drive 2, Singapore 117543}
\author{Dimitris G. Angelakis}
\email{dimitris.angelakis@qubit.org}
\affiliation{Centre for Quantum Technologies, National University of Singapore, 3 Science Drive 2, Singapore 117543}
\affiliation{School of Electrical and Computer Engineering, Technical University of Crete, Chania, Greece 73100}


\title{Quantum supremacy with analog quantum processors for \\ material science and machine learning}

\date{\today}

\begin{abstract}
Quantum supremacy is the ability of quantum processors to outperform classical computers at certain tasks. In digital random quantum circuit approaches for supremacy, the output distribution produced is described by the Porter-Thomas (PT) distribution. In this regime, the system uniformly explores its entire Hilbert space, which makes simulating such quantum dynamics with classical computational resources impossible for large systems. However, the latter has no direct application so far in solving a specific problem. In this work, we show that the same sampling complexity can be achieved from driven analog quantum processors, with less stringent requirements for coherence and control. More importantly, we discuss how to apply this approach to solve problems in quantum simulations of phases of matter and machine learning. Specifically, we consider a simple quantum spin chain with nearest-neighbor interactions driven by a global magnetic field. We show how quantum supremacy is achieved as a consequence of the thermalization due to the interplay between the disorder and the driven many-body dynamics. We analyze how the achieved PT distribution can be used as an accessible reference distribution to probe the many-body localization (MBL) phase transition. In the second part of our work, we show how our setup can be used for generative modeling machine learning tasks. We propose a novel variational hybrid quantum-classical approach, exploiting the system's inherent tunable MBL dynamics, to train the device to learn distributions of complex classical data. The performance of our training protocol depends solely on the phase that the quantum system is in, which makes fine-tuning of local parameters not necessary. The protocol is implementable in a range of driven quantum many-body systems, compatible with noisy intermediate-scale quantum devices. 
\end{abstract}

\maketitle

\section{Introduction} 

Recent experimental advances strongly suggest that noisy intermediate-scale quantum (NISQ) devices with few hundred quantum nodes/qubits are soon to be available \cite{2018_preskill_quantum}. In principle, these devices can outperform classical computers in particular tasks by exploiting their intractably large Hilbert space, achieving the so-called quantum supremacy~\cite{2017_Harrow_Nat}. So far, the first steps toward the experimental characterization of quantum supremacy have been made with linear optical networks for boson sampling~\cite{2013_walmsley1_sci, 2013_Broome_Sci, 2013_walther_nat, 2013_fabio_natpho} and in driven disordered superconducting circuits~\cite{2018_neill_sci}. The latter is inspired by random quantum circuit proposals~\cite{2018_hartmut_natphy}, where the system undergoes chaotic evolution and uniformly explores its entire Hilbert space within its coherence time. 

Among the most promising applications for NISQ devices are hybrid quantum-classical variational approaches, commonly referred to as quantum approximate optimization algorithms
\cite{McClean_2016_njp, 2014_jeremy_natcom, 2017_ibm_nat, 2016_Omalley_PRX, 2018_Hempel_PRX, 2017_Li_PRX, 2018_Kokail_ArXiV, 2017_Otterbach_ArXiV, 2018_Zhu_ArXiV,2019_ibm}. 
The general idea is to use the output state of a quantum system as a variational function that is optimized for a given problem with the help of a classical feedback loop. These approaches have been used to solve problems in quantum chemistry~\cite{2014_jeremy_natcom, 2017_ibm_nat, 2016_Omalley_PRX, 2018_Hempel_PRX}, machine learning~\cite{2019_ibm, 2018_Zhu_ArXiV, 2017_Otterbach_ArXiV}, and high-energy physics \cite{2018_Kokail_ArXiV}. Similar to any variational techniques, their accuracy is tied to having an accurate ansatz, which in general, might not be known or easily implementable.
In those cases, the model should be flexible enough to capture the answer that may be far from the initial guess. In this aspect, quantum systems have been shown to have more expressive power, i.e., the ability to model complex functions, than standard classical tools such as artificial neuron networks \cite{2018_tao_arxiv,2018_lloyd_arxiv,2019_kashfi_arxiv}.
Hence, having access to a large Hilbert space and efficient training protocols are keys to demonstrate quantum supremacy in the learning performance.

In parallel with outstanding efforts in using digital approaches on NISQ platforms, the alternative paradigm of analog quantum simulators (AQSs)~\cite{2012_zoller_natphy, 2012_lewenstein_rpp,2014_dieter_epj} has already shown possible to reach regimes inaccessible to existing numerical techniques~\cite{2016_Choi_Sci}. AQSs are controllable quantum computers built to simulate specific quantum systems that are hard to simulate classically. Unlike universal quantum computers, they do not require error correction nor high-fidelity universal quantum gates, making their implementation easier and more versatile. For example, experimental platforms for AQSs include cold atoms in optical lattices \cite{2016_Choi_Sci, 2017_gross_sci}, trapped ions \cite{2012_blatt_np,2017_monroe_nat}, Rydberg atoms \cite{2017_lukin_nat}, superconducting circuits \cite{2012_koch_natphy}, quantum dots \cite{2017_vadersypen_nat}, defects in solid-state systems~\cite{2012_lukin_natcom}, nuclear magnetic resonance \cite{2017_andrea_natcomm}, and interacting photons \cite{2017_angelakis_rpp,2014_lukin_natphy}. 
Consequently, AQSs stand as a natural computing paradigm for NISQ devices.

In this work, we discuss how to efficiently achieve quantum supremacy in AQSs and its applications in probing phases of matter and machine learning. As in the case of random quantum circuits~\cite{2018_hartmut_natphy}, we characterize signatures of quantum supremacy by measuring the difference between the output distribution and the Porter-Thomas (PT) distribution~\cite{1956_pt}. The latter is a signature of quantum chaos and indicates that the system uniformly explores its intractably large Hilbert space. As a specific example of AQSs, we consider a simple isolated quantum Ising spin chain with nearest-neighbor interactions, periodically driven by a global magnetic field with disordered on-site energies. We show that the system reaches the PT distribution on a shorter timescale than random quantum circuits with the same connectivity~\cite{2018_hartmut_natphy}, as the drive effectively generates longer-range interactions and thus changes the topology of the architectures. 

As a direct application, we show how the achieved PT distribution can be used as a reference distribution to probe the driven many-body localization (MBL) phase transitions. The latter is a quantum phase transition from thermalized to many-body localized phases resulting from the interplay of interactions, the disorder, and the drive~\cite{2015_Ponte_PRL, 2016_Abanin_AoP, 2013_Alessio_AoP, 2014_Rigol_PRX, 2017_Roushan_Sci}. Our protocol bypasses the need to probe level statistics, which is very challenging experimentally, especially in driven systems~\cite{2017_Roushan_Sci}. 

For the second application, we show how AQSs can be trained to learn complex distributions of classical data, in an analogy of the unsupervised task known as generative modeling in machine learning~\cite{2004_tony}. 
We describe how the accuracy of our training protocol solely depends on the phase in which the quantum device operates so that fine-tuning is not required.  Due to its simplicity, our proposal is generic to driven quantum many-body systems and compatible with NISQ devices.


\section{Achieving quantum supremacy: \\ Analog vs Digital }

\subsection{The analog model} 

\begin{figure}
\includegraphics[width=0.95\columnwidth]{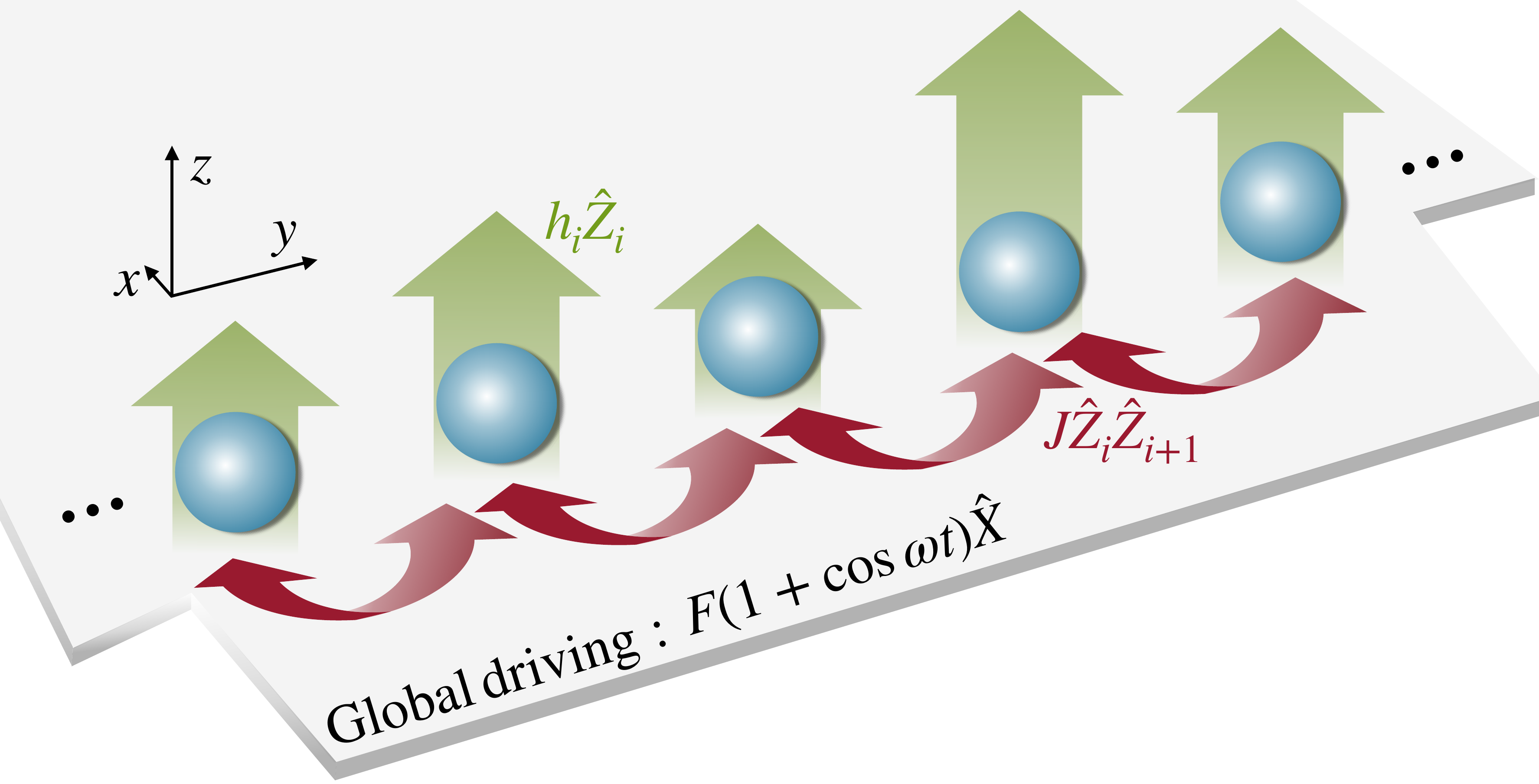}
\caption{Sketch of a driven and disordered quantum Ising chain with nearest-neighbor interactions. The green arrows of different lengths represent the disorder in the local magnetic field along $z$ and the wide gray arrow represents the global time-dependent magnetic field along $x$.}
\label{fig0}
\end{figure}

In this work, we consider driven Ising chains described by the Hamiltonian $\hat{H}(t) = \hat{H}_0 + f(t) \hat{H}_d$, with
\begin{align}
\hat{H}_0=\sum_{i=1}^L h_i \hat{Z}_i + J\sum_{i=1}^{L-1}\hat{Z}_i\hat{Z}_{i+1}, \text{ and } \hat{H}_d = F\sum_{i=1}^L\hat{X}_i
\end{align}
where $L$ is the number of sites, $\{\hat{X}_i,\hat{Y}_i,\hat{Z}_i\}$ are Pauli's spin operators at site $i$, $J$ is the interaction, and $F$ is the driving amplitude, as depicted in Fig. \ref{fig0}. We consider disordered magnetic fields along the $z$-axis with strengths $h_i$ being drawn from a uniform distribution ranging from $0$ to $W$, i.e.~$h_i \in \left[0,W\right]$. We choose a smooth sinusoidal periodic drive of frequency $\omega$
\begin{equation}
f(t)=\left[1-\cos(\omega t)\right]/2, 
\label{eq:ft}
\end{equation}
so that the resulting magnetic field along $x$ remains positive at all time. 
This simple model has been implemented in various quantum platforms, including Rydberg atoms \cite{2017_lukin_nat}, trapped ions \cite{2017_monroe_nat}, and superconducting circuits \cite{2017_Otterbach_ArXiV}. We note that quantum supremacy of two-dimensional Ising spin lattices without the drive, suitable for cold atoms in optical lattices, has been discussed in \cite{2018_eisert_prx,PhysRevLett.118.040502}.

\begin{figure*}
\includegraphics[width=0.95\textwidth]{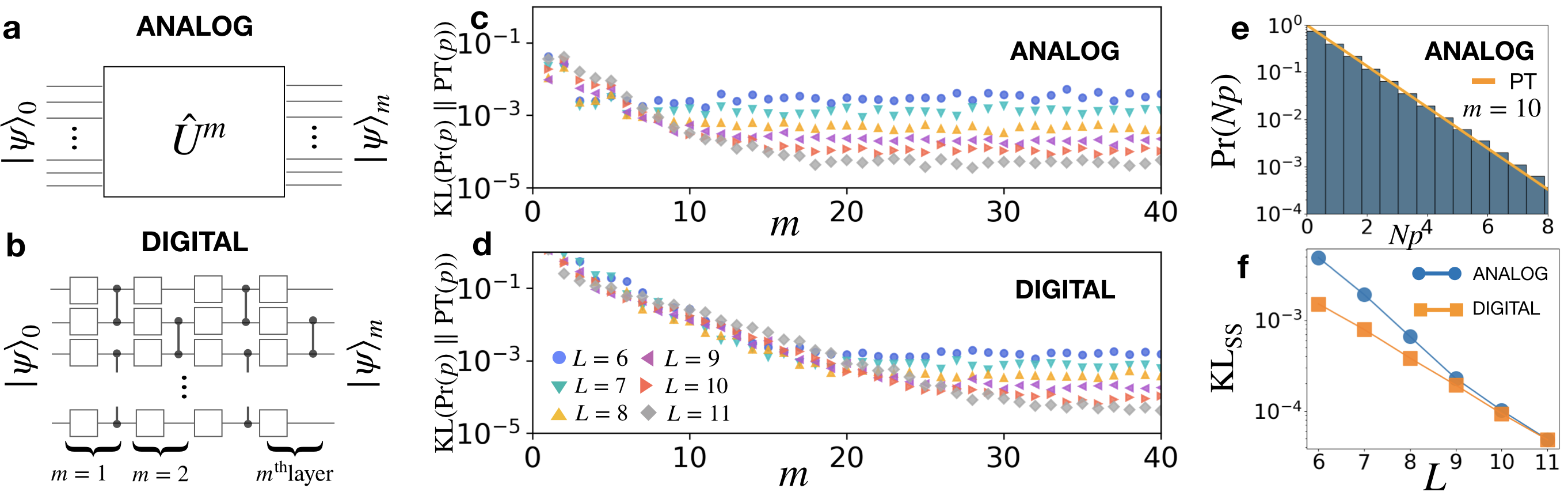}
\caption{\textbf{Benchmarking quantum supremacy with analog and digital approaches:} $\textbf{(a)}$ a circuit diagram of the analog driven quantum Ising chain with $m$ driving cycles. $\textbf{(b)}$ a circuit diagram of random quantum circuits containing $m$ layers of random single qubits in the set $\{\sqrt{X},\sqrt{Y},T\}$ and controlled-Z gates. $\textbf{(c)}$ ($\textbf{(d)}$)  $\text{KL}(\text{Pr}(p)\parallel\text{PT}(p))$ as a function of $m$ and the size of the system $L$ for the analog (digital) approach. $\textbf{(e)}$ The output distribution, weighted by $N=2^L$, in the analog case with $L=11$ and $m=10$. The yellow line is the exact PT distribution. $\textbf{(f)}$ KL divergence at the long time limit of the analog and the digital cases as a function of $L$. ( $W=5J$, $F=2.5J$, $\omega=8J$ and 500 disorder realizations.) }
\label{fig1}
\end{figure*}


The initial state $|\psi\rangle_0$ is chosen to be a product state where every spins point along the $+z$ direction. The state after $m$ driving cycles is then $|\psi\rangle_m = \hat{U}^m |\psi\rangle_0$, with 
\begin{equation}
\hat{U}=\hat{\mathcal{T}}\exp \left[-i\int_0^T \hat{H}(t) dt\right],
\label{eq:u}
\end{equation}
where $\hat{H}(t)=\hat{H}(t+T)$, $T=2\pi/\omega$, and $\hat{\mathcal{T}}$ is the time-ordering operator.  The sketch of the time evolution is depicted in Fig.~\ref{fig1}(a). After evolving, we calculate the output probability $p_m(\textbf{z})$ on the computational basis $\{|\textbf{z}\rangle\}$. Here $\textbf{z}=\left[z_1,z_2,...z_L\right]$, where $z_i\in\{\pm 1\}$ is the spin configuration along the z-direction. The distribution of $p_m(\textbf{z})$ averaged over $D$ disorder realizations is denoted as $\text{Pr}(p)$. 

\subsection{Random quantum circuits}

To compare with digital approaches, we simulate random quantum circuits for quantum supremacy, which is the 1D version of the proposal in Ref.\cite{2018_hartmut_natphy}. They consist of $m$ layers of gates and $L$ qubits, as depicted in Fig. \ref{fig1}(b). Each layer consists of one sub-layer of single-qubit gates and another sub-layer of controlled-Z gates. The first single-qubit layer consists of Hadamard gates. Other single-qubit gates are chosen randomly from the set $\{\sqrt{X},\sqrt{Y},T\}$, where $\sqrt{X}$ ($\sqrt{Y}$) represents a $\pi/2$ rotation around the $X$ ($Y$) axis of the Bloch sphere and $T$ is a non-Clifford gate representing a diagonal matrix $\{1,e^{i\pi/4}\}$. As in the analog case, we measure the output distribution ${\rm Pr}(p)$ averaged over $D$ realizations of the random circuits. 

To make a fair comparison, we set the driving period in the analog case to be the same as the time it takes to implement one layer of gates in the digital case \footnote{Since single-qubit gates can usually be implemented in a much faster time scale in the experiment, we assume that each layer takes $\sim \pi/4J$ which is the time to implement the controlled-Z gate using the $ZZ$ coupling. This condition sets the driving frequency to be $\omega=8J$.}. The entangling gates here are also restricted to nearest-neighbors as the interactions in the analog case. 

\subsection{Characterizing quantum supremacy: \\ analog versus digital}

A key ingredient to demonstrate quantum supremacy, both in the analog and the digital cases, is quantum chaos. Quantum chaotic systems are exponentially sensitive to perturbations \cite{2010_haake} and their evolutions are accurately described by the random matrix theory \cite{PhysRevX.8.021062}. In the limit where all matrix elements are fully random, the output state has an equal probability of being anywhere in the Hilbert space. Its output distribution then follows the Porter-Thomas (PT) distribution $\text{PT}(p)\equiv Ne^{-Np}$~\cite{2018_hartmut_natphy}, where $N=2^L\gg 1$. Simulating such systems on a classical computer requires analyzing its full quantum dynamics with resources that grow exponentially with the size of the system. For random quantum circuits, it is expected that state-of-the-art supercomputers will fail to simulate the circuits with more than approximately a hundred qubits~\cite{2019arXiv190500444V,Haner:2017:PSQ:3126908.3126947,2017arXiv171005867P}. The latter regime is referred to as the quantum supremacy regime.

Signatures of quantum supremacy can be characterized by measuring the difference between the output probabilities ${\rm Pr}(p)$ from the quantum device and the PT distribution. This can be done by using the Kullback-Leibler (KL) divergence defined as
\begin{equation}
\text{KL}(P(x)\parallel Q(x))=\sum_{x}P(x)\log \left( \frac{ P(x) }{ Q(x)}\right)\geq 0,
\end{equation}
where $ P(x)$ and $ Q(x)$ are the two distributions to be compared and $x$ is an input variable. The KL divergence is zero only when $P(x)=Q(x)$ for all $x$. We use this measure to compare various distributions throughout the text.

We simulate the dynamics of the driven Ising chain as described by Eq. \ref{eq:u} using exact diagonalization. In Fig. \ref{fig1}(c), we plot $\text{KL}(\text{Pr}(p)\parallel\text{PT}(p))$ for the analog case as a function of driving cycles $m$ for different $L$. We find that the KL divergence decays exponentially to a constant value. Fig. \ref{fig1}(e) shows an example of ${\rm Pr}(p)$ after 10 driving cycles for visualization. In Fig. \ref{fig1}(f), we plot the KL divergence at the long time limit, $\text{KL}_{\text{SS}}$, as a function of $L$. It shows that $\text{KL}_{\text{SS}}$ also decays exponentially with $L$. Hence, ${\rm Pr}(p)$ tends to the PT distribution in the thermodynamic limit.

In the digital case, ${\rm KL}(\text{Pr}(p)\parallel {\rm PT}(p))$ as a function of $m$ is plotted in Fig. \ref{fig1}(d). It shows that the analog approach converges at a faster rate. Fig. \ref{fig1}(f) shows that in the long-time limit, both approaches give the same value of the KL divergence for $L>9$. In the presence of several tens of qubits required for quantum supremacy, these features make the analog approach more favorable for NISQ devices due to the limited coherence time. Morever, the analog approach does not require precise local control to implement the gates.


\subsection{Driving-induced long-range interactions and computational complexity}

\begin{figure}
\includegraphics[width=0.47\textwidth]{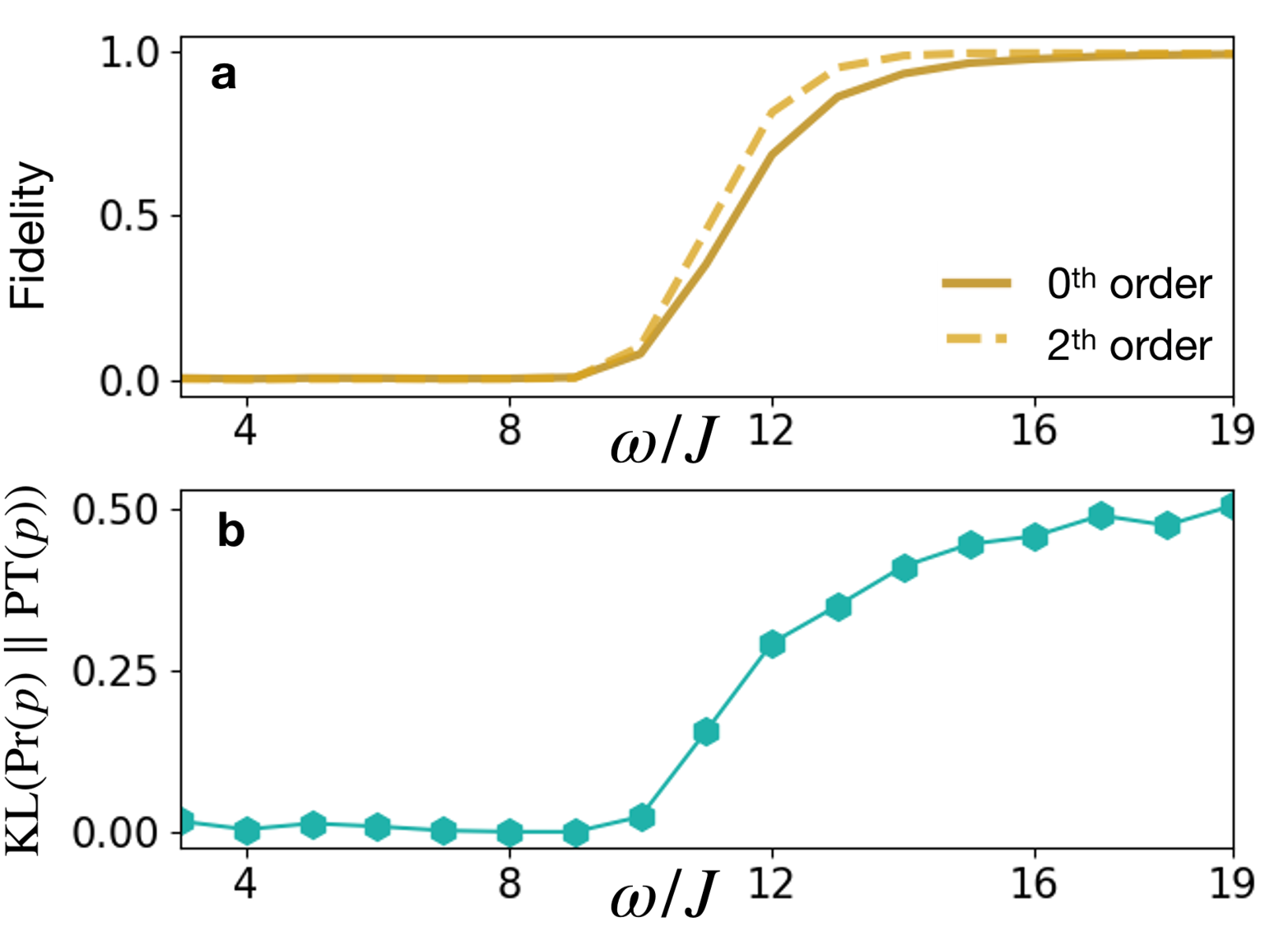}
\caption{\textbf{Magnus expansion and Porter-Thomas distribution} \textbf{(a)} Fidelity of the state evolved from the first three orders of the Magnus expansion after one driving cycle, compared to the exact time evolution. \textbf{(b)} KL divergence between the output distributions from the exact time evolution and the PT distribution. ($F=2.5J, W=2J, L=9$ and 500 disorder realizations.) }
\label{fig2}
\end{figure}

Achieving the PT distribution with a driven analog quantum processor can be viewed as a result of effective long-range interactions generated by the drive. To see this, it is insightful to investigate the unitary operator $\hat{U}$ that describes the evolution of a quantum state. The dynamics at stroboscopic times, $t_m=mT$, can be described by a time-independent Floquet Hamiltonian $\hat{H}_F$, defined as $\hat{U} \equiv \exp \left(-i \hat{H}_F T\right)$~\cite{2015_Bukov_AiP}.

For interacting systems, it is generally impossible to find an analytic form for $\hat H_F$, and one needs to resort to approximations. The most common approach is using the Magnus expansion to expand $\exp \left[-i H_F T\right]$ in a power series of $E_c/\omega$, where $E_c$ is a characteristic energy of the Hamiltonian $\hat H(t)$ which, in our case, depends on $J$, $W$, and $F$. 
Doing so, one can write the Floquet Hamiltonian as $\hat{H}_{F}=\sum_{l=0}^{\infty} \hat{H}_{F}^{(l)}$, where the two first terms, for example, read
\begin{align} \label{eq:ME}
\begin{split}
    \hat H_F^{(0)} &= \frac{1}{T}\int_{0}^{T}d\tau_1  \hat H(\tau_1) =\hat{H}_0+0.5\hat{H}_d, \\
    \hat H_F^{(1)} &= \frac{1}{2iT}\int_{0}^{T}d\tau_1\int_{0}^{\tau_1}d\tau_2  [\hat H(\tau_1),\hat H(\tau_2)].
\end{split}
\end{align}
The first term is simply the time average of $\hat{H}(t)$. The second term is zero as a consequence of the sinusoidal driving $f(t)$. The first correction to $\hat{H}_F^{(0)}$ is $H_F^{(2)}$, which is computed in Appendix~\ref{App:ME}. In the limit of infinite frequencies, the Floquet Hamiltonian formally coincides with $\hat{H}_F^{(0)}$, while the series diverges in the low frequency regime $\omega < E_c$.

The crucial point is that higher-order contributions generally include multi-body and longer- (but finite) range effective interactions. For example, $\hat H_F^{(2)}$ includes three-body interaction terms of the form $\hat H_i \sim \frac{J^2}{\omega^2}F\hat Z_{i-1}\hat X_{i}\hat Z_{i+1}$  [cf.~Eq.~\eqref{eq:hf2} in Appendix~\ref{App:ME}]. This tendency suggests that in the limit where the series diverges, the dynamics is governed by infinitely-long range multi-body interactions. This observation sheds light on the advantages of periodically driven systems to explore larger regions of their Hilbert space by highlighting the effectively enhanced connectivity of the model, which leads to faster growth of entanglement. 

In Fig.~\ref{fig2}, we picture the convergence of the Magnus expansion by plotting in panel (a) the fidelity $|\langle\psi_m|\psi^{(n)}\rangle|^2$  between a state evolved under $\hat H(t)$ after one driving cycle ($m=1$) and a state evolved under the truncated Magnus series $\ket{\psi^{(n)}} = \exp (-i\sum_{l=0}^n\hat H_F^{(l)}T) \ket{\psi}_0$ with $n=0$ and $2$. We see that for small driving frequencies $\omega$, the truncated Magnus series fails to capture the time evolution which suggests the divergence of the series. As a comparison, Fig.~\ref{fig2} (b) shows  $\text{KL}(\text{Pr}(p)\parallel\text{PT}(p))$ as a function of the driving frequency. We see that the convergence of ${\rm Pr}(p)$ toward ${\rm PT}(p)$ coincides with the divergence of the Magnus expansion.  

\begin{figure*}
\includegraphics[width=0.95\textwidth]{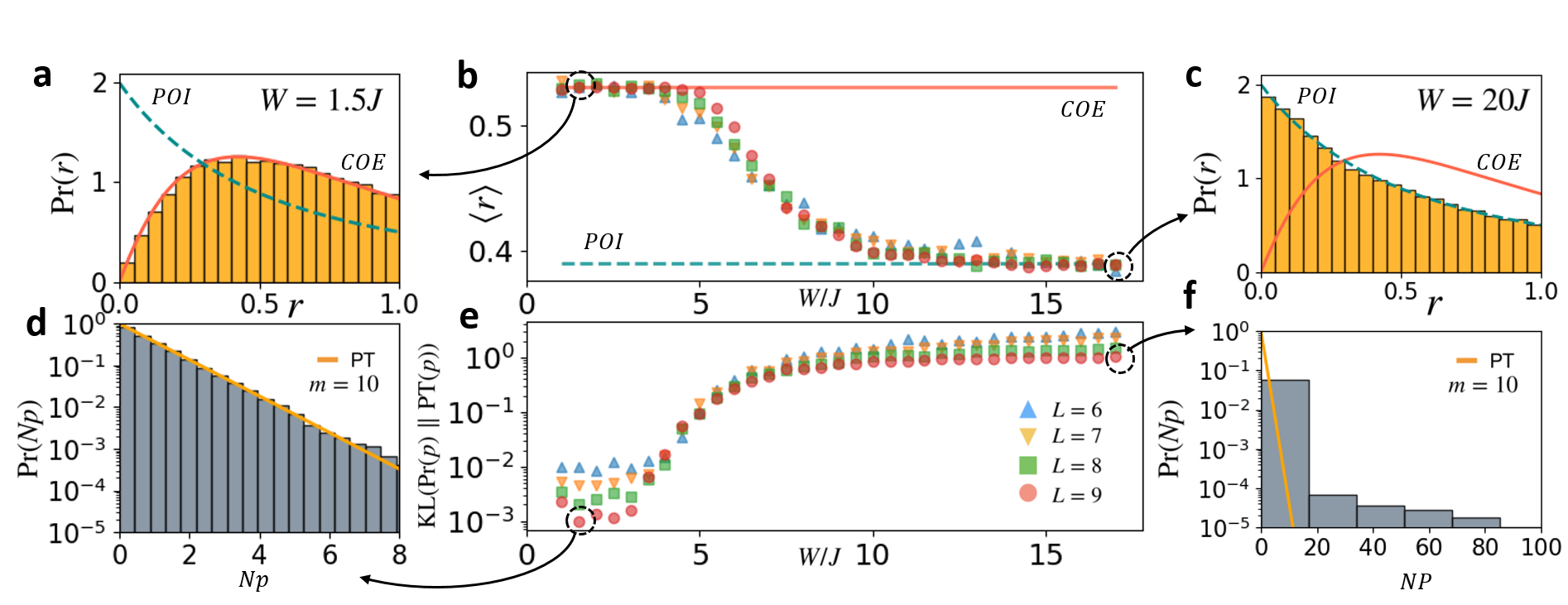}
\caption{\textbf{Probing the MBL phase transition}. \textbf{(a)}  (\textbf{(c)}) The level statistics of the driven Ising chain in the chaotic (MBL) regime with $W=1.5J$ ($W=20J$) and $L=9$. \textbf{(b)} The mean level statistics $\langle r\rangle$ as a function of disorder strength.  \textbf{(d)}  (\textbf{(f)}) The output distribution in the chaotic (MBL) regime with $W=1.5J$ ($W=20J$) and $L=9$. \textbf{(e)} $\text{KL}(\text{Pr}(p)\parallel\text{PT}(p))$ as a function of disorder strength. ( $F=2.5J$, $m=10$, $\omega=8J$ and 100 disorder realizations.)  }
\label{fig3}
\end{figure*}



The appearance of effective long-range interactions and the PT distribution present a significant challenge to simulate the driven disordered Ising chains with approximate numerical methods. For example, the standard matrix-product-state-based techniques are applicable only for weakly-entangled systems with short-range interactions \cite{PhysRevLett.91.147902, SCHOLLWOCK201196}. Moreover, the presence of disorder makes the effective fermionic model, derived using the Jordan-Wigner transformation, non-local and analytically hard to solve \cite{Russomanno_2016}. In the high-frequency limit, one can find analytical solutions for particular driving pulses \cite{PhysRevLett.112.140408}, but no general solutions exist. In general, simulating the driven disordered Ising chain as described by Eq. (\ref{eq:u}) on a classical computer requires exact numerics, which quickly becomes intractable as the number of sites increased. 


\section{Probing quantum phases of matter} 

In the previous section, we have seen the advantages of the driven Ising chain in achieving quantum supremacy, compared to the digital random quantum circuits with the same topology. Here, we discuss a close connection between the achieved quantum supremacy and quantum chaos in periodically-driven systems. This relation strongly suggests that the PT distribution in our system cannot be obtained without the drive. We then show a direct application of our sampling protocol in probing driven many-body localization (MBL) phase transitions, where driven quantum many-body systems fail to thermalize due to disorder. 

The concept of thermalization in an isolated quantum system begins with the eigenstate thermalization hypothesis~\cite{2006_Goldstein_PRL}, which states that, for an initial state $\ket{\psi_0}$ with $\bar E = \bra{\psi_0}\hat{H}_F \ket{\psi_0}$, any generic observable is expected to evolve toward the micro-canonical ensemble prediction associated with the energy $\bar E \pm \Delta E$, where $\Delta E$ is the variance of $\langle\hat{H}_F\rangle$. In the thermodynamic limit, this ensemble is equivalent to the canonical ensemble with a temperature $T = \hbar \bar E/k_B$~\cite{2016_Alessio_AiP}. In the driven case, it has been shown that if $\hat{H}_F$ has a diverging Magnus expansion, the system may thermalize with a corresponding infinite temperature \cite{2014_Rigol_PRX}. In Appendix \ref{app:pt}, we show analytically that such infinite temperature ensemble is equivalent to the ensemble of states that follow the PT distribution. This simple relation implies that obtaining the PT distribution without the drive is not possible in our system due to the conservation of energy. A detailed discussion on thermalization in the undriven case is provided in Appendix \ref{app:chaos}.

In the presence of large disorder, the above description fails, giving rise to the MBL phase, which has been the focus of numerous studies both theoretically and experimentally in recent years \cite{2015_huse_arcmp, 2016_rigol_ap, 2018arXiv180411065A}. Unlike standard quantum phase transitions which happen at the ground state, the MBL transitions occur in the dynamics and involve every eigenstate. Identifying the transition on a classical computer then requires exact diagonalization which is not possible for large systems \cite{2013_Alessio_AoP,2016_Abanin_AoP,2015_Ponte_PRL,2014_Rigol_PRX,PhysRevB.82.174411}. Partial experimental signatures of the MBL transitions have been observed in cold neutral atoms \cite{Schreiber842,PhysRevLett.116.140401,2018arXiv180509819L,2016_Choi_Sci}, superconducting circuits \cite{2017_Roushan_Sci,PhysRevLett.120.050507}, and trapped ions \cite{2016_monroe_nat}. 

\subsection{Probing the MBL transition with level statistics}

One of the standard ways to analyze the dynamics of the driven system is via the notion of the level statistics of the unitary operator $\hat{U}$ \cite{,2016_Abanin_AoP,2014_Rigol_PRX,PhysRevB.82.174411}. Let $|\phi_n\rangle$ be an eigenstate of the Floquet Hamiltonian, \textit{i.e.}, 
\begin{equation}
\hat{H}_F|\phi_n\rangle = \epsilon_n|\phi_n\rangle,
\end{equation}
where $\{\epsilon_n\}$ are eigenenergies of $\hat{H}_F$. It follows that 
\begin{equation}
\hat{U}=\sum_n e^{i\theta_n}|\phi_n\rangle\langle \phi_n|,
\end{equation}
where $\theta_n = \epsilon_n T \text{ modulo 2}\pi$. The level statistics $\text{P}(r)$ is a normalized distribution of the level spacing 
\begin{equation}
r_n= \frac{\text{min}(\delta_n,\delta_{n+1})}{\text{max}(\delta_n,\delta_{n+1})},
\end{equation}
with $\delta_n = \theta_{n+1}-\theta_n$ and $\theta_{n+1}>\theta_n$. $\hat{U}$ is said to be chaotic if it can be described by a random unitary matrix where $\{\theta_n\}$ follow the Circular Orthogonal Ensemble (COE) statistics, 
\begin{align}
\text{P}_{\rm COE}(r)=&\frac{2}{3}\left(\frac{\sin\left(\frac{2\pi r}{r+1}\right)}{2\pi r^2}+\frac{1}{(r+1)^2}+\frac{\sin\left(\frac{2\pi}{r+1}\right)}{2\pi} \right.\nonumber \\ &\left.-\frac{\cos\left(\frac{2\pi}{r+1}\right)}{r+1}-\frac{\cos\left(\frac{2\pi r}{r+1}\right)}{r(r+1)}\right).
\end{align}
The plot of $\text{P}_{\rm COE}(r)$ is depicted in Fig. \ref{fig3}(a) (red solid curve), showing a peak near $r=0.5$ with $\langle r\rangle_{\rm COE}\approx0.527$. This statistics implies level repulsion as $\text{P}_{\rm COE}(0)=0$, meaning that $\{\theta_n\}$ are correlated.

Notice that in contrast to the eigenenergy $\epsilon_n$, the phase $\theta_n$ or the `quasi-energy' only takes the value within the range $\left[0,2\pi\right]$.  This effect is referred to as energy folding. In the high-frequency limit $\epsilon_n T\ll 2\pi$, $\{\theta_n\}$ and $\{\epsilon_n\}$ follow the same statistics as they are linearly related. However, in the low-frequency limit $\epsilon_n T\gg2\pi$, the statistics of $\{\theta_n\}$ can be vastly different from that of $\{\epsilon_n\}$ due to the folding. For example, if $\hat{H}_F$ contains only short-range interactions and $\{\epsilon_n\}$ shows level repulsion, $\{\theta_n\}$ can appear to be uncorrelated \cite{2016_rigol_ap}. When $\hat{H}_F$ contains long-range interactions, i.e., the Magnus expansion diverges, the repulsion among $\{\epsilon_n\}$ can be so strong that the folded phases remain correlated. This relation stresses the important role of the drive in generating the COE statistics in our system. 

When the system is in the MBL phase with localized energy eigenstates, the energies $\{ \epsilon_n \}$ are uncorrelated and so are the phases $\{ \theta_n \}$. In this phase, the statistics of $\{\theta_n\}$ follows the Poisson (POI) statistics \cite{,2016_Abanin_AoP,PhysRevB.82.174411},
\begin{equation}
P_{\rm POI}(r) = \frac{2}{(1+r)^2}.
\end{equation}
The plot of $\text{P}_{\rm POI}(r)$ is depicted in Fig. \ref{fig3}(a) (green dashed curve), showing a peak at $r=0$ with $\langle r \rangle_{\rm POI} \approx 0.386$. 

In Fig. \ref{fig3}(b), we plot the averaged level spacing $\langle r\rangle$ calculated using the level statistics of a driven quantum Ising chain as a function of disorder strength. We see the transition between the COE at $W< 5J$ and the POI statistics at $W>10 J$. The level statistics in the chaotic and the MBL phase are plotted in Fig. \ref{fig3}(a) and \ref{fig3}(c) showing a good fit to the COE and the POI statistics respectively.

Even though the level statistics have been widely used in theoretical studies of MBL, its experimental realizations are limited. In the undriven case, signatures of level statistics have been measured using two interacting photons in a nine-site superconducting circuit \cite{2017_Roushan_Sci}. The spectroscopy method proposed in Ref. \cite{2017_Roushan_Sci} is based on the Fourier transform of the time evolution. Its resolution is limited by the coherence time of the system. In the driven case, the quasi-energies become significantly denser due to the energy-folding when $\epsilon_nT\gg 2\pi$, making the measurement even more challenging. 

\subsection{Probing the MBL transition with the Porter-Thomas distribution}
 
Since probing the level statistics in the experiment is very challenging, we propose here to use ${\rm KL}({\rm Pr}(p)\parallel {\rm PT}(p))$ as an order parameter to probe the MBL phase transition. This order parameter captures a direct signature of the COE statistics: the latter implies that $\hat{U}$ is a fully random unitary matrix, hence a state evolving under $\hat{U}$ is a random vector in the Hilbert space which leads to the PT distribution. 

In Fig. \ref{fig3}(e), we plot $\text{KL}(\text{Pr}(p)\parallel\text{PT}(p))$ as a function of disorder strength $W$. For $W<5J$, we see that the output probability converges to the PT distribution, as expected since the system is chaotic. The KL divergence starts to increase at $W\gtrsim5J$ and saturates at $W\sim 10 J$. This trend is consistent with $\langle r\rangle$ shown Fig. \ref{fig3}(b). In the MBL phase, the output distribution, shown in Fig. \ref{fig3}(f), is far from the PT distribution as expected because the system is localized near the initial state and does not explore the whole Hilbert space. We explore this relation for different values of the driving-frequencies over the coupling, $\omega/J$, in Appendix \ref{app:pd}. 

We note that in Refs \cite{Schreiber842,PhysRevLett.116.140401,2018arXiv180509819L,2016_Choi_Sci}, the spreading of an initially localized state is used to observe the MBL transition. However, such order parameter only gives informations about the spreading in real space, but not in the Hilbert space. The latter can be measured by $\text{KL}(\text{Pr}(p)\parallel\text{PT}(p))$.


\begin{figure}
\includegraphics[width=0.47\textwidth]{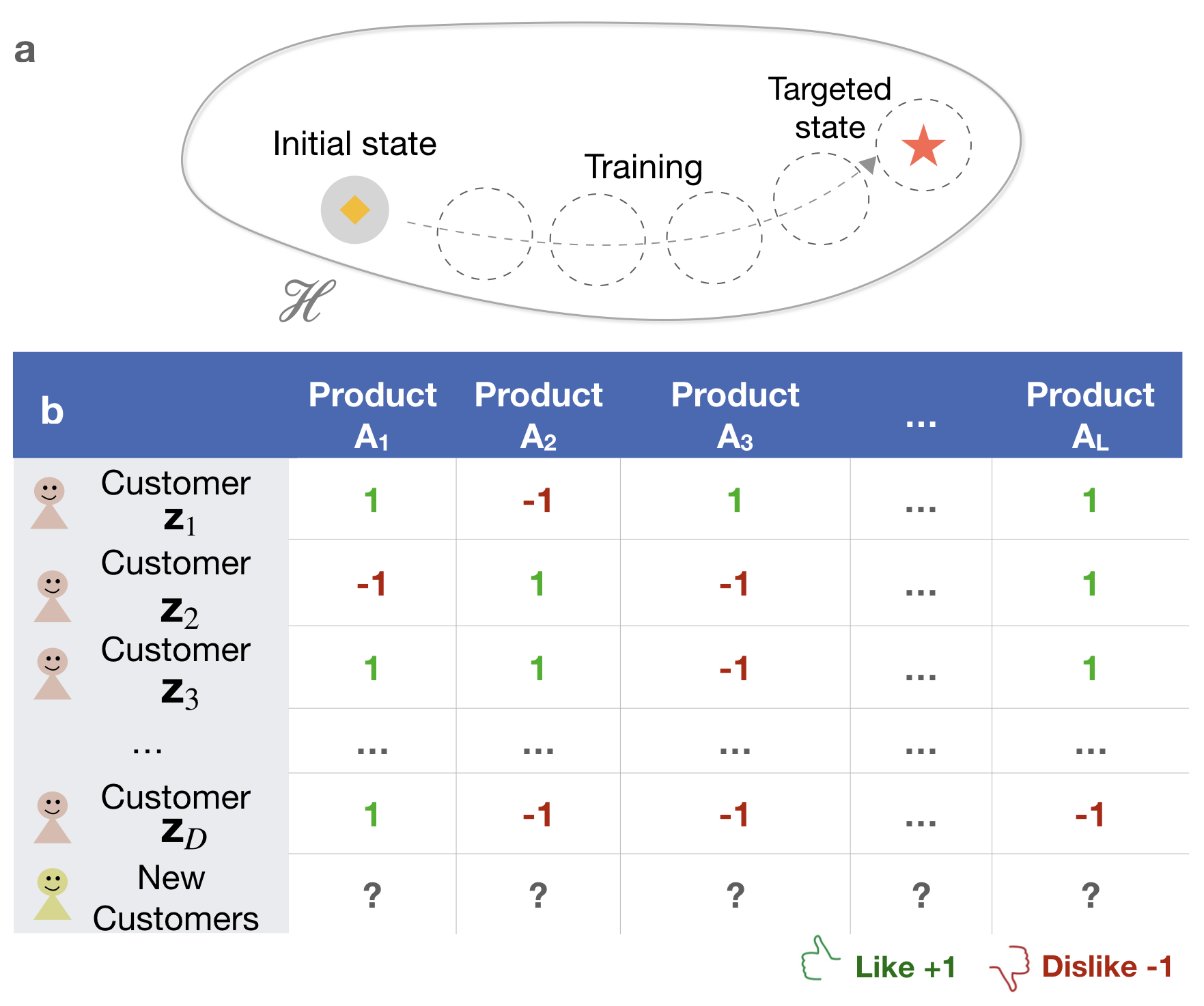}
\caption{\textbf{Quantum machine learning for product recommendation.} \textbf{(a)} Diagrams showing the dynamics of the system in the Hilbert space during the training in the driven MBL phase. \textbf{(b)} A table demonstrating a real-world application of generative modeling tasks in machine learning. Each customer $i\in \left[1,D\right],$ is asked to rate the product $ j\in\left[1,L\right]$ whether he/she likes ($+1$) or dislike the product ($-1$).}
\label{fig4}
\end{figure}

\section{Machine learning in the Hilbert space with an analog quantum processor} 

So far, we have shown that there are two phases associated with the driven disorder quantum Ising chain. In this section, we show how the same setup, with additional feedback loops, can be used to solve problems in machine learning, where the accuracy of the protocol solely depends on which phases the system operates in. Our goal here is to guide or `train' the quantum system to a specific point in the Hilbert space which represents the answer to a particular problem, as depicted in Fig.~\ref{fig4}(a). We show that in the chaotic phase, although the system is capable of probing the entire Hilbert space, training is impossible due to the chaotic nature of the dynamics. In the MBL phase, on the other hand, while the system only explores a small regime in the Hilbert space after each iteration, the dynamics can still efficiently probe the Hilbert space by quenching the evolution. However, in this regime, the more controlled evolution allows a successful optimization process.

\subsection{Generative modeling tasks}

We demonstrate our argument by tackling a generative modeling problem. The latter is an unsupervised machine-learning task, meaning that the training data are unlabelled. The goal is to find the unknown probability distribution, $q(\textbf{z})$, of the training data. Here, the data is a set of binary vectors $\{\textbf{z}\}_{\rm data}=\{\textbf{z}_1,\textbf{z}_2,...\}$. As a real-world example, this data can represent the opinions of a group of customers on a set of $L$ different products, as depicted in Fig.~\ref{fig4}(b). The opinion of the customer $i$ is represented by a binary vector $\textbf{z}_i=\left[z_{i1},z_{i2},...,z_{iL}\right]$ where $z_{ij}=1$ if he/she likes the product $j$ and $-1$ otherwise. After knowing $q(\textbf{z})$, the company can generate new data from this distribution and recommends products with $+1$ score to new customers. 

In this section we use an artificial dataset as a working example. To assure the generality of the data, we assume that $q(\textbf{z})$ is the Boltzmann distribution of classical Ising spins with all-to-all connectivity, \textit{i.e.},
\begin{equation}
q(\textbf{z}) = \frac{1}{Z}e^{- E(\textbf{z}) / k_B T}, 
\label{eq:q}
\end{equation}
where $Z=\sum_{\textbf{z}} \exp{(- E(\textbf{z}) / k_B T)}$ is the partition function, $k_B$ is the Boltzmann constant, $T$ plays the role of a temperature, and  
\begin{equation}
E(\textbf{z}) = \sum_{i=1}^La_i z_i + \sum_{\langle i,j\rangle} b_{ij}z_iz_{j}
\label{eq:bm}
\end{equation}
with $a_i$, $b_{ij}$ being random numbers between $\pm J/2$. This model is known as the Boltzmann machine which is one of the standard types of artificial neuron networks used in machine learning and has been shown to capture a wide range of real-world data \cite{ACKLEY1987522}. Its quantum version has been studied in \cite{PhysRevX.8.021050, 2019_yudong}.

\begin{figure*}
\includegraphics[width=0.95\textwidth]{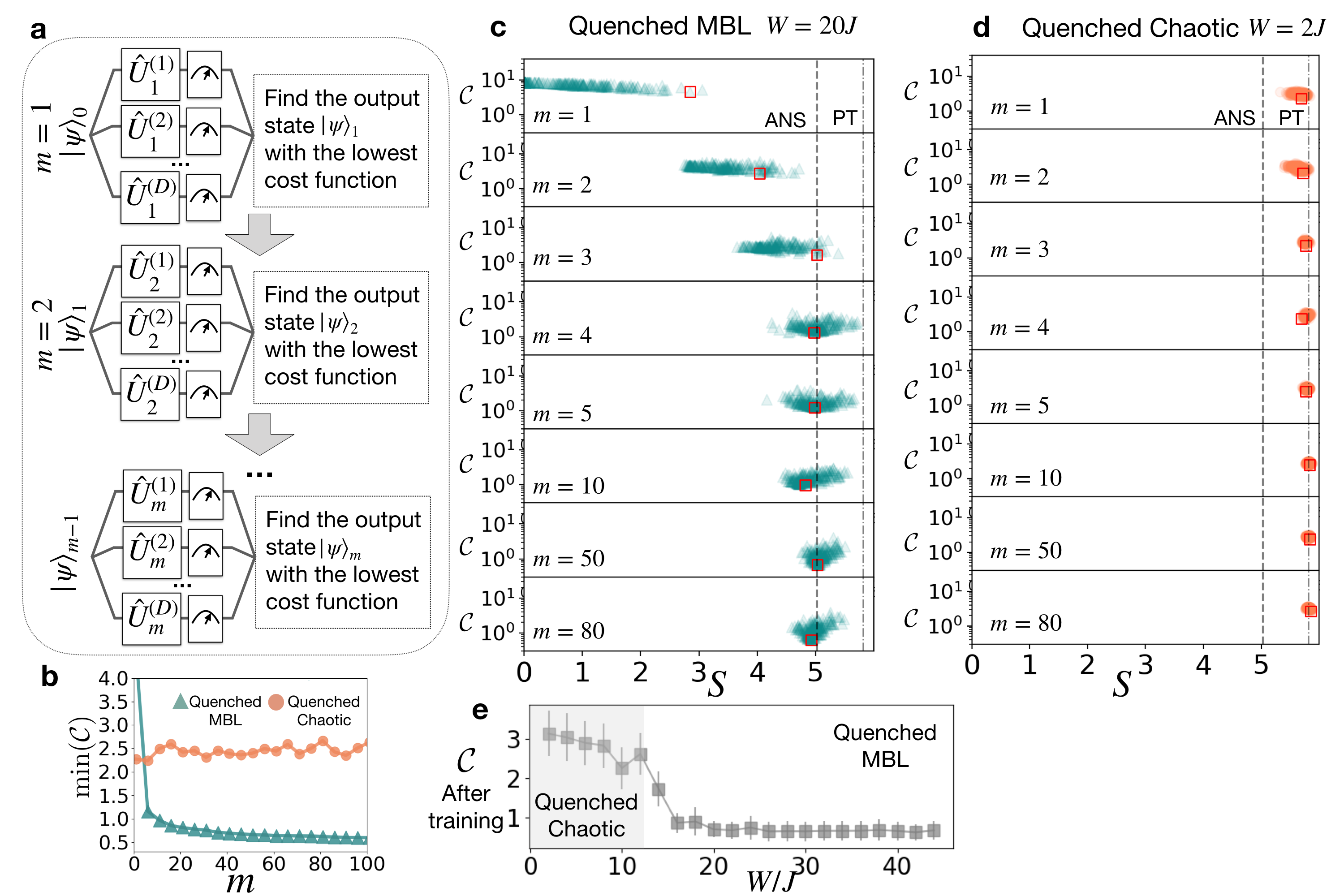}
\caption{\textbf{Training a driven analog quantum processor:} \textbf{(a)} A sketch of optimization loops used in the training protocol. The superscribed index is a label for a disorder realization. \textbf{(b)} The lowest cost function at each driving cycle as a function of $m$ for the quenched MBL (chaotic) regime with $W=20J$ ($W=2J$). \textbf{(c) ((d))} The scattering plots of the cost function $\mathcal{C}$ and the entropy $S$ of the all outputs from each driving cycle in both phases. The red square marks the point with the lowest cost function in each driving cycle. The vertical dashed line is the entropy $S_{\tilde{q}}$ of the targeted distribution $\tilde{q}(\textbf{z})$. The semi-dashed is the entropy $S_{\rm PT}$ of the states that follow the PT distribution. \textbf{(e)}The cost function at $m=500$ as a function of disorder strength $W$. The results are averaged over 50 dataset, \textit{i.e.}, 50 realizations of $\{a_i,b_i\}$ in Eq. \ref{eq:bm}. Each dataset consists of 3000 samples. ($F=2.5J$, $\omega=8J$, $k_BT=J$, $D = 140$, and $L=9$).}
\label{fig6}
\end{figure*}

\subsection{Training driven analog quantum processors}

Classically, the distribution of $\{\textbf{z}\}_{\rm data}$ can be obtained by first guessing a model $p_{\rm model}^{(\pmb{\theta})}(\textbf{z})$, such as the Poisson or the Boltzmann distribution, which has some variational parameters $\pmb{\theta}$. `Training' is done by minimizing the cost function ${{\rm KL}(p_{\rm model}^{(\pmb{\theta})}(\textbf{z}) \parallel q(\textbf{z}))}$ using either gradient descent or gradient-free optimization algorithms. 

In our case, we show how the distribution of $\{\textbf{z}\}_{\rm data}$ can be recovered as the output probability $p_m(\textbf{z})$ of the driven quantum Ising chain. Instead of the static disorder as before, the disorder is quenched after each driving cycle. The time evolution is described as
\begin{equation}
|\psi\rangle_m = \hat{U}_m...\hat{U}_3\hat{U}_2\hat{U}_1|\psi\rangle_0,
\label{eq:quenched}
\end{equation}
where $\hat{U}_1$, $\hat{U}_2$,  $\hat{U}_3$,..., $\hat{U}_m$ are propagators as defined in Eq. (\ref{eq:u}) with different disorder realizations but the same $W,J,F$ and $\omega$ \footnote{Quenched disorder can be done in various platforms including trapped ions \cite{2017_monroe_nat2} and superconducting circuits \cite{2017_Roushan_Sci}}. The subscript indicates the order in which the propagator is applied. Our cost function is defined as $\mathcal{C}={\rm KL}\left(p_m(\textbf{z})\parallel\tilde{q}(\textbf{z})\right)$, where $\tilde{q}(\textbf{z})$ is the normalized histogram of $\{\textbf{z}\}_{\rm data}$. `Training' is done by using feedback loops to find a set of disorder realizations that minimizes the cost function via a gradient-free approach. 

Our training protocol, depicted in Fig. \ref{fig6}(a) takes place as follows:
\begin{enumerate}
\item Initialize the system at $|\psi\rangle_m=|\psi\rangle_0$ with $m=0$.
\item Evolve the system by one driving cycle $|\psi\rangle_{m+1}=\hat{U}|\psi\rangle_m$, and then measure $\mathcal{C}$.
\item Repeat the step (2) $D$ times with different disorder realization. 
\item Choose the disorder realization that has the lowest $\mathcal{C}$, label the corresponding evolution as $U_m$, and update $m\to m+1$.
\item Repeat the step (3)-(4) until convergence.
\end{enumerate}
Note that, as shown below, our optimization is done in the Hilbert space, which is also an important difference to the usual optimization approaches which are done in the parameter space \cite{2018_hartmut_natcom, PhysRevX.8.021050}.

\subsection{Results and discussions}

Results of the training, depicted in Fig.~\ref{fig6}(b), compare the convergence of the protocol for systems in the quenched MBL and chaotic regimes. It shows that the protocol only converges in the quenched MBL regime. To visualize the dynamics of the system during the training, we choose to calculate the classical entropy associated with the output distribution in the step (2),
\begin{equation}
S_p=-\sum_{\textbf{z}} p_m (\textbf{z}) \log (p_m(\textbf{z})). 
\end{equation}
In Fig.~\ref{fig6}(c)-(d), we plot $\{S_p\}$ and $\{\mathcal{C}\}$ for all disorder realizations at different $m$. The vertical dashed line is the entropy of the training data, 
\begin{equation}
S_{\tilde{q}}=-\sum_{\textbf{z}}\tilde{q}(\textbf{z})\log (\tilde{q}(\textbf{z})). 
\end{equation}
The semi-dashed line is the entropy of the states that follow the PT distribution \cite{2018_hartmut_natphy}, 
\begin{equation}
S_{\rm PT}=-L\log 2+\gamma. 
\end{equation}
As shown in \ref{fig6}(c), in the MBL phase, although the system starts from a state with low entropy, the quenched disorder quickly drives the system into states with an entropy closer to $S_{\tilde{q}}$. As shown in Fig. \ref{fig6}(b), the convergence rate slows down when the entropy of the system becomes comparable to $S_{\tilde{q}}$ at approximately $m=5$. On the other hand, the system in the quenched chaotic regime is trapped near $S_{\rm PT}$, as shown in Fig.~\ref{fig6}(d), meaning that no optimization can be efficiently implemented due to the chaotic dynamics. 

In Fig. \ref{fig6}(e), we plot the cost function after the training for different disorder strengths $W$. It confirms that the protocol does not give an accurate result when the system is in the quenched chaotic regime. In the quenched MBL regime, the protocol converges to the same $C$ independent of $W$. With larger $W/J$, we find that the system takes a longer time to reach the same entropy as $S_{\tilde{q}}$. However, the total convergent time is approximately the same throughout the range of $W/J$ considered in Fig. \ref{fig6}(e). This shows that the performance of our training protocol depends solely on the phase of the quantum device and fine-tuning is not required. 

\begin{figure}
\includegraphics[width=0.45\textwidth]{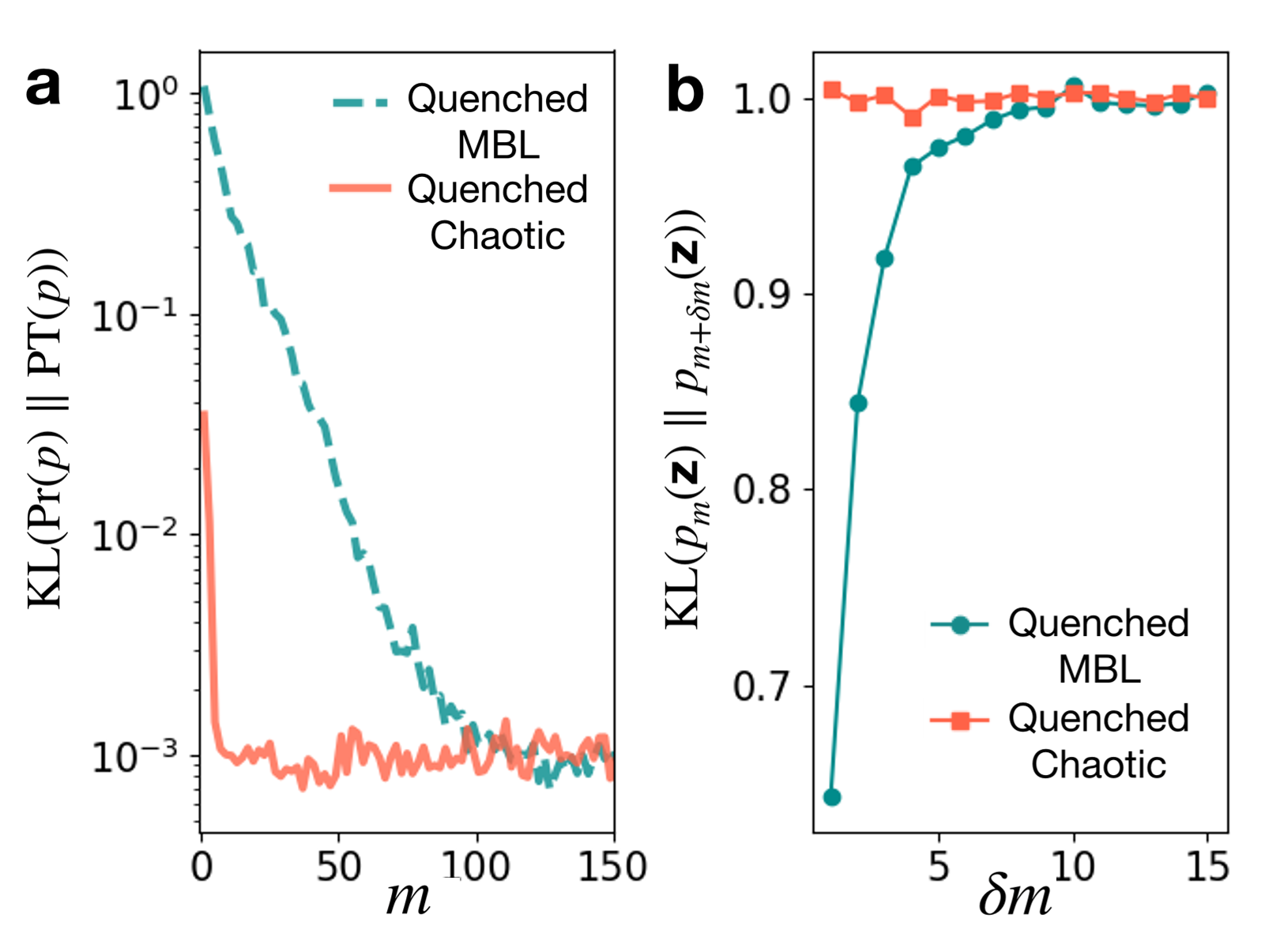}
\caption{\textbf{Quenched MBL and chaotic dynamics without feedback loops:} \textbf{(a)} ${\rm KL} ({\rm Pr}(p)\parallel {\rm PT} (p))$ as a function of $m$ in the quenched chaotic regime with $W=2J$ (orange yellow line) and the quenched MBL regime with $W=20J$ (dashed blue line). \textbf{(b)} ${\rm KL}(p_{m+\delta m}(\textbf{z})\parallel p_m(\textbf{z}))$, averaged over $m\in\{80,81,...,90\}$, as a function of $\delta m$ in the quenched chaotic regime (orange square dots) and the quenched MBL regime (blue dots).($F=2.5J$, $\omega=8J$, $L=9$).}
\label{fig5}
\end{figure}

The better learning performance in the quenched MBL regime stems from two factors. First, unlike the static disorder case, the system during the quenched MBL dynamics can go to anywhere in the Hilbert space. Second, the system has a `short-term memory' which enables a more controlled evolution. To show the former, in Fig. \ref{fig5}(a), we plot ${\rm KL}(\text{Pr}(p)\parallel {\rm PT}(p))$ calculated from the quenched dynamics \textit{without the feedback loops}, as a function of $m$. It shows that the system in both phases reaches the PT distribution over time. To show the memory effect, in Fig. \ref{fig5}(b), we plot ${\rm KL}(p_{m+\delta m }(\textbf{z})\parallel p_m(\textbf{z}))$ which is the difference between the output distributions at $m$ and $m+\delta m$ cycles as a function of $\delta m$. It shows that, in the quenched MBL regime, the memory decays exponentially with $\delta m$, while there is no memory in the quenched chaotic case.

We note that although the modulation of the magnetic field along the $x$-axis is the key to achieving quantum supremacy in the previous sections, it is not strictly required here as the quenched disorder acts as a drive. However, as shown in Appendix \ref{app:nodrive}, we find a significantly degraded learning performance when the modulation is removed, i.e. $f(t)=0.5$.

\section{Conclusions and outlooks} 

In this work, we have shown how to achieve quantum supremacy efficiently in analog quantum processors. Compared to the digital random circuit approaches with the same connectivity, analog methods require shorter coherence time and less control to reach the quantum supremacy regime. As a first direct application, we show that the difference between the output distribution and the PT distribution, $\text{KL}(\text{Pr}(p)\parallel\text{PT}(p))$, is a direct signature of the COE statistics and can be used as an accessible order parameter to probe driven MBL phase transitions. For machine learning applications, we show that quenched MBL dynamics and feedback loops can be used to train the system to learn complex classical data. We show that the performance of our training protocol depends solely on the phase of the quantum device and fine-tuning is not required.

Our results pave a way on how NISQ processors can be used to solve real-world problems beyond the reach of classical computers. Note that, in our machine-learning example, the number of products or `features' is the same as the number of qubits/sites. The latter can vary from a few tens to a few hundred for NISQ devices. However, for real-world datasets, the number of features can be far exceeding this number. Nevertheless, in practice, dimension reduction techniques such as principal component analysis (PCA) can be used to reduce the number of features drastically \cite{clustering}. For example, in single-cell RNA sequencing data, each cell can have several ten thousand features, which are usually reduced to 30-50 features before clustering algorithms were applied \cite{rna}. This number provides a promising avenue to benchmark NISQ devices with real-world data against classical methods. Similar to other hybrid quantum-classical or classical gradient-free optimization algorithms, the scalability and the convergence of our protocol have not been proven mathematically in a general case. Its applicability has to be tested on a case-by-case basis. 

While this work sheds light on an essential aspect for training quantum devices based on their phases, there is still room for the improvement of the optimization of the protocol. For example, it would be interesting to investigate how to implement the more efficient use of all the information measured during the training and how it could be applied to quantum chemistry problems.

\section{Acknowledgement}
We thank Pedram Roushan and Thiparat Chotibut for fruitful discussion. This research is supported by the National Research Foundation, Prime Minister's Office, Singapore and the Ministry of Education, Singapore under the Research Centres of Excellence programme. It was also partially funded by Polisimulator project co-financed by Greece and the EU Regional Development Fund, the European Research Council under the European Union's Seventh Framework Programme(FP7/2007-2013)/

\bibliography{ref}

\begin{thebibliography}{74}%
\makeatletter
\providecommand \@ifxundefined [1]{%
 \@ifx{#1\undefined}
}%
\providecommand \@ifnum [1]{%
 \ifnum #1\expandafter \@firstoftwo
 \else \expandafter \@secondoftwo
 \fi
}%
\providecommand \@ifx [1]{%
 \ifx #1\expandafter \@firstoftwo
 \else \expandafter \@secondoftwo
 \fi
}%
\providecommand \natexlab [1]{#1}%
\providecommand \enquote  [1]{``#1''}%
\providecommand \bibnamefont  [1]{#1}%
\providecommand \bibfnamefont [1]{#1}%
\providecommand \citenamefont [1]{#1}%
\providecommand \href@noop [0]{\@secondoftwo}%
\providecommand \href [0]{\begingroup \@sanitize@url \@href}%
\providecommand \@href[1]{\@@startlink{#1}\@@href}%
\providecommand \@@href[1]{\endgroup#1\@@endlink}%
\providecommand \@sanitize@url [0]{\catcode `\\12\catcode `\$12\catcode
  `\&12\catcode `\#12\catcode `\^12\catcode `\_12\catcode `\%12\relax}%
\providecommand \@@startlink[1]{}%
\providecommand \@@endlink[0]{}%
\providecommand \url  [0]{\begingroup\@sanitize@url \@url }%
\providecommand \@url [1]{\endgroup\@href {#1}{\urlprefix }}%
\providecommand \urlprefix  [0]{URL }%
\providecommand \Eprint [0]{\href }%
\providecommand \doibase [0]{http://dx.doi.org/}%
\providecommand \selectlanguage [0]{\@gobble}%
\providecommand \bibinfo  [0]{\@secondoftwo}%
\providecommand \bibfield  [0]{\@secondoftwo}%
\providecommand \translation [1]{[#1]}%
\providecommand \BibitemOpen [0]{}%
\providecommand \bibitemStop [0]{}%
\providecommand \bibitemNoStop [0]{.\EOS\space}%
\providecommand \EOS [0]{\spacefactor3000\relax}%
\providecommand \BibitemShut  [1]{\csname bibitem#1\endcsname}%
\let\auto@bib@innerbib\@empty
\bibitem [{\citenamefont {Preskill}(2018)}]{2018_preskill_quantum}%
  \BibitemOpen
  \bibfield  {author} {\bibinfo {author} {\bibfnamefont {John}\ \bibnamefont
  {Preskill}},\ }\bibfield  {title} {\enquote {\bibinfo {title} {Quantum
  {C}omputing in the {NISQ} era and beyond},}\ }\href {\doibase
  10.22331/q-2018-08-06-79} {\bibfield  {journal} {\bibinfo  {journal}
  {{Quantum}}\ }\textbf {\bibinfo {volume} {2}},\ \bibinfo {pages} {79}
  (\bibinfo {year} {2018})}\BibitemShut {NoStop}%
\bibitem [{\citenamefont {Harrow}\ and\ \citenamefont
  {Montanaro}(2017)}]{2017_Harrow_Nat}%
  \BibitemOpen
  \bibfield  {author} {\bibinfo {author} {\bibfnamefont {A.~W.}\ \bibnamefont
  {Harrow}}\ and\ \bibinfo {author} {\bibfnamefont {A.}~\bibnamefont
  {Montanaro}},\ }\bibfield  {title} {\enquote {\bibinfo {title} {Quantum
  computational supremacy},}\ }\href {https://doi.org/10.1038/nature23458}
  {\bibfield  {journal} {\bibinfo  {journal} {Nature}\ }\textbf {\bibinfo
  {volume} {549}},\ \bibinfo {pages} {203} (\bibinfo {year}
  {2017})}\BibitemShut {NoStop}%
\bibitem [{\citenamefont {Spring}\ \emph {et~al.}(2013)\citenamefont {Spring},
  \citenamefont {Metcalf}, \citenamefont {Humphreys}, \citenamefont
  {Kolthammer}, \citenamefont {Jin}, \citenamefont {Barbieri}, \citenamefont
  {Datta}, \citenamefont {Thomas-Peter}, \citenamefont {Langford},
  \citenamefont {Kundys}, \citenamefont {Gates}, \citenamefont {Smith},
  \citenamefont {Smith},\ and\ \citenamefont {Walmsley}}]{2013_walmsley1_sci}%
  \BibitemOpen
  \bibfield  {author} {\bibinfo {author} {\bibfnamefont {J.~B.}\ \bibnamefont
  {Spring}}, \bibinfo {author} {\bibfnamefont {B.~J.}\ \bibnamefont {Metcalf}},
  \bibinfo {author} {\bibfnamefont {P.~C.}\ \bibnamefont {Humphreys}}, \bibinfo
  {author} {\bibfnamefont {W.~S.}\ \bibnamefont {Kolthammer}}, \bibinfo
  {author} {\bibfnamefont {X.-M.}\ \bibnamefont {Jin}}, \bibinfo {author}
  {\bibfnamefont {M.}~\bibnamefont {Barbieri}}, \bibinfo {author}
  {\bibfnamefont {A.}~\bibnamefont {Datta}}, \bibinfo {author} {\bibfnamefont
  {N.}~\bibnamefont {Thomas-Peter}}, \bibinfo {author} {\bibfnamefont {N.~K.}\
  \bibnamefont {Langford}}, \bibinfo {author} {\bibfnamefont {D.}~\bibnamefont
  {Kundys}}, \bibinfo {author} {\bibfnamefont {J.~C.}\ \bibnamefont {Gates}},
  \bibinfo {author} {\bibfnamefont {B.~J.}\ \bibnamefont {Smith}}, \bibinfo
  {author} {\bibfnamefont {P.~G.~R.}\ \bibnamefont {Smith}}, \ and\ \bibinfo
  {author} {\bibfnamefont {I.~A.}\ \bibnamefont {Walmsley}},\ }\bibfield
  {title} {\enquote {\bibinfo {title} {Boson sampling on a photonic chip},}\
  }\href {\doibase 10.1126/science.1231692} {\bibfield  {journal} {\bibinfo
  {journal} {Science}\ }\textbf {\bibinfo {volume} {339}},\ \bibinfo {pages}
  {798--801} (\bibinfo {year} {2013})}\BibitemShut {NoStop}%
\bibitem [{\citenamefont {Broome}\ \emph {et~al.}(2013)\citenamefont {Broome},
  \citenamefont {Fedrizzi}, \citenamefont {Rahimi-Keshari}, \citenamefont
  {Dove}, \citenamefont {Aaronson}, \citenamefont {Ralph},\ and\ \citenamefont
  {White}}]{2013_Broome_Sci}%
  \BibitemOpen
  \bibfield  {author} {\bibinfo {author} {\bibfnamefont {M.~A.}\ \bibnamefont
  {Broome}}, \bibinfo {author} {\bibfnamefont {A.}~\bibnamefont {Fedrizzi}},
  \bibinfo {author} {\bibfnamefont {S.}~\bibnamefont {Rahimi-Keshari}},
  \bibinfo {author} {\bibfnamefont {J.}~\bibnamefont {Dove}}, \bibinfo {author}
  {\bibfnamefont {S.}~\bibnamefont {Aaronson}}, \bibinfo {author}
  {\bibfnamefont {T.~C.}\ \bibnamefont {Ralph}}, \ and\ \bibinfo {author}
  {\bibfnamefont {A.~G.}\ \bibnamefont {White}},\ }\bibfield  {title} {\enquote
  {\bibinfo {title} {Photonic boson sampling in a tunable circuit},}\ }\href
  {\doibase 10.1126/science.1231440} {\bibfield  {journal} {\bibinfo  {journal}
  {Science}\ }\textbf {\bibinfo {volume} {339}},\ \bibinfo {pages} {794--798}
  (\bibinfo {year} {2013})}\BibitemShut {NoStop}%
\bibitem [{\citenamefont {Tillmann}\ \emph {et~al.}(2013)\citenamefont
  {Tillmann}, \citenamefont {Daki{\'c}}, \citenamefont {Heilmann},
  \citenamefont {Nolte}, \citenamefont {Szameit},\ and\ \citenamefont
  {Walther}}]{2013_walther_nat}%
  \BibitemOpen
  \bibfield  {author} {\bibinfo {author} {\bibfnamefont {M.}~\bibnamefont
  {Tillmann}}, \bibinfo {author} {\bibfnamefont {B.}~\bibnamefont {Daki{\'c}}},
  \bibinfo {author} {\bibfnamefont {R.}~\bibnamefont {Heilmann}}, \bibinfo
  {author} {\bibfnamefont {S.}~\bibnamefont {Nolte}}, \bibinfo {author}
  {\bibfnamefont {A.}~\bibnamefont {Szameit}}, \ and\ \bibinfo {author}
  {\bibfnamefont {P.}~\bibnamefont {Walther}},\ }\bibfield  {title} {\enquote
  {\bibinfo {title} {Experimental boson sampling},}\ }\href
  {https://doi.org/10.1038/nphoton.2013.102} {\bibfield  {journal} {\bibinfo
  {journal} {Nature Photonics}\ }\textbf {\bibinfo {volume} {7}},\ \bibinfo
  {pages} {540} (\bibinfo {year} {2013})}\BibitemShut {NoStop}%
\bibitem [{\citenamefont {Crespi}\ \emph {et~al.}(2013)\citenamefont {Crespi},
  \citenamefont {Osellame}, \citenamefont {Ramponi}, \citenamefont {Brod},
  \citenamefont {Galv{\~a}o}, \citenamefont {Spagnolo}, \citenamefont
  {Vitelli}, \citenamefont {Maiorino}, \citenamefont {Mataloni},\ and\
  \citenamefont {Sciarrino}}]{2013_fabio_natpho}%
  \BibitemOpen
  \bibfield  {author} {\bibinfo {author} {\bibfnamefont {A.}~\bibnamefont
  {Crespi}}, \bibinfo {author} {\bibfnamefont {R.}~\bibnamefont {Osellame}},
  \bibinfo {author} {\bibfnamefont {R.}~\bibnamefont {Ramponi}}, \bibinfo
  {author} {\bibfnamefont {D.~J.}\ \bibnamefont {Brod}}, \bibinfo {author}
  {\bibfnamefont {E.~F.}\ \bibnamefont {Galv{\~a}o}}, \bibinfo {author}
  {\bibfnamefont {N.}~\bibnamefont {Spagnolo}}, \bibinfo {author}
  {\bibfnamefont {C.}~\bibnamefont {Vitelli}}, \bibinfo {author} {\bibfnamefont
  {E.}~\bibnamefont {Maiorino}}, \bibinfo {author} {\bibfnamefont
  {P.}~\bibnamefont {Mataloni}}, \ and\ \bibinfo {author} {\bibfnamefont
  {F.}~\bibnamefont {Sciarrino}},\ }\bibfield  {title} {\enquote {\bibinfo
  {title} {Integrated multimode interferometers with arbitrary designs for
  photonic boson sampling},}\ }\href {https://doi.org/10.1038/nphoton.2013.112}
  {\bibfield  {journal} {\bibinfo  {journal} {Nature Photonics}\ }\textbf
  {\bibinfo {volume} {7}},\ \bibinfo {pages} {545} (\bibinfo {year}
  {2013})}\BibitemShut {NoStop}%
\bibitem [{\citenamefont {Neill}\ \emph {et~al.}(2018)\citenamefont {Neill},
  \citenamefont {Roushan}, \citenamefont {Kechedzhi}, \citenamefont {Boixo},
  \citenamefont {Isakov}, \citenamefont {Smelyanskiy}, \citenamefont {Megrant},
  \citenamefont {Chiaro}, \citenamefont {Dunsworth}, \citenamefont {Arya},
  \citenamefont {Barends}, \citenamefont {Burkett}, \citenamefont {Chen},
  \citenamefont {Chen}, \citenamefont {Fowler}, \citenamefont {Foxen},
  \citenamefont {Giustina}, \citenamefont {Graff}, \citenamefont {Jeffrey},
  \citenamefont {Huang}, \citenamefont {Kelly}, \citenamefont {Klimov},
  \citenamefont {Lucero}, \citenamefont {Mutus}, \citenamefont {Neeley},
  \citenamefont {Quintana}, \citenamefont {Sank}, \citenamefont {Vainsencher},
  \citenamefont {Wenner}, \citenamefont {White}, \citenamefont {Neven},\ and\
  \citenamefont {Martinis}}]{2018_neill_sci}%
  \BibitemOpen
  \bibfield  {author} {\bibinfo {author} {\bibfnamefont {C.}~\bibnamefont
  {Neill}}, \bibinfo {author} {\bibfnamefont {P.}~\bibnamefont {Roushan}},
  \bibinfo {author} {\bibfnamefont {K.}~\bibnamefont {Kechedzhi}}, \bibinfo
  {author} {\bibfnamefont {S.}~\bibnamefont {Boixo}}, \bibinfo {author}
  {\bibfnamefont {S.~V.}\ \bibnamefont {Isakov}}, \bibinfo {author}
  {\bibfnamefont {V.}~\bibnamefont {Smelyanskiy}}, \bibinfo {author}
  {\bibfnamefont {A.}~\bibnamefont {Megrant}}, \bibinfo {author} {\bibfnamefont
  {B.}~\bibnamefont {Chiaro}}, \bibinfo {author} {\bibfnamefont
  {A.}~\bibnamefont {Dunsworth}}, \bibinfo {author} {\bibfnamefont
  {K.}~\bibnamefont {Arya}}, \bibinfo {author} {\bibfnamefont {R.}~\bibnamefont
  {Barends}}, \bibinfo {author} {\bibfnamefont {B.}~\bibnamefont {Burkett}},
  \bibinfo {author} {\bibfnamefont {Y.}~\bibnamefont {Chen}}, \bibinfo {author}
  {\bibfnamefont {Z.}~\bibnamefont {Chen}}, \bibinfo {author} {\bibfnamefont
  {A.}~\bibnamefont {Fowler}}, \bibinfo {author} {\bibfnamefont
  {B.}~\bibnamefont {Foxen}}, \bibinfo {author} {\bibfnamefont
  {M.}~\bibnamefont {Giustina}}, \bibinfo {author} {\bibfnamefont
  {R.}~\bibnamefont {Graff}}, \bibinfo {author} {\bibfnamefont
  {E.}~\bibnamefont {Jeffrey}}, \bibinfo {author} {\bibfnamefont
  {T.}~\bibnamefont {Huang}}, \bibinfo {author} {\bibfnamefont
  {J.}~\bibnamefont {Kelly}}, \bibinfo {author} {\bibfnamefont
  {P.}~\bibnamefont {Klimov}}, \bibinfo {author} {\bibfnamefont
  {E.}~\bibnamefont {Lucero}}, \bibinfo {author} {\bibfnamefont
  {J.}~\bibnamefont {Mutus}}, \bibinfo {author} {\bibfnamefont
  {M.}~\bibnamefont {Neeley}}, \bibinfo {author} {\bibfnamefont
  {C.}~\bibnamefont {Quintana}}, \bibinfo {author} {\bibfnamefont
  {D.}~\bibnamefont {Sank}}, \bibinfo {author} {\bibfnamefont {A.}~\bibnamefont
  {Vainsencher}}, \bibinfo {author} {\bibfnamefont {J.}~\bibnamefont {Wenner}},
  \bibinfo {author} {\bibfnamefont {T.~C.}\ \bibnamefont {White}}, \bibinfo
  {author} {\bibfnamefont {H.}~\bibnamefont {Neven}}, \ and\ \bibinfo {author}
  {\bibfnamefont {J.~M.}\ \bibnamefont {Martinis}},\ }\bibfield  {title}
  {\enquote {\bibinfo {title} {A blueprint for demonstrating quantum supremacy
  with superconducting qubits},}\ }\href {\doibase 10.1126/science.aao4309}
  {\bibfield  {journal} {\bibinfo  {journal} {Science}\ }\textbf {\bibinfo
  {volume} {360}},\ \bibinfo {pages} {195--199} (\bibinfo {year}
  {2018})}\BibitemShut {NoStop}%
\bibitem [{\citenamefont {Boixo}\ \emph {et~al.}(2018)\citenamefont {Boixo},
  \citenamefont {Isakov}, \citenamefont {Smelyanskiy}, \citenamefont {Babbush},
  \citenamefont {Ding}, \citenamefont {Jiang}, \citenamefont {Bremner},
  \citenamefont {Martinis},\ and\ \citenamefont {Neven}}]{2018_hartmut_natphy}%
  \BibitemOpen
  \bibfield  {author} {\bibinfo {author} {\bibfnamefont {S.}~\bibnamefont
  {Boixo}}, \bibinfo {author} {\bibfnamefont {S.~V.}\ \bibnamefont {Isakov}},
  \bibinfo {author} {\bibfnamefont {V.~N.}\ \bibnamefont {Smelyanskiy}},
  \bibinfo {author} {\bibfnamefont {R.}~\bibnamefont {Babbush}}, \bibinfo
  {author} {\bibfnamefont {N.}~\bibnamefont {Ding}}, \bibinfo {author}
  {\bibfnamefont {Z.}~\bibnamefont {Jiang}}, \bibinfo {author} {\bibfnamefont
  {M.~J.}\ \bibnamefont {Bremner}}, \bibinfo {author} {\bibfnamefont {J.~M.}\
  \bibnamefont {Martinis}}, \ and\ \bibinfo {author} {\bibfnamefont
  {H.}~\bibnamefont {Neven}},\ }\bibfield  {title} {\enquote {\bibinfo {title}
  {Characterizing quantum supremacy in near-term devices},}\ }\href {\doibase
  10.1038/s41567-018-0124-x} {\bibfield  {journal} {\bibinfo  {journal} {Nature
  Physics}\ }\textbf {\bibinfo {volume} {14}},\ \bibinfo {pages} {595--600}
  (\bibinfo {year} {2018})}\BibitemShut {NoStop}%
\bibitem [{\citenamefont {R~McClean}\ \emph {et~al.}(2016)\citenamefont
  {R~McClean}, \citenamefont {Romero}, \citenamefont {Babbush},\ and\
  \citenamefont {Aspuru-Guzik}}]{McClean_2016_njp}%
  \BibitemOpen
  \bibfield  {author} {\bibinfo {author} {\bibfnamefont {J.}~\bibnamefont
  {R~McClean}}, \bibinfo {author} {\bibfnamefont {J.}~\bibnamefont {Romero}},
  \bibinfo {author} {\bibfnamefont {R.}~\bibnamefont {Babbush}}, \ and\
  \bibinfo {author} {\bibfnamefont {A.}~\bibnamefont {Aspuru-Guzik}},\
  }\bibfield  {title} {\enquote {\bibinfo {title} {The theory of variational
  hybrid quantum-classical algorithms},}\ }\href {\doibase
  10.1088/1367-2630/18/2/023023} {\bibfield  {journal} {\bibinfo  {journal}
  {New Journal of Physics}\ }\textbf {\bibinfo {volume} {18}},\ \bibinfo
  {pages} {023023} (\bibinfo {year} {2016})}\BibitemShut {NoStop}%
\bibitem [{\citenamefont {Peruzzo}\ \emph {et~al.}(2014)\citenamefont
  {Peruzzo}, \citenamefont {McClean}, \citenamefont {Shadbolt}, \citenamefont
  {Yung}, \citenamefont {Zhou}, \citenamefont {Love}, \citenamefont
  {Aspuru-Guzik},\ and\ \citenamefont {O'Brien}}]{2014_jeremy_natcom}%
  \BibitemOpen
  \bibfield  {author} {\bibinfo {author} {\bibfnamefont {A.}~\bibnamefont
  {Peruzzo}}, \bibinfo {author} {\bibfnamefont {J.}~\bibnamefont {McClean}},
  \bibinfo {author} {\bibfnamefont {P.}~\bibnamefont {Shadbolt}}, \bibinfo
  {author} {\bibfnamefont {M.-H.}\ \bibnamefont {Yung}}, \bibinfo {author}
  {\bibfnamefont {X.-Q.}\ \bibnamefont {Zhou}}, \bibinfo {author}
  {\bibfnamefont {P.~J.}\ \bibnamefont {Love}}, \bibinfo {author}
  {\bibfnamefont {A.}~\bibnamefont {Aspuru-Guzik}}, \ and\ \bibinfo {author}
  {\bibfnamefont {J.~L.}\ \bibnamefont {O'Brien}},\ }\bibfield  {title}
  {\enquote {\bibinfo {title} {A variational eigenvalue solver on a photonic
  quantum processor},}\ }\href {https://doi.org/10.1038/ncomms5213} {\bibfield
  {journal} {\bibinfo  {journal} {Nature Communications}\ }\textbf {\bibinfo
  {volume} {5}},\ \bibinfo {pages} {4213} (\bibinfo {year} {2014})}\BibitemShut
  {NoStop}%
\bibitem [{\citenamefont {Kandala}\ \emph {et~al.}(2017)\citenamefont
  {Kandala}, \citenamefont {Mezzacapo}, \citenamefont {Temme}, \citenamefont
  {Takita}, \citenamefont {Brink}, \citenamefont {Chow},\ and\ \citenamefont
  {Gambetta}}]{2017_ibm_nat}%
  \BibitemOpen
  \bibfield  {author} {\bibinfo {author} {\bibfnamefont {A.}~\bibnamefont
  {Kandala}}, \bibinfo {author} {\bibfnamefont {A.}~\bibnamefont {Mezzacapo}},
  \bibinfo {author} {\bibfnamefont {K.}~\bibnamefont {Temme}}, \bibinfo
  {author} {\bibfnamefont {M.}~\bibnamefont {Takita}}, \bibinfo {author}
  {\bibfnamefont {M.}~\bibnamefont {Brink}}, \bibinfo {author} {\bibfnamefont
  {J.~M.}\ \bibnamefont {Chow}}, \ and\ \bibinfo {author} {\bibfnamefont
  {J.~M.}\ \bibnamefont {Gambetta}},\ }\bibfield  {title} {\enquote {\bibinfo
  {title} {Hardware-efficient variational quantum eigensolver for small
  molecules and quantum magnets},}\ }\href
  {https://doi.org/10.1038/nature23879} {\bibfield  {journal} {\bibinfo
  {journal} {Nature}\ }\textbf {\bibinfo {volume} {549}},\ \bibinfo {pages}
  {242} (\bibinfo {year} {2017})}\BibitemShut {NoStop}%
\bibitem [{\citenamefont {O'Malley}\ \emph {et~al.}(2016)\citenamefont
  {O'Malley}, \citenamefont {Babbush}, \citenamefont {Kivlichan}, \citenamefont
  {Romero}, \citenamefont {McClean}, \citenamefont {Barends}, \citenamefont
  {Kelly}, \citenamefont {Roushan}, \citenamefont {Tranter}, \citenamefont
  {Ding}, \citenamefont {Campbell}, \citenamefont {Chen}, \citenamefont {Chen},
  \citenamefont {Chiaro}, \citenamefont {Dunsworth}, \citenamefont {Fowler},
  \citenamefont {Jeffrey}, \citenamefont {Lucero}, \citenamefont {Megrant},
  \citenamefont {Mutus}, \citenamefont {Neeley}, \citenamefont {Neill},
  \citenamefont {Quintana}, \citenamefont {Sank}, \citenamefont {Vainsencher},
  \citenamefont {Wenner}, \citenamefont {White}, \citenamefont {Coveney},
  \citenamefont {Love}, \citenamefont {Neven}, \citenamefont {Aspuru-Guzik},\
  and\ \citenamefont {Martinis}}]{2016_Omalley_PRX}%
  \BibitemOpen
  \bibfield  {author} {\bibinfo {author} {\bibfnamefont {P.~J.~J.}\
  \bibnamefont {O'Malley}}, \bibinfo {author} {\bibfnamefont {R.}~\bibnamefont
  {Babbush}}, \bibinfo {author} {\bibfnamefont {I.~D.}\ \bibnamefont
  {Kivlichan}}, \bibinfo {author} {\bibfnamefont {J.}~\bibnamefont {Romero}},
  \bibinfo {author} {\bibfnamefont {J.~R.}\ \bibnamefont {McClean}}, \bibinfo
  {author} {\bibfnamefont {R.}~\bibnamefont {Barends}}, \bibinfo {author}
  {\bibfnamefont {J.}~\bibnamefont {Kelly}}, \bibinfo {author} {\bibfnamefont
  {P.}~\bibnamefont {Roushan}}, \bibinfo {author} {\bibfnamefont
  {A.}~\bibnamefont {Tranter}}, \bibinfo {author} {\bibfnamefont
  {N.}~\bibnamefont {Ding}}, \bibinfo {author} {\bibfnamefont {B.}~\bibnamefont
  {Campbell}}, \bibinfo {author} {\bibfnamefont {Y.}~\bibnamefont {Chen}},
  \bibinfo {author} {\bibfnamefont {Z.}~\bibnamefont {Chen}}, \bibinfo {author}
  {\bibfnamefont {B.}~\bibnamefont {Chiaro}}, \bibinfo {author} {\bibfnamefont
  {A.}~\bibnamefont {Dunsworth}}, \bibinfo {author} {\bibfnamefont {A.~G.}\
  \bibnamefont {Fowler}}, \bibinfo {author} {\bibfnamefont {E.}~\bibnamefont
  {Jeffrey}}, \bibinfo {author} {\bibfnamefont {E.}~\bibnamefont {Lucero}},
  \bibinfo {author} {\bibfnamefont {A.}~\bibnamefont {Megrant}}, \bibinfo
  {author} {\bibfnamefont {J.~Y.}\ \bibnamefont {Mutus}}, \bibinfo {author}
  {\bibfnamefont {M.}~\bibnamefont {Neeley}}, \bibinfo {author} {\bibfnamefont
  {C.}~\bibnamefont {Neill}}, \bibinfo {author} {\bibfnamefont
  {C.}~\bibnamefont {Quintana}}, \bibinfo {author} {\bibfnamefont
  {D.}~\bibnamefont {Sank}}, \bibinfo {author} {\bibfnamefont {A.}~\bibnamefont
  {Vainsencher}}, \bibinfo {author} {\bibfnamefont {J.}~\bibnamefont {Wenner}},
  \bibinfo {author} {\bibfnamefont {T.~C.}\ \bibnamefont {White}}, \bibinfo
  {author} {\bibfnamefont {P.~V.}\ \bibnamefont {Coveney}}, \bibinfo {author}
  {\bibfnamefont {P.~J.}\ \bibnamefont {Love}}, \bibinfo {author}
  {\bibfnamefont {H.}~\bibnamefont {Neven}}, \bibinfo {author} {\bibfnamefont
  {A.}~\bibnamefont {Aspuru-Guzik}}, \ and\ \bibinfo {author} {\bibfnamefont
  {J.~M.}\ \bibnamefont {Martinis}},\ }\bibfield  {title} {\enquote {\bibinfo
  {title} {Scalable quantum simulation of molecular energies},}\ }\href
  {\doibase 10.1103/PhysRevX.6.031007} {\bibfield  {journal} {\bibinfo
  {journal} {Phys. Rev. X}\ }\textbf {\bibinfo {volume} {6}},\ \bibinfo {pages}
  {031007} (\bibinfo {year} {2016})}\BibitemShut {NoStop}%
\bibitem [{\citenamefont {Hempel}\ \emph {et~al.}(2018)\citenamefont {Hempel},
  \citenamefont {Maier}, \citenamefont {Romero}, \citenamefont {McClean},
  \citenamefont {Monz}, \citenamefont {Shen}, \citenamefont {Jurcevic},
  \citenamefont {Lanyon}, \citenamefont {Love}, \citenamefont {Babbush},
  \citenamefont {Aspuru-Guzik}, \citenamefont {Blatt},\ and\ \citenamefont
  {Roos}}]{2018_Hempel_PRX}%
  \BibitemOpen
  \bibfield  {author} {\bibinfo {author} {\bibfnamefont {C.}~\bibnamefont
  {Hempel}}, \bibinfo {author} {\bibfnamefont {C.}~\bibnamefont {Maier}},
  \bibinfo {author} {\bibfnamefont {J.}~\bibnamefont {Romero}}, \bibinfo
  {author} {\bibfnamefont {J.}~\bibnamefont {McClean}}, \bibinfo {author}
  {\bibfnamefont {T.}~\bibnamefont {Monz}}, \bibinfo {author} {\bibfnamefont
  {H.}~\bibnamefont {Shen}}, \bibinfo {author} {\bibfnamefont {P.}~\bibnamefont
  {Jurcevic}}, \bibinfo {author} {\bibfnamefont {B.~P.}\ \bibnamefont
  {Lanyon}}, \bibinfo {author} {\bibfnamefont {P.}~\bibnamefont {Love}},
  \bibinfo {author} {\bibfnamefont {R.}~\bibnamefont {Babbush}}, \bibinfo
  {author} {\bibfnamefont {A.}~\bibnamefont {Aspuru-Guzik}}, \bibinfo {author}
  {\bibfnamefont {R.}~\bibnamefont {Blatt}}, \ and\ \bibinfo {author}
  {\bibfnamefont {C.~F.}\ \bibnamefont {Roos}},\ }\bibfield  {title} {\enquote
  {\bibinfo {title} {{Quantum Chemistry Calculations on a Trapped-Ion Quantum
  Simulator}},}\ }\href {\doibase 10.1103/PhysRevX.8.031022} {\bibfield
  {journal} {\bibinfo  {journal} {Physical Review X}\ }\textbf {\bibinfo
  {volume} {8}},\ \bibinfo {eid} {031022} (\bibinfo {year} {2018})}\BibitemShut
  {NoStop}%
\bibitem [{\citenamefont {Li}\ and\ \citenamefont
  {Benjamin}(2017)}]{2017_Li_PRX}%
  \BibitemOpen
  \bibfield  {author} {\bibinfo {author} {\bibfnamefont {Y.}~\bibnamefont
  {Li}}\ and\ \bibinfo {author} {\bibfnamefont {S.~C.}\ \bibnamefont
  {Benjamin}},\ }\bibfield  {title} {\enquote {\bibinfo {title} {Efficient
  variational quantum simulator incorporating active error minimization},}\
  }\href {\doibase 10.1103/PhysRevX.7.021050} {\bibfield  {journal} {\bibinfo
  {journal} {Phys. Rev. X}\ }\textbf {\bibinfo {volume} {7}},\ \bibinfo {pages}
  {021050} (\bibinfo {year} {2017})}\BibitemShut {NoStop}%
\bibitem [{\citenamefont {Kokail}\ \emph {et~al.}(2019)\citenamefont {Kokail},
  \citenamefont {Maier}, \citenamefont {van Bijnen}, \citenamefont {Brydges},
  \citenamefont {Joshi}, \citenamefont {Jurcevic}, \citenamefont {Muschik},
  \citenamefont {Silvi}, \citenamefont {Blatt}, \citenamefont {Roos},\ and\
  \citenamefont {Zoller}}]{2018_Kokail_ArXiV}%
  \BibitemOpen
  \bibfield  {author} {\bibinfo {author} {\bibfnamefont {C.}~\bibnamefont
  {Kokail}}, \bibinfo {author} {\bibfnamefont {C.}~\bibnamefont {Maier}},
  \bibinfo {author} {\bibfnamefont {R.}~\bibnamefont {van Bijnen}}, \bibinfo
  {author} {\bibfnamefont {T.}~\bibnamefont {Brydges}}, \bibinfo {author}
  {\bibfnamefont {M.~K.}\ \bibnamefont {Joshi}}, \bibinfo {author}
  {\bibfnamefont {P.}~\bibnamefont {Jurcevic}}, \bibinfo {author}
  {\bibfnamefont {C.~A.}\ \bibnamefont {Muschik}}, \bibinfo {author}
  {\bibfnamefont {P.}~\bibnamefont {Silvi}}, \bibinfo {author} {\bibfnamefont
  {R.}~\bibnamefont {Blatt}}, \bibinfo {author} {\bibfnamefont {C.~F.}\
  \bibnamefont {Roos}}, \ and\ \bibinfo {author} {\bibfnamefont
  {P.}~\bibnamefont {Zoller}},\ }\bibfield  {title} {\enquote {\bibinfo {title}
  {Self-verifying variational quantum simulation of lattice models},}\ }\href
  {\doibase 10.1038/s41586-019-1177-4} {\bibfield  {journal} {\bibinfo
  {journal} {Nature}\ }\textbf {\bibinfo {volume} {569}},\ \bibinfo {pages}
  {355--360} (\bibinfo {year} {2019})}\BibitemShut {NoStop}%
\bibitem [{\citenamefont {{Otterbach}}\ \emph {et~al.}()\citenamefont
  {{Otterbach}}, \citenamefont {{Manenti}}, \citenamefont {{Alidoust}},
  \citenamefont {{Bestwick}}, \citenamefont {{Block}}, \citenamefont {{Bloom}},
  \citenamefont {{Caldwell}}, \citenamefont {{Didier}}, \citenamefont
  {{Schuyler Fried}}, \citenamefont {{Hong}}, \citenamefont {{Karalekas}},
  \citenamefont {{Osborn}}, \citenamefont {{Papageorge}}, \citenamefont
  {{Peterson}}, \citenamefont {{Prawiroatmodjo}}, \citenamefont {{Rubin}},
  \citenamefont {{Ryan}}, \citenamefont {{Scarabelli}}, \citenamefont
  {{Scheer}}, \citenamefont {{Sete}}, \citenamefont {{Sivarajah}},
  \citenamefont {{Smith}}, \citenamefont {{Staley}}, \citenamefont {{Tezak}},
  \citenamefont {{Zeng}}, \citenamefont {{Hudson}}, \citenamefont {{Johnson}},
  \citenamefont {{Reagor}}, \citenamefont {{da Silva}},\ and\ \citenamefont
  {{Rigetti}}}]{2017_Otterbach_ArXiV}%
  \BibitemOpen
  \bibfield  {author} {\bibinfo {author} {\bibfnamefont {J.~S.}\ \bibnamefont
  {{Otterbach}}}, \bibinfo {author} {\bibfnamefont {R.}~\bibnamefont
  {{Manenti}}}, \bibinfo {author} {\bibfnamefont {N.}~\bibnamefont
  {{Alidoust}}}, \bibinfo {author} {\bibfnamefont {A.}~\bibnamefont
  {{Bestwick}}}, \bibinfo {author} {\bibfnamefont {M.}~\bibnamefont {{Block}}},
  \bibinfo {author} {\bibfnamefont {B.}~\bibnamefont {{Bloom}}}, \bibinfo
  {author} {\bibfnamefont {S.}~\bibnamefont {{Caldwell}}}, \bibinfo {author}
  {\bibfnamefont {N.}~\bibnamefont {{Didier}}}, \bibinfo {author}
  {\bibfnamefont {E.}~\bibnamefont {{Schuyler Fried}}}, \bibinfo {author}
  {\bibfnamefont {S.}~\bibnamefont {{Hong}}}, \bibinfo {author} {\bibfnamefont
  {P.}~\bibnamefont {{Karalekas}}}, \bibinfo {author} {\bibfnamefont {C.~B.}\
  \bibnamefont {{Osborn}}}, \bibinfo {author} {\bibfnamefont {A.}~\bibnamefont
  {{Papageorge}}}, \bibinfo {author} {\bibfnamefont {E.~C.}\ \bibnamefont
  {{Peterson}}}, \bibinfo {author} {\bibfnamefont {G.}~\bibnamefont
  {{Prawiroatmodjo}}}, \bibinfo {author} {\bibfnamefont {N.}~\bibnamefont
  {{Rubin}}}, \bibinfo {author} {\bibfnamefont {Colm~A.}\ \bibnamefont
  {{Ryan}}}, \bibinfo {author} {\bibfnamefont {D.}~\bibnamefont
  {{Scarabelli}}}, \bibinfo {author} {\bibfnamefont {M.}~\bibnamefont
  {{Scheer}}}, \bibinfo {author} {\bibfnamefont {E.~A.}\ \bibnamefont
  {{Sete}}}, \bibinfo {author} {\bibfnamefont {P.}~\bibnamefont {{Sivarajah}}},
  \bibinfo {author} {\bibfnamefont {Robert~S.}\ \bibnamefont {{Smith}}},
  \bibinfo {author} {\bibfnamefont {A.}~\bibnamefont {{Staley}}}, \bibinfo
  {author} {\bibfnamefont {N.}~\bibnamefont {{Tezak}}}, \bibinfo {author}
  {\bibfnamefont {W.~J.}\ \bibnamefont {{Zeng}}}, \bibinfo {author}
  {\bibfnamefont {A.}~\bibnamefont {{Hudson}}}, \bibinfo {author}
  {\bibfnamefont {Blake~R.}\ \bibnamefont {{Johnson}}}, \bibinfo {author}
  {\bibfnamefont {M.}~\bibnamefont {{Reagor}}}, \bibinfo {author}
  {\bibfnamefont {M.~P.}\ \bibnamefont {{da Silva}}}, \ and\ \bibinfo {author}
  {\bibfnamefont {C.}~\bibnamefont {{Rigetti}}},\ }\bibfield  {title} {\enquote
  {\bibinfo {title} {{Unsupervised Machine Learning on a Hybrid Quantum
  Computer}},}\ }\href@noop {} {\ }\Eprint {http://arxiv.org/abs/1712.05771}
  {arXiv:1712.05771} \BibitemShut {NoStop}%
\bibitem [{\citenamefont {{Zhu}}\ \emph {et~al.}()\citenamefont {{Zhu}},
  \citenamefont {{Linke}}, \citenamefont {{Benedetti}}, \citenamefont
  {{Landsman}}, \citenamefont {{Nguyen}}, \citenamefont {{Alderete}},
  \citenamefont {{Perdomo-Ortiz}}, \citenamefont {{Korda}}, \citenamefont
  {{Garfoot}}, \citenamefont {{Brecque}}, \citenamefont {{Egan}}, \citenamefont
  {{Perdomo}},\ and\ \citenamefont {{Monroe}}}]{2018_Zhu_ArXiV}%
  \BibitemOpen
  \bibfield  {author} {\bibinfo {author} {\bibfnamefont {D.}~\bibnamefont
  {{Zhu}}}, \bibinfo {author} {\bibfnamefont {N.~M.}\ \bibnamefont {{Linke}}},
  \bibinfo {author} {\bibfnamefont {M.}~\bibnamefont {{Benedetti}}}, \bibinfo
  {author} {\bibfnamefont {K.~A.}\ \bibnamefont {{Landsman}}}, \bibinfo
  {author} {\bibfnamefont {N.~H.}\ \bibnamefont {{Nguyen}}}, \bibinfo {author}
  {\bibfnamefont {C.~H.}\ \bibnamefont {{Alderete}}}, \bibinfo {author}
  {\bibfnamefont {A.}~\bibnamefont {{Perdomo-Ortiz}}}, \bibinfo {author}
  {\bibfnamefont {N.}~\bibnamefont {{Korda}}}, \bibinfo {author} {\bibfnamefont
  {A.}~\bibnamefont {{Garfoot}}}, \bibinfo {author} {\bibfnamefont
  {C.}~\bibnamefont {{Brecque}}}, \bibinfo {author} {\bibfnamefont
  {L.}~\bibnamefont {{Egan}}}, \bibinfo {author} {\bibfnamefont
  {O.}~\bibnamefont {{Perdomo}}}, \ and\ \bibinfo {author} {\bibfnamefont
  {C.}~\bibnamefont {{Monroe}}},\ }\bibfield  {title} {\enquote {\bibinfo
  {title} {{Training of Quantum Circuits on a Hybrid Quantum Computer}},}\
  }\href@noop {} {\ }\Eprint {http://arxiv.org/abs/1812.08862}
  {arXiv:1812.08862} \BibitemShut {NoStop}%
\bibitem [{\citenamefont {Havl{\'\i}{\v c}ek}\ \emph
  {et~al.}(2019)\citenamefont {Havl{\'\i}{\v c}ek}, \citenamefont
  {C{\'o}rcoles}, \citenamefont {Temme}, \citenamefont {Harrow}, \citenamefont
  {Kandala}, \citenamefont {Chow},\ and\ \citenamefont {Gambetta}}]{2019_ibm}%
  \BibitemOpen
  \bibfield  {author} {\bibinfo {author} {\bibfnamefont {Vojt{\v e}ch}\
  \bibnamefont {Havl{\'\i}{\v c}ek}}, \bibinfo {author} {\bibfnamefont
  {Antonio~D.}\ \bibnamefont {C{\'o}rcoles}}, \bibinfo {author} {\bibfnamefont
  {Kristan}\ \bibnamefont {Temme}}, \bibinfo {author} {\bibfnamefont {Aram~W.}\
  \bibnamefont {Harrow}}, \bibinfo {author} {\bibfnamefont {Abhinav}\
  \bibnamefont {Kandala}}, \bibinfo {author} {\bibfnamefont {Jerry~M.}\
  \bibnamefont {Chow}}, \ and\ \bibinfo {author} {\bibfnamefont {Jay~M.}\
  \bibnamefont {Gambetta}},\ }\bibfield  {title} {\enquote {\bibinfo {title}
  {Supervised learning with quantum-enhanced feature spaces},}\ }\href
  {\doibase 10.1038/s41586-019-0980-2} {\bibfield  {journal} {\bibinfo
  {journal} {Nature}\ }\textbf {\bibinfo {volume} {567}},\ \bibinfo {pages}
  {209--212} (\bibinfo {year} {2019})}\BibitemShut {NoStop}%
\bibitem [{\citenamefont {{Du}}\ \emph {et~al.}()\citenamefont {{Du}},
  \citenamefont {{Hsieh}}, \citenamefont {{Liu}},\ and\ \citenamefont
  {{Tao}}}]{2018_tao_arxiv}%
  \BibitemOpen
  \bibfield  {author} {\bibinfo {author} {\bibfnamefont {Y.}~\bibnamefont
  {{Du}}}, \bibinfo {author} {\bibfnamefont {M.-H.}\ \bibnamefont {{Hsieh}}},
  \bibinfo {author} {\bibfnamefont {T.}~\bibnamefont {{Liu}}}, \ and\ \bibinfo
  {author} {\bibfnamefont {D.}~\bibnamefont {{Tao}}},\ }\bibfield  {title}
  {\enquote {\bibinfo {title} {{The Expressive Power of Parameterized Quantum
  Circuits}},}\ }\href@noop {} {\ }\Eprint {http://arxiv.org/abs/1810.11922}
  {arXiv:1810.11922} \BibitemShut {NoStop}%
\bibitem [{\citenamefont {{Killoran}}\ \emph {et~al.}()\citenamefont
  {{Killoran}}, \citenamefont {{Bromley}}, \citenamefont {{Arrazola}},
  \citenamefont {{Schuld}}, \citenamefont {{Quesada}},\ and\ \citenamefont
  {{Lloyd}}}]{2018_lloyd_arxiv}%
  \BibitemOpen
  \bibfield  {author} {\bibinfo {author} {\bibfnamefont {N.}~\bibnamefont
  {{Killoran}}}, \bibinfo {author} {\bibfnamefont {T.~R.}\ \bibnamefont
  {{Bromley}}}, \bibinfo {author} {\bibfnamefont {J.~M.}\ \bibnamefont
  {{Arrazola}}}, \bibinfo {author} {\bibfnamefont {M.}~\bibnamefont
  {{Schuld}}}, \bibinfo {author} {\bibfnamefont {N.}~\bibnamefont {{Quesada}}},
  \ and\ \bibinfo {author} {\bibfnamefont {S.}~\bibnamefont {{Lloyd}}},\
  }\bibfield  {title} {\enquote {\bibinfo {title} {{Continuous-variable quantum
  neural networks}},}\ }\href@noop {} {\ }\Eprint
  {http://arxiv.org/abs/1806.06871} {arXiv:1806.06871} \BibitemShut {NoStop}%
\bibitem [{\citenamefont {{Coyle}}\ \emph {et~al.}()\citenamefont {{Coyle}},
  \citenamefont {{Mills}}, \citenamefont {{Danos}},\ and\ \citenamefont
  {{Kashefi}}}]{2019_kashfi_arxiv}%
  \BibitemOpen
  \bibfield  {author} {\bibinfo {author} {\bibfnamefont {B.}~\bibnamefont
  {{Coyle}}}, \bibinfo {author} {\bibfnamefont {D.}~\bibnamefont {{Mills}}},
  \bibinfo {author} {\bibfnamefont {V.}~\bibnamefont {{Danos}}}, \ and\
  \bibinfo {author} {\bibfnamefont {E.}~\bibnamefont {{Kashefi}}},\ }\bibfield
  {title} {\enquote {\bibinfo {title} {{The Born Supremacy: Quantum Advantage
  and Training of an Ising Born Machine}},}\ }\href@noop {} {\ }\Eprint
  {http://arxiv.org/abs/1904.02214} {arXiv:1904.02214} \BibitemShut {NoStop}%
\bibitem [{\citenamefont {Cirac}\ and\ \citenamefont
  {Zoller}(2012)}]{2012_zoller_natphy}%
  \BibitemOpen
  \bibfield  {author} {\bibinfo {author} {\bibfnamefont {J.~I.}\ \bibnamefont
  {Cirac}}\ and\ \bibinfo {author} {\bibfnamefont {P.}~\bibnamefont {Zoller}},\
  }\bibfield  {title} {\enquote {\bibinfo {title} {Goals and opportunities in
  quantum simulation},}\ }\href {http://dx.doi.org/10.1038/nphys2275}
  {\bibfield  {journal} {\bibinfo  {journal} {Nat Phys}\ }\textbf {\bibinfo
  {volume} {8}},\ \bibinfo {pages} {264} (\bibinfo {year} {2012})}\BibitemShut
  {NoStop}%
\bibitem [{\citenamefont {Hauke}\ \emph {et~al.}(2012)\citenamefont {Hauke},
  \citenamefont {Cucchietti}, \citenamefont {Tagliacozzo}, \citenamefont
  {Deutsch},\ and\ \citenamefont {Lewenstein}}]{2012_lewenstein_rpp}%
  \BibitemOpen
  \bibfield  {author} {\bibinfo {author} {\bibfnamefont {P.}~\bibnamefont
  {Hauke}}, \bibinfo {author} {\bibfnamefont {F.~M.}\ \bibnamefont
  {Cucchietti}}, \bibinfo {author} {\bibfnamefont {L.}~\bibnamefont
  {Tagliacozzo}}, \bibinfo {author} {\bibfnamefont {I.}~\bibnamefont
  {Deutsch}}, \ and\ \bibinfo {author} {\bibfnamefont {M.}~\bibnamefont
  {Lewenstein}},\ }\bibfield  {title} {\enquote {\bibinfo {title} {Can one
  trust quantum simulators?}}\ }\href
  {http://stacks.iop.org/0034-4885/75/i=8/a=082401} {\bibfield  {journal}
  {\bibinfo  {journal} {Reports on Progress in Physics}\ }\textbf {\bibinfo
  {volume} {75}},\ \bibinfo {pages} {082401} (\bibinfo {year}
  {2012})}\BibitemShut {NoStop}%
\bibitem [{\citenamefont {Johnson}\ \emph {et~al.}(2014)\citenamefont
  {Johnson}, \citenamefont {Clark},\ and\ \citenamefont
  {Jaksch}}]{2014_dieter_epj}%
  \BibitemOpen
  \bibfield  {author} {\bibinfo {author} {\bibfnamefont {T.~H.}\ \bibnamefont
  {Johnson}}, \bibinfo {author} {\bibfnamefont {S.~R.}\ \bibnamefont {Clark}},
  \ and\ \bibinfo {author} {\bibfnamefont {D.}~\bibnamefont {Jaksch}},\
  }\bibfield  {title} {\enquote {\bibinfo {title} {What is a quantum
  simulator?}}\ }\href {\doibase 10.1140/epjqt10} {\bibfield  {journal}
  {\bibinfo  {journal} {EPJ Quantum Technology}\ }\textbf {\bibinfo {volume}
  {1}},\ \bibinfo {pages} {10} (\bibinfo {year} {2014})}\BibitemShut {NoStop}%
\bibitem [{\citenamefont {Choi}\ \emph {et~al.}(2016)\citenamefont {Choi},
  \citenamefont {Hild}, \citenamefont {Zeiher}, \citenamefont {Schau{\ss}},
  \citenamefont {Rubio-Abadal}, \citenamefont {Yefsah}, \citenamefont
  {Khemani}, \citenamefont {Huse}, \citenamefont {Bloch},\ and\ \citenamefont
  {Gross}}]{2016_Choi_Sci}%
  \BibitemOpen
  \bibfield  {author} {\bibinfo {author} {\bibfnamefont {J.-Y.}\ \bibnamefont
  {Choi}}, \bibinfo {author} {\bibfnamefont {S.}~\bibnamefont {Hild}}, \bibinfo
  {author} {\bibfnamefont {J.}~\bibnamefont {Zeiher}}, \bibinfo {author}
  {\bibfnamefont {P.}~\bibnamefont {Schau{\ss}}}, \bibinfo {author}
  {\bibfnamefont {A.}~\bibnamefont {Rubio-Abadal}}, \bibinfo {author}
  {\bibfnamefont {T.}~\bibnamefont {Yefsah}}, \bibinfo {author} {\bibfnamefont
  {V.}~\bibnamefont {Khemani}}, \bibinfo {author} {\bibfnamefont {D.~A.}\
  \bibnamefont {Huse}}, \bibinfo {author} {\bibfnamefont {I.}~\bibnamefont
  {Bloch}}, \ and\ \bibinfo {author} {\bibfnamefont {C.}~\bibnamefont
  {Gross}},\ }\bibfield  {title} {\enquote {\bibinfo {title} {Exploring the
  many-body localization transition in two dimensions},}\ }\href {\doibase
  10.1126/science.aaf8834} {\bibfield  {journal} {\bibinfo  {journal}
  {Science}\ }\textbf {\bibinfo {volume} {352}},\ \bibinfo {pages} {1547--1552}
  (\bibinfo {year} {2016})}\BibitemShut {NoStop}%
\bibitem [{\citenamefont {Gross}\ and\ \citenamefont
  {Bloch}(2017)}]{2017_gross_sci}%
  \BibitemOpen
  \bibfield  {author} {\bibinfo {author} {\bibfnamefont {C.}~\bibnamefont
  {Gross}}\ and\ \bibinfo {author} {\bibfnamefont {I.}~\bibnamefont {Bloch}},\
  }\bibfield  {title} {\enquote {\bibinfo {title} {Quantum simulations with
  ultracold atoms in optical lattices},}\ }\href {\doibase
  10.1126/science.aal3837} {\bibfield  {journal} {\bibinfo  {journal}
  {Science}\ }\textbf {\bibinfo {volume} {357}},\ \bibinfo {pages} {995--1001}
  (\bibinfo {year} {2017})}\BibitemShut {NoStop}%
\bibitem [{\citenamefont {Blatt}\ and\ \citenamefont
  {Roos}(2012)}]{2012_blatt_np}%
  \BibitemOpen
  \bibfield  {author} {\bibinfo {author} {\bibfnamefont {R.}~\bibnamefont
  {Blatt}}\ and\ \bibinfo {author} {\bibfnamefont {C.~F.}\ \bibnamefont
  {Roos}},\ }\bibfield  {title} {\enquote {\bibinfo {title} {Quantum
  simulations with trapped ions},}\ }\href
  {http://dx.doi.org/10.1038/nphys2252} {\bibfield  {journal} {\bibinfo
  {journal} {Nature Physics}\ }\textbf {\bibinfo {volume} {8}},\ \bibinfo
  {pages} {277} (\bibinfo {year} {2012})}\BibitemShut {NoStop}%
\bibitem [{\citenamefont {Zhang}\ \emph
  {et~al.}(2017{\natexlab{a}})\citenamefont {Zhang}, \citenamefont {Pagano},
  \citenamefont {Hess}, \citenamefont {Kyprianidis}, \citenamefont {Becker},
  \citenamefont {Kaplan}, \citenamefont {Gorshkov}, \citenamefont {Gong},\ and\
  \citenamefont {Monroe}}]{2017_monroe_nat}%
  \BibitemOpen
  \bibfield  {author} {\bibinfo {author} {\bibfnamefont {J.}~\bibnamefont
  {Zhang}}, \bibinfo {author} {\bibfnamefont {G.}~\bibnamefont {Pagano}},
  \bibinfo {author} {\bibfnamefont {P.~W.}\ \bibnamefont {Hess}}, \bibinfo
  {author} {\bibfnamefont {A.}~\bibnamefont {Kyprianidis}}, \bibinfo {author}
  {\bibfnamefont {P.}~\bibnamefont {Becker}}, \bibinfo {author} {\bibfnamefont
  {H.}~\bibnamefont {Kaplan}}, \bibinfo {author} {\bibfnamefont {A.~V.}\
  \bibnamefont {Gorshkov}}, \bibinfo {author} {\bibfnamefont {Z.~X.}\
  \bibnamefont {Gong}}, \ and\ \bibinfo {author} {\bibfnamefont
  {C.}~\bibnamefont {Monroe}},\ }\bibfield  {title} {\enquote {\bibinfo {title}
  {Observation of a many-body dynamical phase transition with a 53-qubit
  quantum simulator},}\ }\href {https://doi.org/10.1038/nature24654} {\bibfield
   {journal} {\bibinfo  {journal} {Nature}\ }\textbf {\bibinfo {volume}
  {551}},\ \bibinfo {pages} {601} (\bibinfo {year}
  {2017}{\natexlab{a}})}\BibitemShut {NoStop}%
\bibitem [{\citenamefont {Bernien}\ \emph {et~al.}(2017)\citenamefont
  {Bernien}, \citenamefont {Schwartz}, \citenamefont {Keesling}, \citenamefont
  {Levine}, \citenamefont {Omran}, \citenamefont {Pichler}, \citenamefont
  {Choi}, \citenamefont {Zibrov}, \citenamefont {Endres}, \citenamefont
  {Greiner}, \citenamefont {Vuleti{\'c}},\ and\ \citenamefont
  {Lukin}}]{2017_lukin_nat}%
  \BibitemOpen
  \bibfield  {author} {\bibinfo {author} {\bibfnamefont {H.}~\bibnamefont
  {Bernien}}, \bibinfo {author} {\bibfnamefont {S.}~\bibnamefont {Schwartz}},
  \bibinfo {author} {\bibfnamefont {A.}~\bibnamefont {Keesling}}, \bibinfo
  {author} {\bibfnamefont {H.}~\bibnamefont {Levine}}, \bibinfo {author}
  {\bibfnamefont {A.}~\bibnamefont {Omran}}, \bibinfo {author} {\bibfnamefont
  {H.}~\bibnamefont {Pichler}}, \bibinfo {author} {\bibfnamefont
  {S.}~\bibnamefont {Choi}}, \bibinfo {author} {\bibfnamefont {A.~S.}\
  \bibnamefont {Zibrov}}, \bibinfo {author} {\bibfnamefont {M.}~\bibnamefont
  {Endres}}, \bibinfo {author} {\bibfnamefont {M.}~\bibnamefont {Greiner}},
  \bibinfo {author} {\bibfnamefont {V.}~\bibnamefont {Vuleti{\'c}}}, \ and\
  \bibinfo {author} {\bibfnamefont {M.~D.}\ \bibnamefont {Lukin}},\ }\bibfield
  {title} {\enquote {\bibinfo {title} {Probing many-body dynamics on a 51-atom
  quantum simulator},}\ }\href {https://doi.org/10.1038/nature24622} {\bibfield
   {journal} {\bibinfo  {journal} {Nature}\ }\textbf {\bibinfo {volume}
  {551}},\ \bibinfo {pages} {579} (\bibinfo {year} {2017})}\BibitemShut
  {NoStop}%
\bibitem [{\citenamefont {Houck}\ \emph {et~al.}(2012)\citenamefont {Houck},
  \citenamefont {T{\"u}reci},\ and\ \citenamefont {Koch}}]{2012_koch_natphy}%
  \BibitemOpen
  \bibfield  {author} {\bibinfo {author} {\bibfnamefont {A.~A.}\ \bibnamefont
  {Houck}}, \bibinfo {author} {\bibfnamefont {H.~E.}\ \bibnamefont
  {T{\"u}reci}}, \ and\ \bibinfo {author} {\bibfnamefont {J.}~\bibnamefont
  {Koch}},\ }\bibfield  {title} {\enquote {\bibinfo {title} {On-chip quantum
  simulation with superconducting circuits},}\ }\href
  {http://dx.doi.org/10.1038/nphys2251} {\bibfield  {journal} {\bibinfo
  {journal} {Nature Physics}\ }\textbf {\bibinfo {volume} {8}},\ \bibinfo
  {pages} {292} (\bibinfo {year} {2012})}\BibitemShut {NoStop}%
\bibitem [{\citenamefont {Hensgens}\ \emph {et~al.}(2017)\citenamefont
  {Hensgens}, \citenamefont {Fujita}, \citenamefont {Janssen}, \citenamefont
  {Li}, \citenamefont {Van~Diepen}, \citenamefont {Reichl}, \citenamefont
  {Wegscheider}, \citenamefont {Das~Sarma},\ and\ \citenamefont
  {Vandersypen}}]{2017_vadersypen_nat}%
  \BibitemOpen
  \bibfield  {author} {\bibinfo {author} {\bibfnamefont {T.}~\bibnamefont
  {Hensgens}}, \bibinfo {author} {\bibfnamefont {T.}~\bibnamefont {Fujita}},
  \bibinfo {author} {\bibfnamefont {L.}~\bibnamefont {Janssen}}, \bibinfo
  {author} {\bibfnamefont {Xiao}\ \bibnamefont {Li}}, \bibinfo {author}
  {\bibfnamefont {C.~J.}\ \bibnamefont {Van~Diepen}}, \bibinfo {author}
  {\bibfnamefont {C.}~\bibnamefont {Reichl}}, \bibinfo {author} {\bibfnamefont
  {W.}~\bibnamefont {Wegscheider}}, \bibinfo {author} {\bibfnamefont
  {S.}~\bibnamefont {Das~Sarma}}, \ and\ \bibinfo {author} {\bibfnamefont
  {L.~M.~K.}\ \bibnamefont {Vandersypen}},\ }\bibfield  {title} {\enquote
  {\bibinfo {title} {Quantum simulation of a fermi--hubbard model using a
  semiconductor quantum dot array},}\ }\href
  {http://dx.doi.org/10.1038/nature23022} {\bibfield  {journal} {\bibinfo
  {journal} {Nature}\ }\textbf {\bibinfo {volume} {548}},\ \bibinfo {pages}
  {70} (\bibinfo {year} {2017})}\BibitemShut {NoStop}%
\bibitem [{\citenamefont {Yao}\ \emph {et~al.}(2012)\citenamefont {Yao},
  \citenamefont {Jiang}, \citenamefont {Gorshkov}, \citenamefont {Maurer},
  \citenamefont {Giedke}, \citenamefont {Cirac},\ and\ \citenamefont
  {Lukin}}]{2012_lukin_natcom}%
  \BibitemOpen
  \bibfield  {author} {\bibinfo {author} {\bibfnamefont {N.~Y.}\ \bibnamefont
  {Yao}}, \bibinfo {author} {\bibfnamefont {L.}~\bibnamefont {Jiang}}, \bibinfo
  {author} {\bibfnamefont {A.~V.}\ \bibnamefont {Gorshkov}}, \bibinfo {author}
  {\bibfnamefont {P.~C.}\ \bibnamefont {Maurer}}, \bibinfo {author}
  {\bibfnamefont {G.}~\bibnamefont {Giedke}}, \bibinfo {author} {\bibfnamefont
  {J.~I.}\ \bibnamefont {Cirac}}, \ and\ \bibinfo {author} {\bibfnamefont
  {M.~D.}\ \bibnamefont {Lukin}},\ }\bibfield  {title} {\enquote {\bibinfo
  {title} {Scalable architecture for a room temperature solid-state quantum
  information processor},}\ }\href {http://dx.doi.org/10.1038/ncomms1788}
  {\bibfield  {journal} {\bibinfo  {journal} {Nature Communications}\ }\textbf
  {\bibinfo {volume} {3}},\ \bibinfo {pages} {800} (\bibinfo {year}
  {2012})}\BibitemShut {NoStop}%
\bibitem [{\citenamefont {Tosi}\ \emph {et~al.}(2017)\citenamefont {Tosi},
  \citenamefont {Mohiyaddin}, \citenamefont {Schmitt}, \citenamefont {Tenberg},
  \citenamefont {Rahman}, \citenamefont {Klimeck},\ and\ \citenamefont
  {Morello}}]{2017_andrea_natcomm}%
  \BibitemOpen
  \bibfield  {author} {\bibinfo {author} {\bibfnamefont {G.}~\bibnamefont
  {Tosi}}, \bibinfo {author} {\bibfnamefont {F.~A.}\ \bibnamefont
  {Mohiyaddin}}, \bibinfo {author} {\bibfnamefont {V.}~\bibnamefont {Schmitt}},
  \bibinfo {author} {\bibfnamefont {S.}~\bibnamefont {Tenberg}}, \bibinfo
  {author} {\bibfnamefont {R.}~\bibnamefont {Rahman}}, \bibinfo {author}
  {\bibfnamefont {G.}~\bibnamefont {Klimeck}}, \ and\ \bibinfo {author}
  {\bibfnamefont {A.}~\bibnamefont {Morello}},\ }\bibfield  {title} {\enquote
  {\bibinfo {title} {Silicon quantum processor with robust long-distance qubit
  couplings},}\ }\href {\doibase 10.1038/s41467-017-00378-x} {\bibfield
  {journal} {\bibinfo  {journal} {Nature Communications}\ }\textbf {\bibinfo
  {volume} {8}},\ \bibinfo {pages} {450} (\bibinfo {year} {2017})}\BibitemShut
  {NoStop}%
\bibitem [{\citenamefont {Noh}\ and\ \citenamefont
  {Angelakis}(2017)}]{2017_angelakis_rpp}%
  \BibitemOpen
  \bibfield  {author} {\bibinfo {author} {\bibfnamefont {C.}~\bibnamefont
  {Noh}}\ and\ \bibinfo {author} {\bibfnamefont {D.~G}\ \bibnamefont
  {Angelakis}},\ }\bibfield  {title} {\enquote {\bibinfo {title} {Quantum
  simulations and many-body physics with light},}\ }\href
  {http://stacks.iop.org/0034-4885/80/i=1/a=016401} {\bibfield  {journal}
  {\bibinfo  {journal} {Reports on Progress in Physics}\ }\textbf {\bibinfo
  {volume} {80}},\ \bibinfo {pages} {016401} (\bibinfo {year}
  {2017})}\BibitemShut {NoStop}%
\bibitem [{\citenamefont {Chang}\ \emph {et~al.}(2014)\citenamefont {Chang},
  \citenamefont {Vuleti{\'c}},\ and\ \citenamefont
  {Lukin}}]{2014_lukin_natphy}%
  \BibitemOpen
  \bibfield  {author} {\bibinfo {author} {\bibfnamefont {D.~E.}\ \bibnamefont
  {Chang}}, \bibinfo {author} {\bibfnamefont {V.}~\bibnamefont {Vuleti{\'c}}},
  \ and\ \bibinfo {author} {\bibfnamefont {M.~D.}\ \bibnamefont {Lukin}},\
  }\bibfield  {title} {\enquote {\bibinfo {title} {Quantum nonlinear optics
  ---photon by photon},}\ }\href {http://dx.doi.org/10.1038/nphoton.2014.192}
  {\bibfield  {journal} {\bibinfo  {journal} {Nature Photonics}\ }\textbf
  {\bibinfo {volume} {8}},\ \bibinfo {pages} {685} (\bibinfo {year}
  {2014})}\BibitemShut {NoStop}%
\bibitem [{\citenamefont {Porter}\ and\ \citenamefont
  {Thomas}(1956)}]{1956_pt}%
  \BibitemOpen
  \bibfield  {author} {\bibinfo {author} {\bibfnamefont {C.~E.}\ \bibnamefont
  {Porter}}\ and\ \bibinfo {author} {\bibfnamefont {R.~G.}\ \bibnamefont
  {Thomas}},\ }\bibfield  {title} {\enquote {\bibinfo {title} {Fluctuations of
  nuclear reaction widths},}\ }\href {\doibase 10.1103/PhysRev.104.483}
  {\bibfield  {journal} {\bibinfo  {journal} {Phys. Rev.}\ }\textbf {\bibinfo
  {volume} {104}},\ \bibinfo {pages} {483--491} (\bibinfo {year}
  {1956})}\BibitemShut {NoStop}%
\bibitem [{\citenamefont {Ponte}\ \emph {et~al.}(2015)\citenamefont {Ponte},
  \citenamefont {Papi\'{c}}, \citenamefont {Huveneers},\ and\ \citenamefont
  {Abanin}}]{2015_Ponte_PRL}%
  \BibitemOpen
  \bibfield  {author} {\bibinfo {author} {\bibfnamefont {P.}~\bibnamefont
  {Ponte}}, \bibinfo {author} {\bibfnamefont {Z.}~\bibnamefont {Papi\'{c}}},
  \bibinfo {author} {\bibfnamefont {F.}~\bibnamefont {Huveneers}}, \ and\
  \bibinfo {author} {\bibfnamefont {D.~A.}\ \bibnamefont {Abanin}},\ }\bibfield
   {title} {\enquote {\bibinfo {title} {Many-body localization in periodically
  driven systems},}\ }\href {\doibase 10.1103/PhysRevLett.114.140401}
  {\bibfield  {journal} {\bibinfo  {journal} {Phys. Rev. Lett.}\ }\textbf
  {\bibinfo {volume} {114}},\ \bibinfo {pages} {140401} (\bibinfo {year}
  {2015})}\BibitemShut {NoStop}%
\bibitem [{\citenamefont {Abanin}\ \emph {et~al.}(2016)\citenamefont {Abanin},
  \citenamefont {Roeck},\ and\ \citenamefont {Huveneers}}]{2016_Abanin_AoP}%
  \BibitemOpen
  \bibfield  {author} {\bibinfo {author} {\bibfnamefont {D.~A.}\ \bibnamefont
  {Abanin}}, \bibinfo {author} {\bibfnamefont {W.~D.}\ \bibnamefont {Roeck}}, \
  and\ \bibinfo {author} {\bibfnamefont {F.}~\bibnamefont {Huveneers}},\
  }\bibfield  {title} {\enquote {\bibinfo {title} {Theory of many-body
  localization in periodically driven systems},}\ }\href {\doibase
  https://doi.org/10.1016/j.aop.2016.03.010} {\bibfield  {journal} {\bibinfo
  {journal} {Annals of Physics}\ }\textbf {\bibinfo {volume} {372}},\ \bibinfo
  {pages} {1 -- 11} (\bibinfo {year} {2016})}\BibitemShut {NoStop}%
\bibitem [{\citenamefont {D’Alessio}\ and\ \citenamefont
  {Polkovnikov}(2013)}]{2013_Alessio_AoP}%
  \BibitemOpen
  \bibfield  {author} {\bibinfo {author} {\bibfnamefont {L.}~\bibnamefont
  {D’Alessio}}\ and\ \bibinfo {author} {\bibfnamefont {A.}~\bibnamefont
  {Polkovnikov}},\ }\bibfield  {title} {\enquote {\bibinfo {title} {Many-body
  energy localization transition in periodically driven systems},}\ }\href
  {\doibase https://doi.org/10.1016/j.aop.2013.02.011} {\bibfield  {journal}
  {\bibinfo  {journal} {Annals of Physics}\ }\textbf {\bibinfo {volume}
  {333}},\ \bibinfo {pages} {19 -- 33} (\bibinfo {year} {2013})}\BibitemShut
  {NoStop}%
\bibitem [{\citenamefont {D'Alessio}\ and\ \citenamefont
  {Rigol}(2014)}]{2014_Rigol_PRX}%
  \BibitemOpen
  \bibfield  {author} {\bibinfo {author} {\bibfnamefont {L.}~\bibnamefont
  {D'Alessio}}\ and\ \bibinfo {author} {\bibfnamefont {M.}~\bibnamefont
  {Rigol}},\ }\bibfield  {title} {\enquote {\bibinfo {title} {Long-time
  behavior of isolated periodically driven interacting lattice systems},}\
  }\href {\doibase 10.1103/PhysRevX.4.041048} {\bibfield  {journal} {\bibinfo
  {journal} {Phys. Rev. X}\ }\textbf {\bibinfo {volume} {4}},\ \bibinfo {pages}
  {041048} (\bibinfo {year} {2014})}\BibitemShut {NoStop}%
\bibitem [{\citenamefont {Roushan}\ \emph {et~al.}(2017)\citenamefont
  {Roushan}, \citenamefont {Neill}, \citenamefont {Tangpanitanon},
  \citenamefont {Bastidas}, \citenamefont {Megrant}, \citenamefont {Barends},
  \citenamefont {Chen}, \citenamefont {Chen}, \citenamefont {Chiaro},
  \citenamefont {Dunsworth}, \citenamefont {Fowler}, \citenamefont {Foxen},
  \citenamefont {Giustina}, \citenamefont {Jeffrey}, \citenamefont {Kelly},
  \citenamefont {Lucero}, \citenamefont {Mutus}, \citenamefont {Neeley},
  \citenamefont {Quintana}, \citenamefont {Sank}, \citenamefont {Vainsencher},
  \citenamefont {Wenner}, \citenamefont {White}, \citenamefont {Neven},
  \citenamefont {Angelakis},\ and\ \citenamefont
  {Martinis}}]{2017_Roushan_Sci}%
  \BibitemOpen
  \bibfield  {author} {\bibinfo {author} {\bibfnamefont {P.}~\bibnamefont
  {Roushan}}, \bibinfo {author} {\bibfnamefont {C.}~\bibnamefont {Neill}},
  \bibinfo {author} {\bibfnamefont {J.}~\bibnamefont {Tangpanitanon}}, \bibinfo
  {author} {\bibfnamefont {V.~M.}\ \bibnamefont {Bastidas}}, \bibinfo {author}
  {\bibfnamefont {A.}~\bibnamefont {Megrant}}, \bibinfo {author} {\bibfnamefont
  {R.}~\bibnamefont {Barends}}, \bibinfo {author} {\bibfnamefont
  {Y.}~\bibnamefont {Chen}}, \bibinfo {author} {\bibfnamefont {Z.}~\bibnamefont
  {Chen}}, \bibinfo {author} {\bibfnamefont {B.}~\bibnamefont {Chiaro}},
  \bibinfo {author} {\bibfnamefont {A.}~\bibnamefont {Dunsworth}}, \bibinfo
  {author} {\bibfnamefont {A.}~\bibnamefont {Fowler}}, \bibinfo {author}
  {\bibfnamefont {B.}~\bibnamefont {Foxen}}, \bibinfo {author} {\bibfnamefont
  {M.}~\bibnamefont {Giustina}}, \bibinfo {author} {\bibfnamefont
  {E.}~\bibnamefont {Jeffrey}}, \bibinfo {author} {\bibfnamefont
  {J.}~\bibnamefont {Kelly}}, \bibinfo {author} {\bibfnamefont
  {E.}~\bibnamefont {Lucero}}, \bibinfo {author} {\bibfnamefont
  {J.}~\bibnamefont {Mutus}}, \bibinfo {author} {\bibfnamefont
  {M.}~\bibnamefont {Neeley}}, \bibinfo {author} {\bibfnamefont
  {C.}~\bibnamefont {Quintana}}, \bibinfo {author} {\bibfnamefont
  {D.}~\bibnamefont {Sank}}, \bibinfo {author} {\bibfnamefont {A.}~\bibnamefont
  {Vainsencher}}, \bibinfo {author} {\bibfnamefont {J.}~\bibnamefont {Wenner}},
  \bibinfo {author} {\bibfnamefont {T.}~\bibnamefont {White}}, \bibinfo
  {author} {\bibfnamefont {H.}~\bibnamefont {Neven}}, \bibinfo {author}
  {\bibfnamefont {D.~G.}\ \bibnamefont {Angelakis}}, \ and\ \bibinfo {author}
  {\bibfnamefont {J.}~\bibnamefont {Martinis}},\ }\bibfield  {title} {\enquote
  {\bibinfo {title} {Spectroscopic signatures of localization with interacting
  photons in superconducting qubits},}\ }\href {\doibase
  10.1126/science.aao1401} {\bibfield  {journal} {\bibinfo  {journal}
  {Science}\ }\textbf {\bibinfo {volume} {358}},\ \bibinfo {pages} {1175--1179}
  (\bibinfo {year} {2017})}\BibitemShut {NoStop}%
\bibitem [{\citenamefont {Jebara}(2004)}]{2004_tony}%
  \BibitemOpen
  \bibfield  {author} {\bibinfo {author} {\bibfnamefont {Tony}\ \bibnamefont
  {Jebara}},\ }\href@noop {} {\emph {\bibinfo {title} {Machine Learning}}}\
  (\bibinfo  {publisher} {Springer International Publishing, US},\ \bibinfo
  {year} {2004})\BibitemShut {NoStop}%
\bibitem [{\citenamefont {Bermejo-Vega}\ \emph {et~al.}(2018)\citenamefont
  {Bermejo-Vega}, \citenamefont {Hangleiter}, \citenamefont {Schwarz},
  \citenamefont {Raussendorf},\ and\ \citenamefont {Eisert}}]{2018_eisert_prx}%
  \BibitemOpen
  \bibfield  {author} {\bibinfo {author} {\bibfnamefont {Juan}\ \bibnamefont
  {Bermejo-Vega}}, \bibinfo {author} {\bibfnamefont {Dominik}\ \bibnamefont
  {Hangleiter}}, \bibinfo {author} {\bibfnamefont {Martin}\ \bibnamefont
  {Schwarz}}, \bibinfo {author} {\bibfnamefont {Robert}\ \bibnamefont
  {Raussendorf}}, \ and\ \bibinfo {author} {\bibfnamefont {Jens}\ \bibnamefont
  {Eisert}},\ }\bibfield  {title} {\enquote {\bibinfo {title} {Architectures
  for quantum simulation showing a quantum speedup},}\ }\href {\doibase
  10.1103/PhysRevX.8.021010} {\bibfield  {journal} {\bibinfo  {journal} {Phys.
  Rev. X}\ }\textbf {\bibinfo {volume} {8}},\ \bibinfo {pages} {021010}
  (\bibinfo {year} {2018})}\BibitemShut {NoStop}%
\bibitem [{\citenamefont {Gao}\ \emph {et~al.}(2017)\citenamefont {Gao},
  \citenamefont {Wang},\ and\ \citenamefont {Duan}}]{PhysRevLett.118.040502}%
  \BibitemOpen
  \bibfield  {author} {\bibinfo {author} {\bibfnamefont {Xun}\ \bibnamefont
  {Gao}}, \bibinfo {author} {\bibfnamefont {Sheng-Tao}\ \bibnamefont {Wang}}, \
  and\ \bibinfo {author} {\bibfnamefont {L.-M.}\ \bibnamefont {Duan}},\
  }\bibfield  {title} {\enquote {\bibinfo {title} {Quantum supremacy for
  simulating a translation-invariant ising spin model},}\ }\href {\doibase
  10.1103/PhysRevLett.118.040502} {\bibfield  {journal} {\bibinfo  {journal}
  {Phys. Rev. Lett.}\ }\textbf {\bibinfo {volume} {118}},\ \bibinfo {pages}
  {040502} (\bibinfo {year} {2017})}\BibitemShut {NoStop}%
\bibitem [{Note1()}]{Note1}%
  \BibitemOpen
  \bibinfo {note} {Since single-qubit gates can usually be implemented in a
  much faster time scale in the experiment, we assume that each layer takes
  $\sim \pi /4J$ which is the time to implement the controlled-Z gate using the
  $ZZ$ coupling. This condition sets the driving frequency to be $\omega
  =8J$.}\BibitemShut {Stop}%
\bibitem [{\citenamefont {Haake}(2010)}]{2010_haake}%
  \BibitemOpen
  \bibfield  {author} {\bibinfo {author} {\bibfnamefont {Fritz}\ \bibnamefont
  {Haake}},\ }\href@noop {} {\emph {\bibinfo {title} {Quantum Signatures of
  Chaos}}}\ (\bibinfo  {publisher} {Springer International Publishing, US},\
  \bibinfo {year} {2010})\BibitemShut {NoStop}%
\bibitem [{\citenamefont {Kos}\ \emph {et~al.}(2018)\citenamefont {Kos},
  \citenamefont {Ljubotina},\ and\ \citenamefont {Prosen}}]{PhysRevX.8.021062}%
  \BibitemOpen
  \bibfield  {author} {\bibinfo {author} {\bibfnamefont {Pavel}\ \bibnamefont
  {Kos}}, \bibinfo {author} {\bibfnamefont {Marko}\ \bibnamefont {Ljubotina}},
  \ and\ \bibinfo {author} {\bibfnamefont {Toma\ifmmode
  \check{z}\else~\v{z}\fi{}}\ \bibnamefont {Prosen}},\ }\bibfield  {title}
  {\enquote {\bibinfo {title} {Many-body quantum chaos: Analytic connection to
  random matrix theory},}\ }\href {\doibase 10.1103/PhysRevX.8.021062}
  {\bibfield  {journal} {\bibinfo  {journal} {Phys. Rev. X}\ }\textbf {\bibinfo
  {volume} {8}},\ \bibinfo {pages} {021062} (\bibinfo {year}
  {2018})}\BibitemShut {NoStop}%
\bibitem [{\citenamefont {{Villalonga}}\ \emph {et~al.}(2019)\citenamefont
  {{Villalonga}}, \citenamefont {{Lyakh}}, \citenamefont {{Boixo}},
  \citenamefont {{Neven}}, \citenamefont {{Humble}}, \citenamefont {{Biswas}},
  \citenamefont {{Rieffel}}, \citenamefont {{Ho}},\ and\ \citenamefont
  {{Mandr{\`a}}}}]{2019arXiv190500444V}%
  \BibitemOpen
  \bibfield  {author} {\bibinfo {author} {\bibfnamefont {Benjamin}\
  \bibnamefont {{Villalonga}}}, \bibinfo {author} {\bibfnamefont {Dmitry}\
  \bibnamefont {{Lyakh}}}, \bibinfo {author} {\bibfnamefont {Sergio}\
  \bibnamefont {{Boixo}}}, \bibinfo {author} {\bibfnamefont {Hartmut}\
  \bibnamefont {{Neven}}}, \bibinfo {author} {\bibfnamefont {Travis~S.}\
  \bibnamefont {{Humble}}}, \bibinfo {author} {\bibfnamefont {Rupak}\
  \bibnamefont {{Biswas}}}, \bibinfo {author} {\bibfnamefont {Eleanor~G.}\
  \bibnamefont {{Rieffel}}}, \bibinfo {author} {\bibfnamefont {Alan}\
  \bibnamefont {{Ho}}}, \ and\ \bibinfo {author} {\bibfnamefont {Salvatore}\
  \bibnamefont {{Mandr{\`a}}}},\ }\bibfield  {title} {\enquote {\bibinfo
  {title} {{Establishing the Quantum Supremacy Frontier with a 281 Pflop/s
  Simulation}},}\ }\href@noop {} {\bibfield  {journal} {\bibinfo  {journal}
  {arXiv e-prints}\ ,\ \bibinfo {eid} {arXiv:1905.00444}} (\bibinfo {year}
  {2019})},\ \Eprint {http://arxiv.org/abs/1905.00444} {arXiv:1905.00444
  [quant-ph]} \BibitemShut {NoStop}%
\bibitem [{\citenamefont {H\"{a}ner}\ and\ \citenamefont
  {Steiger}(2017)}]{Haner:2017:PSQ:3126908.3126947}%
  \BibitemOpen
  \bibfield  {author} {\bibinfo {author} {\bibfnamefont {Thomas}\ \bibnamefont
  {H\"{a}ner}}\ and\ \bibinfo {author} {\bibfnamefont {Damian~S.}\ \bibnamefont
  {Steiger}},\ }\bibfield  {title} {\enquote {\bibinfo {title} {0.5 petabyte
  simulation of a 45-qubit quantum circuit},}\ }in\ \href {\doibase
  10.1145/3126908.3126947} {\emph {\bibinfo {booktitle} {Proceedings of the
  International Conference for High Performance Computing, Networking, Storage
  and Analysis}}},\ \bibinfo {series and number} {SC '17}\ (\bibinfo
  {publisher} {ACM},\ \bibinfo {address} {New York, NY, USA},\ \bibinfo {year}
  {2017})\ pp.\ \bibinfo {pages} {33:1--33:10}\BibitemShut {NoStop}%
\bibitem [{\citenamefont {{Pednault}}\ \emph {et~al.}(2017)\citenamefont
  {{Pednault}}, \citenamefont {{Gunnels}}, \citenamefont {{Nannicini}},
  \citenamefont {{Horesh}}, \citenamefont {{Magerlein}}, \citenamefont
  {{Solomonik}}, \citenamefont {{Draeger}}, \citenamefont {{Holland}},\ and\
  \citenamefont {{Wisnieff}}}]{2017arXiv171005867P}%
  \BibitemOpen
  \bibfield  {author} {\bibinfo {author} {\bibfnamefont {Edwin}\ \bibnamefont
  {{Pednault}}}, \bibinfo {author} {\bibfnamefont {John~A.}\ \bibnamefont
  {{Gunnels}}}, \bibinfo {author} {\bibfnamefont {Giacomo}\ \bibnamefont
  {{Nannicini}}}, \bibinfo {author} {\bibfnamefont {Lior}\ \bibnamefont
  {{Horesh}}}, \bibinfo {author} {\bibfnamefont {Thomas}\ \bibnamefont
  {{Magerlein}}}, \bibinfo {author} {\bibfnamefont {Edgar}\ \bibnamefont
  {{Solomonik}}}, \bibinfo {author} {\bibfnamefont {Erik~W.}\ \bibnamefont
  {{Draeger}}}, \bibinfo {author} {\bibfnamefont {Eric~T.}\ \bibnamefont
  {{Holland}}}, \ and\ \bibinfo {author} {\bibfnamefont {Robert}\ \bibnamefont
  {{Wisnieff}}},\ }\bibfield  {title} {\enquote {\bibinfo {title} {{Breaking
  the 49-Qubit Barrier in the Simulation of Quantum Circuits}},}\ }\href@noop
  {} {\bibfield  {journal} {\bibinfo  {journal} {arXiv e-prints}\ ,\ \bibinfo
  {eid} {arXiv:1710.05867}} (\bibinfo {year} {2017})},\ \Eprint
  {http://arxiv.org/abs/1710.05867} {arXiv:1710.05867 [quant-ph]} \BibitemShut
  {NoStop}%
\bibitem [{\citenamefont {Bukov}\ \emph {et~al.}(2015)\citenamefont {Bukov},
  \citenamefont {D'Alessio},\ and\ \citenamefont
  {Polkovnikov}}]{2015_Bukov_AiP}%
  \BibitemOpen
  \bibfield  {author} {\bibinfo {author} {\bibfnamefont {M.}~\bibnamefont
  {Bukov}}, \bibinfo {author} {\bibfnamefont {L.}~\bibnamefont {D'Alessio}}, \
  and\ \bibinfo {author} {\bibfnamefont {A.}~\bibnamefont {Polkovnikov}},\
  }\bibfield  {title} {\enquote {\bibinfo {title} {Universal high-frequency
  behavior of periodically driven systems: from dynamical stabilization to
  floquet engineering},}\ }\href {\doibase 10.1080/00018732.2015.1055918}
  {\bibfield  {journal} {\bibinfo  {journal} {Advances in Physics}\ }\textbf
  {\bibinfo {volume} {64}},\ \bibinfo {pages} {139--226} (\bibinfo {year}
  {2015})}\BibitemShut {NoStop}%
\bibitem [{\citenamefont {Vidal}(2003)}]{PhysRevLett.91.147902}%
  \BibitemOpen
  \bibfield  {author} {\bibinfo {author} {\bibfnamefont {Guifr\'e}\
  \bibnamefont {Vidal}},\ }\bibfield  {title} {\enquote {\bibinfo {title}
  {Efficient classical simulation of slightly entangled quantum
  computations},}\ }\href {\doibase 10.1103/PhysRevLett.91.147902} {\bibfield
  {journal} {\bibinfo  {journal} {Phys. Rev. Lett.}\ }\textbf {\bibinfo
  {volume} {91}},\ \bibinfo {pages} {147902} (\bibinfo {year}
  {2003})}\BibitemShut {NoStop}%
\bibitem [{\citenamefont {Schollwöck}(2011)}]{SCHOLLWOCK201196}%
  \BibitemOpen
  \bibfield  {author} {\bibinfo {author} {\bibfnamefont {Ulrich}\ \bibnamefont
  {Schollwöck}},\ }\bibfield  {title} {\enquote {\bibinfo {title} {The
  density-matrix renormalization group in the age of matrix product states},}\
  }\href {\doibase https://doi.org/10.1016/j.aop.2010.09.012} {\bibfield
  {journal} {\bibinfo  {journal} {Annals of Physics}\ }\textbf {\bibinfo
  {volume} {326}},\ \bibinfo {pages} {96 -- 192} (\bibinfo {year} {2011})},\
  \bibinfo {note} {january 2011 Special Issue}\BibitemShut {NoStop}%
\bibitem [{\citenamefont {Russomanno}\ \emph {et~al.}(2016)\citenamefont
  {Russomanno}, \citenamefont {Santoro},\ and\ \citenamefont
  {Fazio}}]{Russomanno_2016}%
  \BibitemOpen
  \bibfield  {author} {\bibinfo {author} {\bibfnamefont {Angelo}\ \bibnamefont
  {Russomanno}}, \bibinfo {author} {\bibfnamefont {Giuseppe~E}\ \bibnamefont
  {Santoro}}, \ and\ \bibinfo {author} {\bibfnamefont {Rosario}\ \bibnamefont
  {Fazio}},\ }\bibfield  {title} {\enquote {\bibinfo {title} {Entanglement
  entropy in a periodically driven ising chain},}\ }\href {\doibase
  10.1088/1742-5468/2016/07/073101} {\bibfield  {journal} {\bibinfo  {journal}
  {Journal of Statistical Mechanics: Theory and Experiment}\ }\textbf {\bibinfo
  {volume} {2016}},\ \bibinfo {pages} {073101} (\bibinfo {year}
  {2016})}\BibitemShut {NoStop}%
\bibitem [{\citenamefont {Bastidas}\ \emph {et~al.}(2014)\citenamefont
  {Bastidas}, \citenamefont {P\'erez-Fern\'andez}, \citenamefont {Vogl},\ and\
  \citenamefont {Brandes}}]{PhysRevLett.112.140408}%
  \BibitemOpen
  \bibfield  {author} {\bibinfo {author} {\bibfnamefont {Victor~Manuel}\
  \bibnamefont {Bastidas}}, \bibinfo {author} {\bibfnamefont {Pedro}\
  \bibnamefont {P\'erez-Fern\'andez}}, \bibinfo {author} {\bibfnamefont
  {Malte}\ \bibnamefont {Vogl}}, \ and\ \bibinfo {author} {\bibfnamefont
  {Tobias}\ \bibnamefont {Brandes}},\ }\bibfield  {title} {\enquote {\bibinfo
  {title} {Quantum criticality and dynamical instability in the kicked-top
  model},}\ }\href {\doibase 10.1103/PhysRevLett.112.140408} {\bibfield
  {journal} {\bibinfo  {journal} {Phys. Rev. Lett.}\ }\textbf {\bibinfo
  {volume} {112}},\ \bibinfo {pages} {140408} (\bibinfo {year}
  {2014})}\BibitemShut {NoStop}%
\bibitem [{\citenamefont {Goldstein}\ \emph {et~al.}(2006)\citenamefont
  {Goldstein}, \citenamefont {Lebowitz}, \citenamefont {Tumulka},\ and\
  \citenamefont {Zangh\`{\i}}}]{2006_Goldstein_PRL}%
  \BibitemOpen
  \bibfield  {author} {\bibinfo {author} {\bibfnamefont {S.}~\bibnamefont
  {Goldstein}}, \bibinfo {author} {\bibfnamefont {J.~L.}\ \bibnamefont
  {Lebowitz}}, \bibinfo {author} {\bibfnamefont {R.}~\bibnamefont {Tumulka}}, \
  and\ \bibinfo {author} {\bibfnamefont {N.}~\bibnamefont {Zangh\`{\i}}},\
  }\bibfield  {title} {\enquote {\bibinfo {title} {Canonical typicality},}\
  }\href {\doibase 10.1103/PhysRevLett.96.050403} {\bibfield  {journal}
  {\bibinfo  {journal} {Phys. Rev. Lett.}\ }\textbf {\bibinfo {volume} {96}},\
  \bibinfo {pages} {050403} (\bibinfo {year} {2006})}\BibitemShut {NoStop}%
\bibitem [{\citenamefont {D'Alessio}\ \emph
  {et~al.}(2016{\natexlab{a}})\citenamefont {D'Alessio}, \citenamefont {Kafri},
  \citenamefont {Polkovnikov},\ and\ \citenamefont {Rigol}}]{2016_Alessio_AiP}%
  \BibitemOpen
  \bibfield  {author} {\bibinfo {author} {\bibfnamefont {L.}~\bibnamefont
  {D'Alessio}}, \bibinfo {author} {\bibfnamefont {Y.}~\bibnamefont {Kafri}},
  \bibinfo {author} {\bibfnamefont {A.}~\bibnamefont {Polkovnikov}}, \ and\
  \bibinfo {author} {\bibfnamefont {M.}~\bibnamefont {Rigol}},\ }\bibfield
  {title} {\enquote {\bibinfo {title} {From quantum chaos and eigenstate
  thermalization to statistical mechanics and thermodynamics},}\ }\href
  {\doibase 10.1080/00018732.2016.1198134} {\bibfield  {journal} {\bibinfo
  {journal} {Advances in Physics}\ }\textbf {\bibinfo {volume} {65}},\ \bibinfo
  {pages} {239--362} (\bibinfo {year} {2016}{\natexlab{a}})}\BibitemShut
  {NoStop}%
\bibitem [{\citenamefont {Nandkishore}\ and\ \citenamefont
  {Huse}(2015)}]{2015_huse_arcmp}%
  \BibitemOpen
  \bibfield  {author} {\bibinfo {author} {\bibfnamefont {Rahul}\ \bibnamefont
  {Nandkishore}}\ and\ \bibinfo {author} {\bibfnamefont {David~A.}\
  \bibnamefont {Huse}},\ }\bibfield  {title} {\enquote {\bibinfo {title}
  {Many-body localization and thermalization in quantum statistical
  mechanics},}\ }\href {\doibase 10.1146/annurev-conmatphys-031214-014726}
  {\bibfield  {journal} {\bibinfo  {journal} {Annual Review of Condensed Matter
  Physics}\ }\textbf {\bibinfo {volume} {6}},\ \bibinfo {pages} {15--38}
  (\bibinfo {year} {2015})},\ \Eprint
  {http://arxiv.org/abs/https://doi.org/10.1146/annurev-conmatphys-031214-014726}
  {https://doi.org/10.1146/annurev-conmatphys-031214-014726} \BibitemShut
  {NoStop}%
\bibitem [{\citenamefont {D'Alessio}\ \emph
  {et~al.}(2016{\natexlab{b}})\citenamefont {D'Alessio}, \citenamefont {Kafri},
  \citenamefont {Polkovnikov},\ and\ \citenamefont {Rigol}}]{2016_rigol_ap}%
  \BibitemOpen
  \bibfield  {author} {\bibinfo {author} {\bibfnamefont {Luca}\ \bibnamefont
  {D'Alessio}}, \bibinfo {author} {\bibfnamefont {Yariv}\ \bibnamefont
  {Kafri}}, \bibinfo {author} {\bibfnamefont {Anatoli}\ \bibnamefont
  {Polkovnikov}}, \ and\ \bibinfo {author} {\bibfnamefont {Marcos}\
  \bibnamefont {Rigol}},\ }\bibfield  {title} {\enquote {\bibinfo {title} {From
  quantum chaos and eigenstate thermalization to statistical mechanics and
  thermodynamics},}\ }\href {\doibase 10.1080/00018732.2016.1198134} {\bibfield
   {journal} {\bibinfo  {journal} {Advances in Physics}\ }\textbf {\bibinfo
  {volume} {65}},\ \bibinfo {pages} {239--362} (\bibinfo {year}
  {2016}{\natexlab{b}})},\ \Eprint
  {http://arxiv.org/abs/https://doi.org/10.1080/00018732.2016.1198134}
  {https://doi.org/10.1080/00018732.2016.1198134} \BibitemShut {NoStop}%
\bibitem [{\citenamefont {{Abanin}}\ \emph {et~al.}(2018)\citenamefont
  {{Abanin}}, \citenamefont {{Altman}}, \citenamefont {{Bloch}},\ and\
  \citenamefont {{Serbyn}}}]{2018arXiv180411065A}%
  \BibitemOpen
  \bibfield  {author} {\bibinfo {author} {\bibfnamefont {Dmitry~A.}\
  \bibnamefont {{Abanin}}}, \bibinfo {author} {\bibfnamefont {Ehud}\
  \bibnamefont {{Altman}}}, \bibinfo {author} {\bibfnamefont {Immanuel}\
  \bibnamefont {{Bloch}}}, \ and\ \bibinfo {author} {\bibfnamefont {Maksym}\
  \bibnamefont {{Serbyn}}},\ }\bibfield  {title} {\enquote {\bibinfo {title}
  {{Many-body localization, thermalization, and entanglement}},}\ }\href@noop
  {} {\bibfield  {journal} {\bibinfo  {journal} {arXiv e-prints}\ ,\ \bibinfo
  {eid} {arXiv:1804.11065}} (\bibinfo {year} {2018})},\ \Eprint
  {http://arxiv.org/abs/1804.11065} {arXiv:1804.11065 [cond-mat.dis-nn]}
  \BibitemShut {NoStop}%
\bibitem [{\citenamefont {Pal}\ and\ \citenamefont
  {Huse}(2010)}]{PhysRevB.82.174411}%
  \BibitemOpen
  \bibfield  {author} {\bibinfo {author} {\bibfnamefont {Arijeet}\ \bibnamefont
  {Pal}}\ and\ \bibinfo {author} {\bibfnamefont {David~A.}\ \bibnamefont
  {Huse}},\ }\bibfield  {title} {\enquote {\bibinfo {title} {Many-body
  localization phase transition},}\ }\href {\doibase
  10.1103/PhysRevB.82.174411} {\bibfield  {journal} {\bibinfo  {journal} {Phys.
  Rev. B}\ }\textbf {\bibinfo {volume} {82}},\ \bibinfo {pages} {174411}
  (\bibinfo {year} {2010})}\BibitemShut {NoStop}%
\bibitem [{\citenamefont {Schreiber}\ \emph {et~al.}(2015)\citenamefont
  {Schreiber}, \citenamefont {Hodgman}, \citenamefont {Bordia}, \citenamefont
  {L{\"u}schen}, \citenamefont {Fischer}, \citenamefont {Vosk}, \citenamefont
  {Altman}, \citenamefont {Schneider},\ and\ \citenamefont
  {Bloch}}]{Schreiber842}%
  \BibitemOpen
  \bibfield  {author} {\bibinfo {author} {\bibfnamefont {Michael}\ \bibnamefont
  {Schreiber}}, \bibinfo {author} {\bibfnamefont {Sean~S.}\ \bibnamefont
  {Hodgman}}, \bibinfo {author} {\bibfnamefont {Pranjal}\ \bibnamefont
  {Bordia}}, \bibinfo {author} {\bibfnamefont {Henrik~P.}\ \bibnamefont
  {L{\"u}schen}}, \bibinfo {author} {\bibfnamefont {Mark~H.}\ \bibnamefont
  {Fischer}}, \bibinfo {author} {\bibfnamefont {Ronen}\ \bibnamefont {Vosk}},
  \bibinfo {author} {\bibfnamefont {Ehud}\ \bibnamefont {Altman}}, \bibinfo
  {author} {\bibfnamefont {Ulrich}\ \bibnamefont {Schneider}}, \ and\ \bibinfo
  {author} {\bibfnamefont {Immanuel}\ \bibnamefont {Bloch}},\ }\bibfield
  {title} {\enquote {\bibinfo {title} {Observation of many-body localization of
  interacting fermions in a quasirandom optical lattice},}\ }\href {\doibase
  10.1126/science.aaa7432} {\bibfield  {journal} {\bibinfo  {journal}
  {Science}\ }\textbf {\bibinfo {volume} {349}},\ \bibinfo {pages} {842--845}
  (\bibinfo {year} {2015})}\BibitemShut {NoStop}%
\bibitem [{\citenamefont {Bordia}\ \emph {et~al.}(2016)\citenamefont {Bordia},
  \citenamefont {L\"uschen}, \citenamefont {Hodgman}, \citenamefont
  {Schreiber}, \citenamefont {Bloch},\ and\ \citenamefont
  {Schneider}}]{PhysRevLett.116.140401}%
  \BibitemOpen
  \bibfield  {author} {\bibinfo {author} {\bibfnamefont {Pranjal}\ \bibnamefont
  {Bordia}}, \bibinfo {author} {\bibfnamefont {Henrik~P.}\ \bibnamefont
  {L\"uschen}}, \bibinfo {author} {\bibfnamefont {Sean~S.}\ \bibnamefont
  {Hodgman}}, \bibinfo {author} {\bibfnamefont {Michael}\ \bibnamefont
  {Schreiber}}, \bibinfo {author} {\bibfnamefont {Immanuel}\ \bibnamefont
  {Bloch}}, \ and\ \bibinfo {author} {\bibfnamefont {Ulrich}\ \bibnamefont
  {Schneider}},\ }\bibfield  {title} {\enquote {\bibinfo {title} {Coupling
  identical one-dimensional many-body localized systems},}\ }\href {\doibase
  10.1103/PhysRevLett.116.140401} {\bibfield  {journal} {\bibinfo  {journal}
  {Phys. Rev. Lett.}\ }\textbf {\bibinfo {volume} {116}},\ \bibinfo {pages}
  {140401} (\bibinfo {year} {2016})}\BibitemShut {NoStop}%
\bibitem [{\citenamefont {{Lukin}}\ \emph {et~al.}(2018)\citenamefont
  {{Lukin}}, \citenamefont {{Rispoli}}, \citenamefont {{Schittko}},
  \citenamefont {{Tai}}, \citenamefont {{Kaufman}}, \citenamefont {{Choi}},
  \citenamefont {{Khemani}}, \citenamefont {{L{\'e}onard}},\ and\ \citenamefont
  {{Greiner}}}]{2018arXiv180509819L}%
  \BibitemOpen
  \bibfield  {author} {\bibinfo {author} {\bibfnamefont {Alexander}\
  \bibnamefont {{Lukin}}}, \bibinfo {author} {\bibfnamefont {Matthew}\
  \bibnamefont {{Rispoli}}}, \bibinfo {author} {\bibfnamefont {Robert}\
  \bibnamefont {{Schittko}}}, \bibinfo {author} {\bibfnamefont {M.~Eric}\
  \bibnamefont {{Tai}}}, \bibinfo {author} {\bibfnamefont {Adam~M.}\
  \bibnamefont {{Kaufman}}}, \bibinfo {author} {\bibfnamefont {Soonwon}\
  \bibnamefont {{Choi}}}, \bibinfo {author} {\bibfnamefont {Vedika}\
  \bibnamefont {{Khemani}}}, \bibinfo {author} {\bibfnamefont {Julian}\
  \bibnamefont {{L{\'e}onard}}}, \ and\ \bibinfo {author} {\bibfnamefont
  {Markus}\ \bibnamefont {{Greiner}}},\ }\bibfield  {title} {\enquote {\bibinfo
  {title} {{Probing entanglement in a many-body-localized system}},}\
  }\href@noop {} {\bibfield  {journal} {\bibinfo  {journal} {arXiv e-prints}\
  ,\ \bibinfo {eid} {arXiv:1805.09819}} (\bibinfo {year} {2018})},\ \Eprint
  {http://arxiv.org/abs/1805.09819} {arXiv:1805.09819 [cond-mat.quant-gas]}
  \BibitemShut {NoStop}%
\bibitem [{\citenamefont {Xu}\ \emph {et~al.}(2018)\citenamefont {Xu},
  \citenamefont {Chen}, \citenamefont {Zeng}, \citenamefont {Zhang},
  \citenamefont {Song}, \citenamefont {Liu}, \citenamefont {Guo}, \citenamefont
  {Zhang}, \citenamefont {Xu}, \citenamefont {Deng}, \citenamefont {Huang},
  \citenamefont {Wang}, \citenamefont {Zhu}, \citenamefont {Zheng},\ and\
  \citenamefont {Fan}}]{PhysRevLett.120.050507}%
  \BibitemOpen
  \bibfield  {author} {\bibinfo {author} {\bibfnamefont {Kai}\ \bibnamefont
  {Xu}}, \bibinfo {author} {\bibfnamefont {Jin-Jun}\ \bibnamefont {Chen}},
  \bibinfo {author} {\bibfnamefont {Yu}~\bibnamefont {Zeng}}, \bibinfo {author}
  {\bibfnamefont {Yu-Ran}\ \bibnamefont {Zhang}}, \bibinfo {author}
  {\bibfnamefont {Chao}\ \bibnamefont {Song}}, \bibinfo {author} {\bibfnamefont
  {Wuxin}\ \bibnamefont {Liu}}, \bibinfo {author} {\bibfnamefont {Qiujiang}\
  \bibnamefont {Guo}}, \bibinfo {author} {\bibfnamefont {Pengfei}\ \bibnamefont
  {Zhang}}, \bibinfo {author} {\bibfnamefont {Da}~\bibnamefont {Xu}}, \bibinfo
  {author} {\bibfnamefont {Hui}\ \bibnamefont {Deng}}, \bibinfo {author}
  {\bibfnamefont {Keqiang}\ \bibnamefont {Huang}}, \bibinfo {author}
  {\bibfnamefont {H.}~\bibnamefont {Wang}}, \bibinfo {author} {\bibfnamefont
  {Xiaobo}\ \bibnamefont {Zhu}}, \bibinfo {author} {\bibfnamefont {Dongning}\
  \bibnamefont {Zheng}}, \ and\ \bibinfo {author} {\bibfnamefont {Heng}\
  \bibnamefont {Fan}},\ }\bibfield  {title} {\enquote {\bibinfo {title}
  {Emulating many-body localization with a superconducting quantum
  processor},}\ }\href {\doibase 10.1103/PhysRevLett.120.050507} {\bibfield
  {journal} {\bibinfo  {journal} {Phys. Rev. Lett.}\ }\textbf {\bibinfo
  {volume} {120}},\ \bibinfo {pages} {050507} (\bibinfo {year}
  {2018})}\BibitemShut {NoStop}%
\bibitem [{\citenamefont {Smith}\ \emph {et~al.}(2016)\citenamefont {Smith},
  \citenamefont {Lee}, \citenamefont {Richerme}, \citenamefont {Neyenhuis},
  \citenamefont {Hess}, \citenamefont {Hauke}, \citenamefont {Heyl},
  \citenamefont {Huse},\ and\ \citenamefont {Monroe}}]{2016_monroe_nat}%
  \BibitemOpen
  \bibfield  {author} {\bibinfo {author} {\bibfnamefont {J.}~\bibnamefont
  {Smith}}, \bibinfo {author} {\bibfnamefont {A.}~\bibnamefont {Lee}}, \bibinfo
  {author} {\bibfnamefont {P.}~\bibnamefont {Richerme}}, \bibinfo {author}
  {\bibfnamefont {B.}~\bibnamefont {Neyenhuis}}, \bibinfo {author}
  {\bibfnamefont {P.~W.}\ \bibnamefont {Hess}}, \bibinfo {author}
  {\bibfnamefont {P.}~\bibnamefont {Hauke}}, \bibinfo {author} {\bibfnamefont
  {M.}~\bibnamefont {Heyl}}, \bibinfo {author} {\bibfnamefont {D.~A.}\
  \bibnamefont {Huse}}, \ and\ \bibinfo {author} {\bibfnamefont
  {C.}~\bibnamefont {Monroe}},\ }\bibfield  {title} {\enquote {\bibinfo {title}
  {Many-body localization in a quantum simulator with programmable random
  disorder},}\ }\href {https://doi.org/10.1038/nphys3783} {\bibfield  {journal}
  {\bibinfo  {journal} {Nature Physics}\ }\textbf {\bibinfo {volume} {12}},\
  \bibinfo {pages} {907 EP --} (\bibinfo {year} {2016})}\BibitemShut {NoStop}%
\bibitem [{\citenamefont {ACKLEY}\ \emph {et~al.}(1987)\citenamefont {ACKLEY},
  \citenamefont {HINTON},\ and\ \citenamefont {SEJNOWSKI}}]{ACKLEY1987522}%
  \BibitemOpen
  \bibfield  {author} {\bibinfo {author} {\bibfnamefont {DAVID~H.}\
  \bibnamefont {ACKLEY}}, \bibinfo {author} {\bibfnamefont {GEOFFREY~E.}\
  \bibnamefont {HINTON}}, \ and\ \bibinfo {author} {\bibfnamefont
  {TERRENCE~J.}\ \bibnamefont {SEJNOWSKI}},\ }\bibfield  {title} {\enquote
  {\bibinfo {title} {A learning algorithm for boltzmann machines},}\ }in\ \href
  {\doibase https://doi.org/10.1016/B978-0-08-051581-6.50053-2} {\emph
  {\bibinfo {booktitle} {Readings in Computer Vision}}},\ \bibinfo {editor}
  {edited by\ \bibinfo {editor} {\bibfnamefont {Martin~A.}\ \bibnamefont
  {Fischler}}\ and\ \bibinfo {editor} {\bibfnamefont {Oscar}\ \bibnamefont
  {Firschein}}}\ (\bibinfo  {publisher} {Morgan Kaufmann},\ \bibinfo {address}
  {San Francisco (CA)},\ \bibinfo {year} {1987})\ pp.\ \bibinfo {pages} {522 --
  533}\BibitemShut {NoStop}%
\bibitem [{\citenamefont {Amin}\ \emph {et~al.}(2018)\citenamefont {Amin},
  \citenamefont {Andriyash}, \citenamefont {Rolfe}, \citenamefont
  {Kulchytskyy},\ and\ \citenamefont {Melko}}]{PhysRevX.8.021050}%
  \BibitemOpen
  \bibfield  {author} {\bibinfo {author} {\bibfnamefont {Mohammad~H.}\
  \bibnamefont {Amin}}, \bibinfo {author} {\bibfnamefont {Evgeny}\ \bibnamefont
  {Andriyash}}, \bibinfo {author} {\bibfnamefont {Jason}\ \bibnamefont
  {Rolfe}}, \bibinfo {author} {\bibfnamefont {Bohdan}\ \bibnamefont
  {Kulchytskyy}}, \ and\ \bibinfo {author} {\bibfnamefont {Roger}\ \bibnamefont
  {Melko}},\ }\bibfield  {title} {\enquote {\bibinfo {title} {Quantum boltzmann
  machine},}\ }\href {\doibase 10.1103/PhysRevX.8.021050} {\bibfield  {journal}
  {\bibinfo  {journal} {Phys. Rev. X}\ }\textbf {\bibinfo {volume} {8}},\
  \bibinfo {pages} {021050} (\bibinfo {year} {2018})}\BibitemShut {NoStop}%
\bibitem [{\citenamefont {{Anschuetz}}\ and\ \citenamefont
  {{Cao}}(2019)}]{2019_yudong}%
  \BibitemOpen
  \bibfield  {author} {\bibinfo {author} {\bibfnamefont {Eric~R.}\ \bibnamefont
  {{Anschuetz}}}\ and\ \bibinfo {author} {\bibfnamefont {Yudong}\ \bibnamefont
  {{Cao}}},\ }\bibfield  {title} {\enquote {\bibinfo {title} {{Realizing
  Quantum Boltzmann Machines Through Eigenstate Thermalization}},}\ }\href@noop
  {} {\bibfield  {journal} {\bibinfo  {journal} {arXiv e-prints}\ ,\ \bibinfo
  {eid} {arXiv:1903.01359}} (\bibinfo {year} {2019})},\ \Eprint
  {http://arxiv.org/abs/1903.01359} {arXiv:1903.01359 [quant-ph]} \BibitemShut
  {NoStop}%
\bibitem [{Note2()}]{Note2}%
  \BibitemOpen
  \bibinfo {note} {Quenched disorder can be done in various platforms including
  trapped ions \cite {2017_monroe_nat2} and superconducting circuits \cite
  {2017_Roushan_Sci}}\BibitemShut {NoStop}%
\bibitem [{\citenamefont {McClean}\ \emph {et~al.}(2018)\citenamefont
  {McClean}, \citenamefont {Boixo}, \citenamefont {Smelyanskiy}, \citenamefont
  {Babbush},\ and\ \citenamefont {Neven}}]{2018_hartmut_natcom}%
  \BibitemOpen
  \bibfield  {author} {\bibinfo {author} {\bibfnamefont {Jarrod~R.}\
  \bibnamefont {McClean}}, \bibinfo {author} {\bibfnamefont {Sergio}\
  \bibnamefont {Boixo}}, \bibinfo {author} {\bibfnamefont {Vadim~N.}\
  \bibnamefont {Smelyanskiy}}, \bibinfo {author} {\bibfnamefont {Ryan}\
  \bibnamefont {Babbush}}, \ and\ \bibinfo {author} {\bibfnamefont {Hartmut}\
  \bibnamefont {Neven}},\ }\bibfield  {title} {\enquote {\bibinfo {title}
  {Barren plateaus in quantum neural network training landscapes},}\ }\href
  {\doibase 10.1038/s41467-018-07090-4} {\bibfield  {journal} {\bibinfo
  {journal} {Nature Communications}\ }\textbf {\bibinfo {volume} {9}},\
  \bibinfo {pages} {4812} (\bibinfo {year} {2018})}\BibitemShut {NoStop}%
\bibitem [{\citenamefont {Saxena}\ \emph {et~al.}(2017)\citenamefont {Saxena},
  \citenamefont {Prasad}, \citenamefont {Gupta}, \citenamefont {Bharill},
  \citenamefont {Patel}, \citenamefont {Tiwari}, \citenamefont {Er},
  \citenamefont {Ding},\ and\ \citenamefont {Lin}}]{clustering}%
  \BibitemOpen
  \bibfield  {author} {\bibinfo {author} {\bibfnamefont {Amit}\ \bibnamefont
  {Saxena}}, \bibinfo {author} {\bibfnamefont {Mukesh}\ \bibnamefont {Prasad}},
  \bibinfo {author} {\bibfnamefont {Akshansh}\ \bibnamefont {Gupta}}, \bibinfo
  {author} {\bibfnamefont {Neha}\ \bibnamefont {Bharill}}, \bibinfo {author}
  {\bibfnamefont {Om~Prakash}\ \bibnamefont {Patel}}, \bibinfo {author}
  {\bibfnamefont {Aruna}\ \bibnamefont {Tiwari}}, \bibinfo {author}
  {\bibfnamefont {Meng~Joo}\ \bibnamefont {Er}}, \bibinfo {author}
  {\bibfnamefont {Weiping}\ \bibnamefont {Ding}}, \ and\ \bibinfo {author}
  {\bibfnamefont {Chin-Teng}\ \bibnamefont {Lin}},\ }\bibfield  {title}
  {\enquote {\bibinfo {title} {A review of clustering techniques and
  developments},}\ }\href {\doibase
  https://doi.org/10.1016/j.neucom.2017.06.053} {\bibfield  {journal} {\bibinfo
   {journal} {Neurocomputing}\ }\textbf {\bibinfo {volume} {267}},\ \bibinfo
  {pages} {664--681} (\bibinfo {year} {2017})}\BibitemShut {NoStop}%
\bibitem [{\citenamefont {Du{\`o}}\ \emph {et~al.}(2018)\citenamefont
  {Du{\`o}}, \citenamefont {Robinson},\ and\ \citenamefont {Soneson}}]{rna}%
  \BibitemOpen
  \bibfield  {author} {\bibinfo {author} {\bibfnamefont {A.}~\bibnamefont
  {Du{\`o}}}, \bibinfo {author} {\bibfnamefont {M.~D}\ \bibnamefont
  {Robinson}}, \ and\ \bibinfo {author} {\bibfnamefont {C.}~\bibnamefont
  {Soneson}},\ }\bibfield  {title} {\enquote {\bibinfo {title} {A systematic
  performance evaluation of clustering methods for single-cell rna-seq data},}\
  }\href {\doibase 10.12688/f1000research.15666.2} {\bibfield  {journal}
  {\bibinfo  {journal} {F1000Research}\ }\textbf {\bibinfo {volume} {7}},\
  \bibinfo {pages} {1141} (\bibinfo {year} {2018})}\BibitemShut {NoStop}%
\bibitem [{\citenamefont {Zhang}\ \emph
  {et~al.}(2017{\natexlab{b}})\citenamefont {Zhang}, \citenamefont {Hess},
  \citenamefont {Kyprianidis}, \citenamefont {Becker}, \citenamefont {Lee},
  \citenamefont {Smith}, \citenamefont {Pagano}, \citenamefont {Potirniche},
  \citenamefont {Potter}, \citenamefont {Vishwanath}, \citenamefont {Yao},\
  and\ \citenamefont {Monroe}}]{2017_monroe_nat2}%
  \BibitemOpen
  \bibfield  {author} {\bibinfo {author} {\bibfnamefont {J.}~\bibnamefont
  {Zhang}}, \bibinfo {author} {\bibfnamefont {P.~W.}\ \bibnamefont {Hess}},
  \bibinfo {author} {\bibfnamefont {A.}~\bibnamefont {Kyprianidis}}, \bibinfo
  {author} {\bibfnamefont {P.}~\bibnamefont {Becker}}, \bibinfo {author}
  {\bibfnamefont {A.}~\bibnamefont {Lee}}, \bibinfo {author} {\bibfnamefont
  {J.}~\bibnamefont {Smith}}, \bibinfo {author} {\bibfnamefont
  {G.}~\bibnamefont {Pagano}}, \bibinfo {author} {\bibfnamefont {I.~D.}\
  \bibnamefont {Potirniche}}, \bibinfo {author} {\bibfnamefont {A.~C.}\
  \bibnamefont {Potter}}, \bibinfo {author} {\bibfnamefont {A.}~\bibnamefont
  {Vishwanath}}, \bibinfo {author} {\bibfnamefont {N.~Y.}\ \bibnamefont {Yao}},
  \ and\ \bibinfo {author} {\bibfnamefont {C.}~\bibnamefont {Monroe}},\
  }\bibfield  {title} {\enquote {\bibinfo {title} {Observation of a discrete
  time crystal},}\ }\href {https://doi.org/10.1038/nature21413} {\bibfield
  {journal} {\bibinfo  {journal} {Nature}\ }\textbf {\bibinfo {volume} {543}},\
  \bibinfo {pages} {217 EP --} (\bibinfo {year}
  {2017}{\natexlab{b}})}\BibitemShut {NoStop}%
\end{thebibliography}%



\pagebreak
\begin{widetext}

\appendix

\section{Magnus expansion of $\hat H(t)$} \label{App:ME}

In this section, we explicitly compute the terms in the Magnus expansion up to the second order correction with an arbitrary reference time $t_0$. For the first order, $\hat H_F^{(1)}$, the analytic form is the following
\begin{align}
\hat H_F^{(1)} &= \frac{1}{2iT}\int_{t_0}^{T+t_0}d\tau_1\int_{t_0}^{\tau_1}d\tau_2  \left[ \hat H(\tau_1),\hat H(\tau_2)\right] \nonumber \\
&= \frac{1}{2iT}\int_{t_0}^{T+t_0}d\tau_1\int_{t_0}^{\tau_1}d\tau_2 \left\{f(\tau_2)-f(\tau_1)\right\}\left[\hat{H}_{ave},\hat{H}_d\right] \nonumber \\
&= \frac{F \sin{\omega t_0}}{\omega} \left\{ \sum_{j=1}^{L} h_j \hat Y_j + J \sum_{j=1}^{L-1}\left(\hat Y_j \hat Z_{j+1}+\hat Z_j \hat Y_{j+1}\right) \right\}.
\end{align}
By specifying $t_0 = 0$ as in the main text, the term disappears.
Now, consider the second order term $\hat{H}_F^{(2)}$
\begin{align}
\hat{H}_F^{(2)} = & -\frac{1}{6T}\int_0^Td\tau_1\int_0^Td\tau_2\int_0^Td\tau_3  \times \left(\left[\hat{H}(\tau_1),\left[\hat{H}(\tau_2),\hat{H}(\tau_3)\right]\right]+(1\Leftrightarrow 3)\right) \nonumber
\\
= & -\frac{1}{6T}\int_{t_0}^{T+t_0}d\tau_1 \int_{t_0}^{\tau_1} d\tau_2\int_{t_0}^{\tau_2}d\tau_3 \left\{f(\tau_1)+f(\tau_3)-2f(\tau_2)\right\} \left[\hat{H}_{ave},\left[\hat{H}_{ave},\hat{H}_d\right]\right] \nonumber  \\
& -\frac{1}{6T}\int_{t_0}^{T+t_0}d\tau_1 \int_{t_0}^{\tau_1} d\tau_2\int_{t_0}^{\tau_2}d\tau_3 \left\{2f(\tau_1)f(\tau_3)-f(\tau_2)f(\tau_1)-f(\tau_2)f(\tau_3)\right\} \left[\hat{H}_d,\left[\hat{H}_{ave},\hat{H}_d\right]\right] \nonumber \\
= & \frac{\cos{\omega t_0}}{2\omega^2} \left[\hat{H}_{ave},\left[\hat{H}_{ave},\hat{H}_d\right]\right]
+ \frac{2+4\cos{\omega t_0}-\cos{2\omega t_0}}{16 \omega^2}\left[\hat{H}_d,\left[\hat{H}_{ave},\hat{H}_d\right]\right] \nonumber \\
= & \frac{2F}{\omega^2} \left\{  \sum_{j=1}^L h_j^2 \hat X_j  + 2J \sum_{j=1}^{L-1} \left(h_j \hat X_j \hat Z_{j+1} + h_{j+1} \hat Z_j \hat X_{j+1}\right) +  J^2 \sum_{j=1}^{L-1} \left(\hat X_j  + \hat X_{j+1}\right) + 2 J^2 \sum_{j=1}^{L-2} \left(\hat Z_j \hat X_{j+1} \hat Z_{j+2}\right) \right\} \nonumber \\
& - \frac{5F^2}{4\omega ^2}\left\{  \sum_{j=1}^L h_j \hat Z_j  + 2J \sum_{j=1}^{L-1} \left(\hat Z_j \hat Z_{j+1} -\hat  Y_j \hat Y_{j+1}\right) \right\},
\label{eq:hf2}
\end{align}
where in the last step we specify $t_0 = 0$. Note that $\hat Z_j \hat X_{j+1} \hat Z_{j+2}$ in the last line is the three-body interaction term which is also mentioned in the main text. The many-body long range interaction terms are common in the higher order corrections. Lastly, the higher order correction terms scale as $\left(\frac{E_c}{\omega}\right)^n$ where $E_c$ is the characteristic energy of the system.


\section{The Porter-Thomas distribution and the infinite-temperature}
\label{app:pt}

In this section, we prove that the ensemble of the states that follows the PT distribution is a mixed state with the infinite temperature. We begin by writing the state as $|\psi\rangle=\sum_{i=1}^Nc_{i}|i\rangle$, where $|i\rangle$ is a basis vector and $N$ is the dimension of the Hilbert space. A random vector in the Hilbert space means that $C_{i}=a_i+ib_i$ is a random complex number, subjected to the constrain $\sum_{i}|c_i|^2=1$. It has been shown that the distribution of the probabilities $p_i=|c_i|^2$ follows the Porter-Thomas distribution, \textit{i.e.} $\text{Pr}(p)=Ne^{-Np}$.

Let $\mathcal{S}_{\rm PT}$ be the ensemble of such states. The density-matrix representation of the ensemble is written as
\begin{align}
\rho &= \frac{1}{D}\sum_{|\psi\rangle_{\nu} \in \mathcal{S}_{\rm PT}}|\psi\rangle_{\nu}\langle \psi |_{\nu} =\frac{1}{D}\sum_{i,j=1}^N\left(\sum_{\nu =1}^{D} c_{\nu,i} c^*_{\nu,j}\right) |i\rangle\langle j |,
\end{align}
where $D$ is the number of states in the ensemble and $\nu$ is the label of the state in the ensemble. Let us first consider the diagonal elements
\begin{align}
\frac{1}{D}\sum_{\nu =1}^{D}p_{\nu,i}&=\int_{p_i=0}^1 {\rm Pr}(p_i)p_i dp_i=\int_{p_i=0}^1 Ne^{-Np_i}p_idp_i \nonumber =-e^{-N}\left(1+\frac{1}{N}\right)+\frac{1}{N} \nonumber \\
&\approx \frac{1}{N} \text{ for } N\gg 1.
\end{align}
To calculate the off-diagonal elements, we assume that $a_i$ and $b_i$ are drawn from a uniform distribution in the range $\left[-\eta/2,\eta/2\right]$ with $\eta\to \infty$. The integral over the Hilbert space is $\int_{-\infty}^{\infty}\left(\prod_{\alpha=1}^Nda_{\alpha}db_{\alpha}\right)\left(\sum_{\alpha=1}^N(a^2_{\alpha}+b^2_{\alpha})-1\right)$. Hence, the ensemble average of the off-diagonal elements is
\begin{align}
\frac{1}{D}\left(\sum_{\nu =1}^{D} c_{\nu,i} c^*_{\nu,j}\right)&=\lim_{\eta\to\infty}\frac{1}{\eta^{2N}}\int_{-\frac{\eta}{2}}^{\frac{\eta}{2}}\left(\prod_{\alpha=1}^Nda_{\alpha}db_{\alpha}\right)\delta \left(\sum_{\alpha=1}^N(a^2_{\alpha}+b^2_{\alpha})-1\right)(a_i+ib_i)(a_j-ib_j ) \nonumber \\
&=\lim_{\eta\to\infty}\frac{1}{\eta^{2N}} \int_{-\frac{\eta}{2}}^{\frac{\eta}{2}}da_i \int_{-\frac{\eta}{2}}^{\frac{\eta}{2}}db_i (a_i+ib_i)\int_{-\frac{\eta}{2}}^{\frac{\eta}{2}}\left(\prod_{\alpha\neq i}da_{\alpha}db_{\alpha}\right)\delta \left(\sum_{\alpha=1}^N(a^2_{\alpha}+b^2_{\alpha})-1\right)(a_j-ib_j )\nonumber \\
&=\lim_{\eta\to\infty}\frac{1}{\eta^{2N}} \int_{-\frac{\eta}{2}}^{\frac{\eta}{2}}da_i \int_{-\frac{\eta}{2}}^{\frac{\eta}{2}}db_i (a_i+ib_i)\mathcal{A},
\end{align}
where $\mathcal{A}=\int_{-\frac{\eta}{2}}^{\frac{\eta}{2}}\left(\prod_{\alpha\neq i}da_{\alpha}db_{\alpha}\right)\delta \left(\sum_{\alpha=1}^N(a^2_{\alpha}+b^2_{\alpha})-1\right)(a_j-ib_j )$. Note that the sign of $A$ is independent of the sign of $a_i$ and $b_i$. We then separate the double integral into four quadrants.
\begin{align}
\frac{1}{D}\left(\sum_{\nu =1}^{D} c_{\nu,i} c^*_{\nu,j}\right)&=\lim_{\eta\to\infty}\frac{1}{\eta^{2N}}\left[ \int_{0}^{\frac{\eta}{2}}da_i \int_{0}^{\frac{\eta}{2}}db_i (a_i+ib_i)+\int_{-\frac{\eta}{2}}^{0}da_i \int_{-\frac{\eta}{2}}^{0}db_i (a_i+ib_i)\right]\mathcal{A} \nonumber \\
&\quad +\lim_{\eta\to\infty}\frac{1}{\eta^{2N}}\left[ \int_{0}^{\frac{\eta}{2}}da_i \int_{-\frac{\eta}{2}}^{0}db_i (a_i+ib_i)+\int_{-\frac{\eta}{2}}^{0}da_i \int_{0}^{\frac{\eta}{2}}db_i (a_i+ib_i)\right]\mathcal{A} \nonumber \\
&=\lim_{\eta\to\infty}\frac{1}{\eta^{2N}}\left[ \int_{0}^{\frac{\eta}{2}}da_i \int_{0}^{\frac{\eta}{2}}db_i (a_i+ib_i)-\int_{0}^{\frac{\eta}{2}}da_i \int_{0}^{\frac{\eta}{2}}db_i (a_i+ib_i)\right]\mathcal{A} \nonumber \\
&\quad +\lim_{\eta\to\infty}\frac{1}{\eta^{2N}}\left[ \int_{0}^{\frac{\eta}{2}}da_i \int_{-\frac{\eta}{2}}^{0}db_i (a_i+ib_i)-\int_{0}^{\frac{\eta}{2}}da_i \int_{-\frac{\eta}{2}}^{0}db_i (a_i+ib_i)\right]\mathcal{A} \nonumber \\
&=0.
\end{align}
Hence, $\rho$ is a diagonal matrix with all diagonal elements equals to $1/N$, consistent with the ensemble with the infinite temperature.


\section{Quantum chaos in time-independent and time-dependent systems}
\label{app:chaos}
Quantum chaotic dynamics are naturally suited to benchmark quantum supremacy as they are exponentially sensitive to any small perturbations, therefore making it hard to simulate classically with approximated methods. 
Furthermore, the PTD can also be interpreted as a signature of quantum chaos as both concepts can be understood within the random matrix theory \cite{PhysRevX.8.021062}.
This link is what mainly motivates the use of the PTD as a signature of quantum supremacy~\cite{2018_hartmut_natphy}.
In the following sections, we discuss the subtle relation between quantum chaos in driven interacting systems and quantum supremacy signatures. 

One of the standard way to observe the emergence of quantum chaos in time-{\it independent} systems is by analyzing the level statistic of the underlying static Hamiltonian~\cite{2016_Alessio_AiP}. For example, writing $\hat H_{\rm ave} = \sum_n \epsilon_n \ket{\phi_n}\bra{\phi_n}$ with eigenstates $\ket{\phi_n}$ of energy $\epsilon_n$, one can define the level spacing as
\begin{equation} \label{eq:rn}
 r_n= \frac{\text{min}(\delta_n,\delta_{n+1})}{\text{max}(\delta_n,\delta_{n+1})}, 
 \end{equation}
where $\delta_n = \epsilon_{n+1}-\epsilon_n$  ($\epsilon_n\leq \epsilon_{n+1}$) is the distance between two adjacent eigenenergies. The level statistic ${\rm Pr}(r)$ is the normalized distribution of $r_n$. The system is said to be chaotic if ${\rm Pr}(r)$ exhibits level repulsion and coincides with the distribution obtained from a random Gaussian matrix. The latter is known as the Gaussian orthogonal ensemble (GOE) and reads ${\rm Pr}_{\rm GOE}(r) = \frac{27}{4}\frac{r+r^2}{(1+r+r^2)^{5/2}}$ with a mean value $\langle r \rangle_{\rm GOE} \approx 0.536$. 

In physical implementations, the Hamiltonian has more structure than a fully random Gaussian matrix.
However, within a small energy window, the eigenstates of a physical Hamiltonian $\hat H$ that exhibits GOE level statistic are expected to be equivalent to random unit vectors (i.e.~no clear structure) when written in a generic basis [e.g.~the $\{\ket{\bf{z}}\}$ basis]. 
As a consequence, for an initial state $\ket{\psi_0}$ with $\bar E = \bra{\psi_0}\hat H \ket{\psi_0}$ and $\Delta E = \sqrt{\bra{\psi_0}\hat H^2 \ket{\psi_0} - |\bra{\psi_0}\hat H \ket{\psi_0}|^2}$, any generic observable is expected to evolve toward the microcanonical ensemble prediction associated to the energy $\bar E \pm \Delta E$, which in the thermodynamic limit is equivalent to the canonical ensemble with a finite temperature $T = \hbar \bar E/k_B$~\cite{2016_Alessio_AiP}.
This result is expected from the eigenstate thermalization hypothesis~\cite{2006_Goldstein_PRL}.
In contrast, for phases of matter where the different energy eigenstates are uncorrelated, as it is the case for MBL phases, the level statistic is expected to follow the Poisson distribution $P_{\rm POI}(r) = \frac{2}{(1+r)^2}$ with $\langle r \rangle_{\rm POI} \approx 0.386$. In this case, the system fails to thermalize, and the eigenstate thermalization hypothesis breaks down.

For periodically driven Hamiltonians, a different point of view has to be adopted. Since it is in general impossible to find a closed form for the time-independent Floquet Hamiltonian $\hat H_F$, one has to focus on the unitary evolution operator $\hat U(T) = \sum_n e^{-i\theta_n} \ket{\psi_n}\bra{\psi_n}$. The same approach follows except that it is the phase statistic that is accessible, i.e.~$\delta_n = \theta_{n+1}-\theta_n$  ($\theta_n\leq \theta_{n+1}$ and $\theta_n \in [0,2\pi]$) in eq.~\eqref{eq:rn}, which leads to qualitatively different signatures. 
For example, in the limit of infinite driving frequencies $\omega\rightarrow\infty$, $\hat H_F$ converges to $\hat H_{\rm ave}$ and $\theta_n = \epsilon_n T \ll 2\pi$, so that the phase $\theta_n$ and energy $\epsilon_n$ statistics are identical.
In the intermediate regime where $\omega$ is finite but $\hat H_F$ is local and extensive (i.e.~convergent Magnus expansion), the folding of the energy spectrum $\theta_n \in [0,2\pi]$ can drastically change the statistic as eigenstates far apart in energy are expected to have little correlations. A $\hat H_F$ that leads to level repulsion (GOE energy statistic) can exhibit Poisson phase statistics within the $[0,2\pi]$ range.

Finally, in the low-frequency limit where the Magnus expansion diverges, $\hat H_F$ can be non-local and so correlations between {\it every} eigenstates can arise, leading to phase repulsion despite the energy spectrum folding. In that case, the corresponding phase statistic tends toward the level statistic of a random unitary matrix, known as the circular orthogonal ensemble (COE).
In analogy with the static scenario, the eigenstates of $\hat U(T)$ are expected to be equivalent to random unit vectors, but this time, over the entire Hilbert space. 
Consequently, any generic observable is expected to evolve toward the micro-canonical ensemble with $\Delta E \rightarrow \infty$ which in the thermodynamic limit corresponds to the canonical ensemble with an infinite temperature~\cite{2014_Rigol_PRX}.


\section{Phase diagram of the driven Ising chain.}\label{app:pd}

In this section, we analyze in more details the level statistic of the unitary evolution $\hat U$ as introduced in Eq.~\eqref{eq:u} in the main text, more precisely we analyze $\langle r \rangle$ and the output state distribution as a function of the driving frequency $\omega$ and the disorder strength $W$. 

In Fig.~\ref{Fig:2DPT} (a), where $\langle r \rangle$ is plotted, we can distinguish three distinct phases: one following the COE statistic in the small frequency and weak disorder regime, one following the GOE statistic for high frequencies and weak disorders and finally, one following Poisson statistic for large disorders. In the high-frequency limit, the Floquet Hamiltonian converges to $\hat H^{(0)}_F$, and the transition between chaotic and MBL phases as the disorder increases is as described in the Appendix~\ref{app:chaos}. In the low-frequency limit, the system responds to the drive, and the Magnus expansion diverges, leading to the COE statistic, as labeled in panel (a).

In Fig.~\ref{Fig:2DPT} (b), where the KL divergence of the output state distribution with the PT distribution is plotted, we can see that for diverging Magnus expansion, the output state follows the PT distribution. Besides, as described in the main text, undriven thermalized phases (high-frequency limit) does not allow to reach the PT distribution. 

Finally, we note that for $\omega \rightarrow 0$, the phase statistic follows the Poisson distribution and the output state diverges from the PT distribution. In this limit, the dynamics becomes approximately adiabatic, and the dynamics is well described by short-range interactions.

 \begin{figure}
\includegraphics[width=0.6\textwidth]{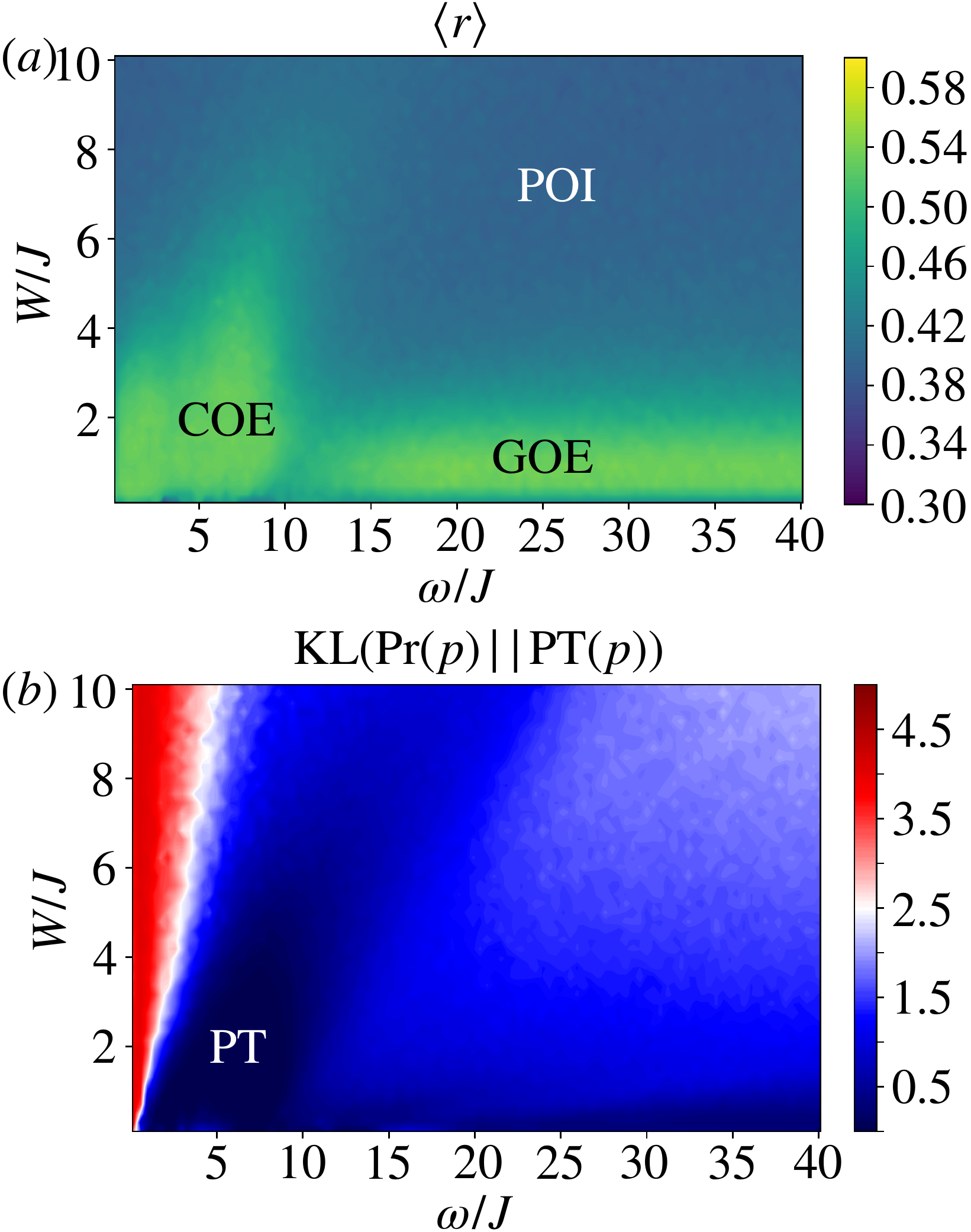}
\caption{\textbf{Phase diagram of the driven Ising chain:} \textbf{(a)} the averaged level spacing $\langle r\rangle$  as a function of $W/J$ and $\omega/J$.\textbf{(b)} $\text{KL}(\text{Pr}(p)\parallel\text{PT}(p))$ as a function of $W/J$ and $\omega/J$. ($m=10, F = 2.5J$, $L=9$) }
\label{Fig:2DPT}
\end{figure}


\section{Learning performance for quenched dynamics without time-dependent driving on $\hat{X}$}
\label{app:nodrive}
The learning performance, characterized by the value of the cost function after the training, as a function of $W$ is shown in Fig.~\ref{fig8} with zero and non-zero global driving along the $X$ axis. It shows that the learning performance degrades in the MBL phase when $f(t)=0.5$. While in the chaotic regime, the system learns better when $f(t)=0.5$. This is expected as the system becomes less chaotic compared to the non-zero driving case.

 \begin{figure}[b]
\includegraphics[width=0.6\textwidth]{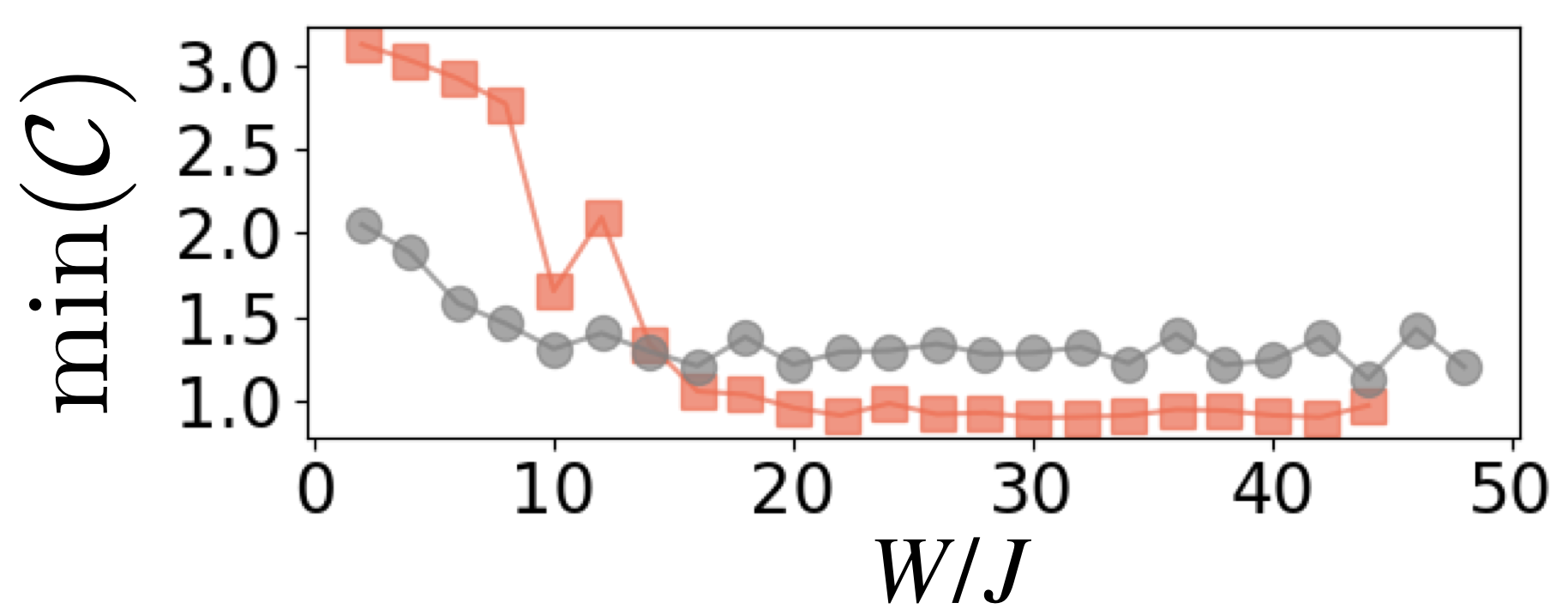}
\caption{\textbf{Phases of matter and the learning performance:} The cost function at $m=100$ as a function of disorder strength $W$ with $f(t)=0.5(1-\cos(\omega t))$ (red squares) and $f(t)=0.5$ (gray circles). The results are averaged over 50 dataset, \textit{i.e.}, 50 realizations of $\{a_i,b_i\}$. ( $\omega=8J$, $F=2.5J$,$k_BT=J$, $D = 140$, and $L=9$).}
\label{fig8}
\end{figure}

\end{widetext}

\end{document}